\documentclass[11pt,a4paper]{article}  
\usepackage{graphics,graphicx,amsfonts,amsmath,amssymb,longtable,xcolor}
\definecolor{darkgreen}{rgb}{0.0, 0.5, 0.0}
\definecolor{aubergine}{rgb}{0.2, 0.0, 0.5}
\usepackage{color}
\usepackage{natbib}
\bibpunct{(}{)}{;}{a}{}{,} %
\usepackage{fancyhdr}
\usepackage[utf8]{inputenc}
\DeclareUnicodeCharacter{223C}{\realtilde}
\PassOptionsToPackage{hyphens}{url}\usepackage[colorlinks=true, linktocpage=true, linktoc=all, citecolor=darkgreen, anchorcolor=aubergine, urlcolor=blue, linkcolor=aubergine]{hyperref}

\defcitealias{SI99}{SI99}
\defcitealias{ALMA-TH}{ALMA-TH}

\title{Self-calibration and improving image fidelity for ALMA and other radio interferometers}
\author{A. M. S. Richards$^1$, E. Moravec$^{2,3}$, S.~Etoka$^1$, E. B. Fomalont$^4$,\\ A. F. P\'{e}rez-S\'{a}nchez$^5$, M. C. Toribio$^6$ and R. A. Laing$^7$\\
  $^1$ JBCA, Department of Physics and Astronomy, University of Manchester, UK\\
  $^2$ Czech ARC Node, Astronomical Institute of the Czech Academy of Sciences,\\ Czech Republic\\
  $^3$ Green Bank Observatory, West Virginia, USA\\
  $^4$ NRAO, Edgemont Road, Charlottesville, VA 22903, USA\\
  $^5$ Leiden Observatory, Leiden University, PO Box 9513, 2300 RA Leiden, The Netherlands.\\
  $^6$ Department of Space, Earth and Environment, Chalmers University of Technology,\\ Onsala Space Observatory, 439 92 Onsala, Sweden\\
  $^7$ SKA Organization, Jodrell Bank, Cheshire, SK11 9DL, UK
}
\begin{document}
\begin{center}\textbf{\large{ALMA Memo 620}} \end{center} \vspace*{-1cm}
{\let\newpage\relax\maketitle}


\tableofcontents
    {\abstract{This manual is intended to help ALMA and other interferometer users improve  images by recognising limitations and how to overcome them and deciding when and how to use self-calibration.        The images provided by the ALMA Science Archive are calibrated using standard observing and data processing routines, including a quality assurance process to make sure that the observations meet the proposer's science requirements. This may not represent the full potential of the data, since any interferometry observation can be imaged with a range of resolutions and surface brightness sensitivity. The separation between phase calibration source and target usually limits the target dynamic range to a few hundred (or 50--100 for challenging conditions) but if the noise in the target field has not reached the thermal limit, improvements may be possible using self-calibration.  This often requires judgements based on the target properties and is not yet automated for all situations. This manual provides background on the instrumental and atmospheric causes of visibility phase and amplitude errors, their effects on imaging and how to improve the signal to noise ratio and image fidelity by self-calibration.  We introduce the conditions for self-calibration to be useful and how to estimate calibration parameter values for a range of observing modes (continuum, spectral line etc.). We also summarise more general error recognition and other techniques to tackle imaging problems. The examples are drawn from ALMA interferometric data processed using CASA, but the principles are generally applicable to most similar cm to sub-mm imaging.
}}

\section{Introduction}
\subsection{Aims and outline of this manual}

Self-calibration involves correcting the visibility phases
and amplitudes of a source by comparing the  visibility data with a model of the source itself  (usually derived in previous imaging of that source), which is used to  estimate the
corrections needed to make the data resemble the model more closely. The term is usually used to describe improving the imaging fidelity of science targets. 

Phase referencing is the technique of deriving phase and amplitude solutions for a source  with an accurate position and well-known (usually point-like) structure, within a small angular separation (typicially a few degrees) from the target, observed every few minutes alternately with the target.  These solutions are then applied to the target, which
assumes that the same atmospheric and instrumental
errors affect the phase calibration source and the target, so the same
corrections work for both (regardless of the source structure). 
In
reality, the difference in time and angular separation between the
phase calibrator and the target means that the atmosphere, and possibly other causes of error, are somewhat
different and the target data are not perfectly corrected by phase
calibrator solutions.  If the target S/N (signal to noise) is
high enough then more accurate corrections can be derived using
self-calibration. This often takes a number of cycles of imaging and
calibration to improve the model and hence the corrections if the
target has a complex structure.

The aim of this manual is to relate practical self-calibration to more
formal explanations of the origins and correction of interferometric
errors, e.g. \cite{TMS}.  As such, we provide references rather than
rigorous derivations.  The relationships developed here are not rigid
recipes but are intended to guide setting parameter values for
self-calibration, or for explaining the techniques in writing up
results. Self-calibration is well explained  in Ch. 10 (Cornwell \& Fomalont) of the proceedings of the
NRAO/NMIMT Synthesis Imaging summer school 1998 (\citet{SI99}, hereafter \citetalias{SI99}), and in a lecture from the 2017 school  \citep{Brogan18} on advanced calibration;
the challenge here is to update/expand the parts relevant to self-calibration.  This manual uses ALMA examples, which are  relevant other arrays, albeit more so for other mm-wave arrays and cm-wave arrays like the VLA in extended configurations.  The principles are similar even for  low-frequency, wide-field dipole arrays or space VLBI but the additional considerations for such extreme situations are not covered, nor are terrestrial applications such as Geodesy.

This Section provides context, summarises why phase (self-)calibration in particular is important and explains some basic jargon.  Section~\ref{sec-causeeffect} describes the causes of errors in radio interferometry, especially relevant to (sub-)mm observations, and quantifies their effects. Section~\ref{sec-afterphref} concentrates on cause and effect of the residual errors affecting a target after phase-reference and other calibration has been applied.  These Sections provide the background to understand how  self-calibration works, but the concepts are best understood through experience so you can start with Section~\ref{sec-practical} which provides a quick-start guide for continuum self-calibration  followed by expressions and examples (in CASA) to help decide when to self-calibrate, how to derive suitable parameter values (such as solution interval), how to assess/improve the quality of the results and when to stop. Section~\ref{sec-sc} covers a wide range of situations such as spectral line self-calibration and Section~\ref{sec-errorrecognition} illustrates image errors and how to tell if self-calibration might reduce these.  The Appendices provide links to other resources and some examples of CASA task settings and code fragments.

\subsection{Interferometric calibration}

Correlated radio interferometry data are recorded for each sample as
complex visibilities, $V$,   given by
\begin{equation}
  V={\mathbf{J}}_{ij}V_0 e^{i \phi}
\label{eq_V0}
\end{equation}
where $V_0$ and  $\phi$ are the `true' amplitude and phase of each visibility 
(per baseline between antennas $i$ and $j$, unit time, spectral channel and polarisation product).  These contain atmospheric and instrumental errors collectively grouped as ${\mathbf{J}}_{ij}$. Antenna-based errors are formed by taking the outer product of the response from antenna $i$ and the conjugate of the response from antenna $j$, so for these terms ${\mathbf{J}}_{ij}$ is shorthand for ${\mathbf{J}}_{i}  \otimes  {\mathbf{J^*}}_{j}$.

Interferometric calibration is described formally by solving appropriate terms in the
Measurement Equation as developed in \citet{Sault1996}, see also e.g.   \citetalias{SI99} Ch. 32 (Sault \& Cornwell) and \citet{TMS} Ch. 4.   The most relevant terms for self-calibration are:
\begin{eqnarray}
  \boldsymbol{V}\!_{ij,\nu}(u,v) &=& \int\! \mathbf{E}_{ij}  \mathbf{T}_{ij} \mathbf{P}_{ij}
  \boldsymbol{I}_{\nu}(l,m)e^{-2{\pi}i(u_{ij}l+v_{ij}m)}dldm  \nonumber \\ 
&&\times\: \mathbf{B}_{ij}\mathbf{G}_{ij}  \mathbf{M}_{ij}   
\label{eq-me}
\end{eqnarray}
$\boldsymbol{V}\!_{ij,\nu}$ represents the complex visibility on a baseline between antennas $i$ and $j$,  at frequency $\nu$, in the $uv$ plane. 

The right hand side represents the Fourier transform of the brightness distribution $\boldsymbol{I}_{\nu}$ in the sky  ($lm$) plane, multiplied by errors.
The field of view of a single pointing of an array is determined by the resolution, termed the primary beam, of the individual antennas (assuming they are all the same size). The effective primary beam (PB) is usually taken as the extent of the response down to some fraction such as 0.5 or 0.2 sensitivity, see  \citetalias{ALMA-TH} ch 3.
The first row of terms are known as direction-dependent errors as the PB response may differ with angle or distance from the pointing centre.
 The next row represents  multiplicative errors which are (usually) independent of the position of the source within the PB. These errors collectively contribute to  ${\mathbf{J}}_{ij}$ in Eq.~\ref{eq_V0}; there are many more but the individual terms shown here are those most relevant to self-calibration:
\begin{itemize}
\item Antenna-based
\begin{itemize}
\item[$\mathbf{E}$] `Electric' Antenna voltage  pattern including primary
  beam and related effects.  
\item[$\mathbf{P}$] Parallactic angle rotation (i.e. the change
  in the orientation of the feeds with respect to the sky as an
  alt-az telescope tracks a target).
\item[$\mathbf{T}$] `Tropospheric' and other atmospheric 
  effects on the gain  manifested as time-dependent scaling of 
  amplitudes and drifts of phases. 
\item[$\mathbf{B}$] Bandpass response affecting amplitude and phase.
\item[$\mathbf{G}$] Gain distortions including scaling due to instrumental effects (e.g. in signal transmission).
\end{itemize}
\item Baseline-based
\begin{itemize}
\item[$\mathbf{M}$] Multiplicative, baseline errors  
  e.g. introduced in the correlator, associated with specific baselines but not all baselines to the antennas involved.  These can affect phase and
  amplitude as a function of time and/or frequency and are sometimes known as non-closing errors (Sec.~\ref{sec-baseline}).
  Similar bandpass-related terms are known as $\mathbf{M}^{\mathrm {f}}$.  
\end{itemize}
\end{itemize}
This omits polarisation-related effects which are normally corrected before self-calibration.  $\mathbf{E}$ and $\mathbf{P}$ are not normally changed by self-calibration either, but the PB correction must be applied (only) at the end of self-calibration (Sec.~\ref{sec-imsc}). 
See Secs.~\ref{sec-practical} and \ref{sec-gaincal} for more explanation and specifics of the implementation in CASA, including Appendix~\ref{ap-gaincal} for the use of {\tt gaincal 'T'} and {\tt 'G'} parameters which are used to denote  corrections derived for the receiver polarisations (e.g. X, Y) averaged and separately, respectively, regardless of origin.

A basic guide to parameter settings for ALMA is
provided by the CASA introduction to
self-calibration, along with  a  recent NRAO Summer School lecture,
\citet{Brogan18} and material for schools ``Self-calibration and advanced imaging (Bologna 2017)'' and
``I-TRAIN 6: Improving image fidelity through self-calibration'' which also
provide links to the VY
CMa data and scripts used for examples in this manual (see Appendix~\ref{ap-schools}).
More
scripts to use for practice or as templates are available in other the CASA
Guides and
other online guides and resources, given in Appendix~\ref{resources}.

Most examples are designed for ALMA data wherein the sky and other causes of signal distortion are the same across the field of view of each antenna (isoplanaticism), so the direction-dependent aspects of the terms given in Eq.~\ref{eq-me} can be ignored.  This is not the case for wide fields at frequencies below a few GHz. Specific problems also arise for ALMA (or any) observations of the Sun and other bright, beam-filling objects. These issues are touched on in Secs.~\ref{sec-wide} and ~\ref{sec-ephemeris}; you should consult specialised (e.g. LOFAR or Solar) guides for more information, see Appendix~\ref{ap-casaguides}.

\subsection{The effects of phase errors on image fidelity}
\label{sec-phase-fidelity}

\begin{figure}
  \includegraphics[width=17cm]{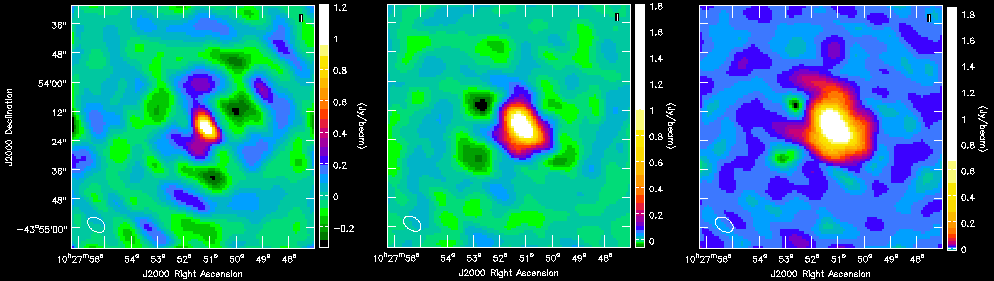}
\caption{\small NGC 3256 (from CASA Guide): Images left to right (note
  different flux scales): After applying phase calibrator solutions,
  phase errors dominate giving anti-symmetric errors with   positive
  artefacts opposite negative artefacts). After phase
  self-calibration, the artefacts are at a reduced level, dominated by amplitude
  errors, producing the symmetric artefacts seen. After phase and amplitude
  self-calibration, small anti-symmetric artefacts remain.}
\label{n_allerrors.png}
\end{figure}

Phase errors generally have the most severe effects for ALMA data. 
Using a simple point source observed at the pointing centre as an example, its visibility  amplitude   $V_0$ should be constant and the same on all baselines and the phase $\phi$  should be flat at zero (see Eq.~\ref{eq_V0}).  If the phase  has a Gaussian error distribution with rms variation
of $\phi_{\mathrm{rms}}$ (in radians) over the interval used for imaging then the
image amplitude is given by
\begin{equation}
\langle V \rangle = V_0 \langle e^{i \phi_{\mathrm{obs}}}\rangle = V_0
e^{-\phi^{2}_{\mathrm {rms}}/2}
\label{phnoise}
\end{equation}
where angle brackets represent the expectation value, see e.g. \citet{TMS} Ch. 7.2.8 for a derivation.
Thus a random phase error of  $\phi_{\mathrm{rms}} = 10^{\circ}$ = 0.1745 rad gives $\langle V \rangle = 0.98 V_0$,
i.e. a 2\% loss of coherence and reduction in amplitude (often referred to as decorrelation); however a
$65^{\circ}$ phase error will reduce the peak by almost 50\%.  The
power has to go somewhere, therefore the off-source noise can be noticeably increased by severe decorrelation.

The complex visibility function of Eq.~\ref{eq_V0} can be expressed
as $V=V_0 (\cos \phi + i \sin \phi)$. The Fourier transform of the
imaginary part, the sine function (corresponding to phase), is
anti-symmetric delta functions, alternately positive and negative (see
\citetalias{SI99}, Ch.~15 (Ekers) fig. 15.1).  This implies that
phase errors produce anti-symmetric image errors, see Fig.~\ref{n_allerrors.png}. Analogously,
amplitude errors produce symmetric image errors.

\subsection{Conventions and terminology}

We  introduce briefly the structure and terminology used for ALMA
observations used in the manual. Please see the ALMA Technical
Handbook for the current cycle (currently 9),  \citet{ALMA-TH}, hereafter \citetalias{ALMA-TH}
for all details of ALMA observing and an overview of
instrumentally-derived corrections and phase referencing.  A spectral
window (spw) is a spectral region selected for observation; up to 4 can be observed at once at the full spw width of 2 GHz. In dual polarisation a full-width spw can be
correlated using Time Domain Multiplexing (TDM) into 128 $\times$ 15.625 MHz
channels. Alternatively, 2 GHz or narrower spw can be correlated using Frequency Domain
Multiplexing (FDM) into up to 4096 channels, usually averaged by factors of at least 2 (or 4, 8, ...). The number of channels (pre-averaging) is halved  for full polarisation observations.  
The first and last 1/32 channels are typically
trimmed. 

An ALMA Execution Block (EB) is a self-contained set of observations
of one or more targets and calibrators, which may be repeated multiple times.  A
spectral scan consists of two or more EBs identical except for offsets
in frequency to produce wider spectral coverage.

We often refer to a field rather than a source, because in imaging and
calibration all detectable emission in the field of view 
should be included, not just the source at the telescope pointing centre. The relevant field of view is usually taken as  the half-power point of the PB (FWHM, full width half maximum) but may be much larger if there are any sources bright enough to produce sidelobes at, say, 10\% or even 1\% sensitivity (such as the parent planet in observations of a moon). 
For
many ALMA observations, especially with the main array of 12-m dishes, there is nothing else detectable but for
observations which are
sensitive to low surface brightness emission (e.g. using the ACA, Atacama Compact Array, also with a wider field of view from 7-m dishes) there
is often much extended emission when looking at nearby galaxies or star-forming regions.

The shortest data averaging time, i.e. the duration of a
single visibility (as defined in Eqs.~\ref{eq-me},~\ref{eq_V0}) is the integration, typically $\sim$6 sec for
calibrated 12-m array ALMA data (2 sec on long baselines) and $\sim$10 sec for the ACA.
A scan is the length of time spent continuously on a particular source e.g. the target. Typically there are source changes every few minutes, e.g. in phase referencing there will be a series of scans of 2 : 5 : 2 mins on the phase calibrator : target : phase calibrator (etc.) which should always start and finish with the phase calibrator.  At sub-mm frequencies and/or on long baselines the scans are shorter.
Occasionally, a very long `stare'
on a single field  (such as for rapidly variable sources or for tests of phase stability etc. as in \citealt{Maud2016}) may be divided into scans without a source change.

Visibility data is customarily plotted on the $uv$ plane, and, for
example, $uv$ distance means distance in the visibility plane, i.e. projected baseline length in wavelength units. A set
of visibility data has at least two `centre's: the pointing centre,
which is where the telescopes pointed, which cannot (for dishes like
ALMA) be changed retrospectively, and the phase centre, where perfectly-calibrated
emission from a specific celestial direction would be exactly
in phase.  By default these are the same but the phase centre can be
changed by rotating the visibility phases (adding the required phase
offsets) to anywhere within the field of view, for example to correct
for a small offset between two observations to be combined.

The receiver system is sensitive to two polarisations, X and Y in the
case of ALMA.  These are combined in the correlator to make
parallel hand correlations XX and YY (occasionally just one correlation) and,
optionally, the cross hands XY and YX (in which case the maximum number of channels per spw is halved to 2048).
Self-calibration is normally performed
using the total intensity (XX+YY) image model, deriving
antenna-based phase corrections for X and Y (these can be averaged, for weak sources or amplitude self-calibration, see Secs.~\ref{sec-ampscal}, \ref{sec-polarisation} and~\ref{sec-avg}).

Historically, the term `gain' referred to the increase in antenna power recorded
on-source.  It is now often used to refer to correction factors derived for
the visibilities, i.e. a gain table, which can contain phase and/or
amplitude calibration (as in the formalism in Eq.~\ref{eq-me}).

All examples in this manual refer to CASA\footnote{\url{https://casa.nrao.edu/}}, in which,
in calibration, the complex visibilities are divided by the complex factors recorded in the gain tables: visibility
amplitudes
are divided by the corrections, and the phase corrections are
subtracted from the visibility phases. Thus, an anomalously low
visibility amplitude should correspond to a low value for the derived
correction (this is the opposite of the AIPS convention).
Imaging examples refer to {\tt tclean} and 
it is assumed that you have some familiarity with CASA and calibrating
and imaging mm- or cm-wave interferometry data; see a basic CASA Guide (links in Appendix~\ref{ap-casaguides})
first if necessary.

Abbreviations and acronyms are listed in Appendix~\ref{acronyms}.
Subscript ${\mathrm{rms}}$ generally denotes an observed error e.g. in
phase, $\phi_{\mathrm{rms}}$, whilst $\epsilon$ generally indicates a
predicted or modelled contribution to the total error
e.g. $\phi_{\epsilon,\mathrm{trop}}$.

\section{Origins and effects of interferometric errors}
\label{sec-causeeffect}

Interferometric calibration is derived and applied at several stages, see \citetalias{ALMA-TH} for more details of the standard processes for ALMA:
\begin{enumerate}
  \item {\bf Calibration during observations:} 
Some calibration is calculated from known or approximated properties of the array or atmosphere in `real time' and applied during observations, such as updates to pointing offsets.
\item{\bf Calibration before/during correlation:}
This includes the bulk delay correction for  the effect of atmospheric refraction which reduces the magnitude of post-correlator phase corrections per solution interval, ideally $<|2\pi|$. ALMA uses known observing parameters (array geometry, elevation,
frequency etc.) to make an approximate prediction.   
\item{\bf Corrections derived from instrumental measurements, applied after correlation:}
Other  measurements made during observations are recorded and used to derive corrections during later processing e.g. in the pipeline. The system temperature
($T_{\mathrm{sys}}$, Eq.~\ref{eq-tsys}) is measured every few minutes.
 For the 12-m array, the  precipitable water vapour (PWV) column above each antenna is measured every few
seconds using water vapour radiometry (WVR) at 183 GHz, and used to estimate the
atmospheric refraction at the observing frequency. A
look-up table of known antenna position corrections is provided. Antenna positions and the flux densities of standard calibrators are measured using separate observations and provided in the recorded data, but may occasionally be updated later.
\item{\bf Calibration derived from astronomical calibration sources:} QSO or other sources are used for bandpass, flux scale, phase reference and sometimes other calibration e.g. polarisation.  These are observed at the phase centre and have good models 
 -- typically point sources of specified flux density, or with a model provided such as derived from from an ephemeris for Solar system objects. The data are calibrated by comparison with the source model, a process analogous to self-calibration but using an a-priori model instead of imaging during calibration (but see Sec.~\ref{sec-nophref} for when this is not adequate).  
These stages are performed using standard calibration (QA2) scripts and/or the ALMA Pipeline; most other observatories have analogous procedures.   
\item{\bf Self-calibration:} This term is normally used to describe further calibration of science targets, after applying the corrections derived from other measurements and sources as above and is usually carried out by the proposer's or archive user's team.   
\end{enumerate}
ALMA correlates data directly as it is observed and calibrations 1. and 2. are applied by ALMA before the data is recorded, i.e. they are irreversible.  Later corrections can be made for small residual errors but not severe  errors which change rapidly (e.g. phase error  $\gg \pi$ radians)  within the shortest solution interval feasible for later calibration.  This is in contrast to some VLBI where data are recorded at each telescope and parameters can be corrected before correction, including adjusting the apparent phase centre (as long as it remains within the PB); this is {\em not} the case for ALMA.  Calibrations 3., 4., 5. can be adjusted and re-applied multiple times if necessary.

Self-calibration primarily corrects residual time-dependent phase and amplitude
errors remaining after calibrations 3. and 4., such as by the pipeline or manual QA2. The dominant cause of errors in ALMA observations is usually
the troposphere, especially on longer baselines and/or at higher
frequencies, or in less good weather. Other atmospheric effects can
become noticeable when seeking high dynamic range/high sensitivity
images and  antenna position and pointing
errors can  be mitigated.  These effects are described below.

\begin{figure}[t]
\includegraphics[width=12cm]{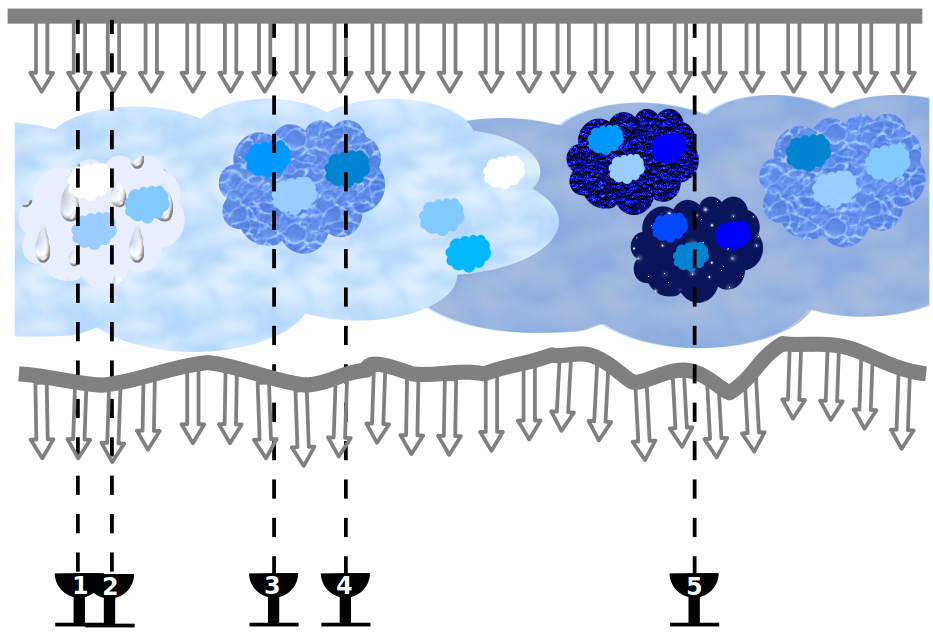}
\caption{\small Cartoon  (inspired by \citetalias{SI99} Ch.~28 Carilli,
  Carlstrom \& Holdaway) of a distant mm-wave wavefront originating above the Earth's atmosphere. Tropospheric water vapour, with structure on multiple scales, refracts the rays, the main effect  being a delay (so rays which were in phase become out of phase) with a much smaller bending effect. Dry air also exerts some refraction which has more impact  on larger scales. Antennas 1 and 2 are so close together that the refraction above them is almost identical. Comparing their location to antennas 3 and 4, there are appreciable differences due to both vertical and horizontal water vapour content fluctuations but with some large scale similarities. Finally, for antenna 5, which is much more distant than the vertical depth of the troposphere, there is no appreciable correlation of atmospheric conditions with respect to the other antennas. }
\label{atmos1.png}
\end{figure}
 
\subsection{Phase errors}

This section provides the background to the origin of phase errors (their practical consequences and quantification are covered in~\ref{sec-qpr}).  These include general atmospheric effects; ALMA avoids observing in high winds, rain and snow, which are fortunately rare but have a  severe effect. For other arrays, clouds have a small effect at frequencies $\nu\lessapprox$15 GHz but snow on the telescope or water/ice on the sub-reflector or receiver cover causes aberration and absorption.  Strong, and especially gusty, winds degrade the pointing accuracy as well as causing rapid atmospheric fluctuations.  The main effects are dealt with in more detail in the following subsections.

  \subsubsection{Tropospheric phase errors}
  \label{sec-trop}
The troposphere extends to 10--15 km above sea level, taken as having a typical height of 5 km above Chajnantur.
Tropospheric phase errors are mainly due to refraction by water vapour although  the structure of dry air important on scales longer than a few km. Figure~\ref{atmos1.png} is a cartoon illustrating how the tropospheric water vapour distribution becomes significantly less correlated the longer the baseline. This is quantified  as explained in \citetalias{SI99} Ch.~28 (Carilli,
Carlstrom \& Holdaway) and summarised by \citet{Brogan18} sec. 3.1.
The typical wind speed above ALMA is 5--10 m s$^{-1}$, or 18--36 km hr$^{-1}$.

  The refractive index of water is constant over almost all the ALMA bands (``non-dispersive''), giving a linear relationship between the phase error and  frequency
\begin{equation}
\phi_{\epsilon,\mathrm{trop}} \propto \nu w/T_{\mathrm{atm}}
\label{eq-linear}
\end{equation}
for a given PWV column $w$ and atmospheric temperature $T_{\mathrm{atm}}$, both of which are measured. The use of WVR-derived corrections relies on this. The water vapour content and thus refraction varies with time and the magnitude of the resulting phase errors  scales with increasing baseline length and increasing
frequency.  
For typical ALMA conditions,
the phase error is given
  by
\begin{equation}
  \phi_{\epsilon,\mathrm{trop}}=KB^{\alpha_{\mathrm K}}/\lambda
  \label{eq-Kolmogorov}
\end{equation}
The
wavelength $\lambda$ is in mm and the baseline length is $B$ km.
$K$ is the Kolmogorov coefficient
for tropospheric turbulence (in the same units as
$\phi_{\epsilon,\mathrm{trop}}$, here, degrees).
$\alpha_{\mathrm K}$ is predicted to be 5/6 for $B<1$ km (where the antenna
separation is less than the depth of the tropospheric layer (baselines 1-3, 1-4, 2-3, 2-4 in Fig.~\ref{atmos1.png}) or 1/3
for $1<B<10$ km (all baselines to antenna 5), falling towards zero on very long baselines if the
atmosphere behaves independently above each telescope.
$K$ was expected to be 100 for raw ALMA
data. Tests at mm-wavelengths on baselines of a few km, under conditions of a few mm PWV,
show that  $K\lesssim50$ can be attained when WVR or
other good
phase corrections are possible.

\begin{figure}[h]
\includegraphics[width=12cm]{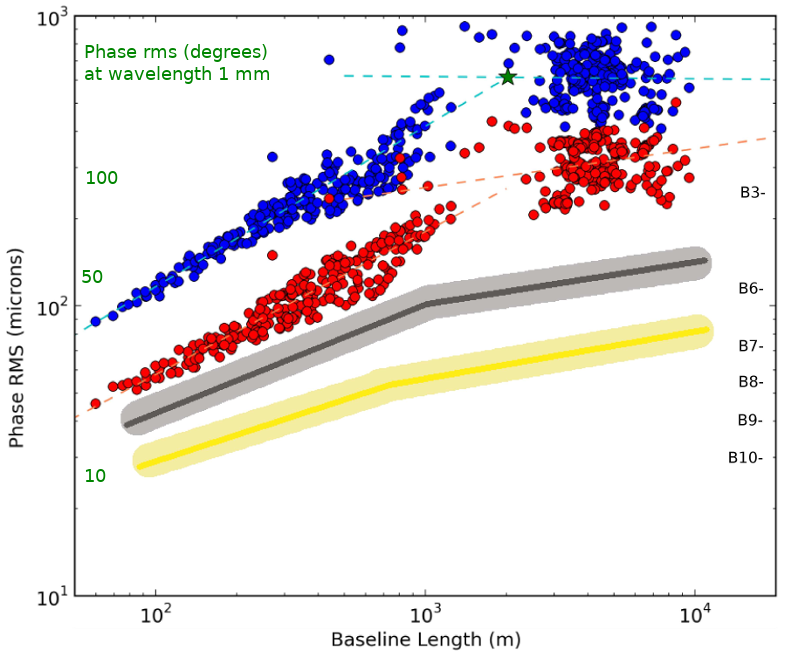}
\caption{\small Temporal phase variation due to atmospheric variations as a function of baseline length,
  adapted from ALMA Memo 606 \citep{Maud2016} fig 1. A QSO was
    observed at PWV $\sim$0.6 mm. The phase rms before and after applying
    WVR corrections is shown in blue (upper circles) and red (lower
    circles), respectively. The dark grey
    and yellow lines are the inferred improvements in phase rms from
    referencing to a phase calibrator source within 3$^{\circ}$ on
    timescales of 2 min and 20 sec, respectively. The band numbers
    listed on the right mark the phase rms values in $\mu$m required
    for 30$^{\circ}$ phase rms in each frequency range, see
    Eq.~\ref{eq-micron-deg} for conversion to phase in degrees.}
\label{ALMAMemo606Fig1.png}
\end{figure}

An ALMA study  \citep{Maud2016}
shows that in these observations the rate of increase
of phase errors is, as predicted,  much less steep for baselines longer than a few km, see Fig.~\ref{ALMAMemo606Fig1.png}. The phase
rms is roughly halved by applying WVR corrections. The phase rms is
given in terms of the path length fluctuation $\phi_{\mathrm{rms
    \mu}}$ in micron, at wavelength $\lambda$ in mm, and assuming that $\phi_{\epsilon,\mathrm{trop}}$ dominates, this is related to the observed phase rms in degrees 
$\phi_{\mathrm{rms}}$ by
\begin{equation}
  \phi_{\mathrm{rms}} =
  360^{\circ}\times \phi_{\mathrm{rms \mu}}\times 10^{-6}/(\lambda \times
  10^{-3}).
  \label{eq-micron-deg} 
  \end{equation}
Thus, from Fig.~\ref{ALMAMemo606Fig1.png}, typical raw phase errors
are 50 -- few 100 degrees and residual phase errors after phase
referencing and applying the WVR correction are a few tens degrees at
$\lambda$ 1 mm.

  A few frequency ranges in ALMA bands may show non-linearity of delay with frequency due  to strong telluric lines.
  The accuracy of corrections derived from WVR measurements is affected by clouds  (i.e. water droplets, not vapour), see  \citet{Maud17} and {\tt Remcloud} in Appendix~\ref{ap-other}. At the other extreme, if the PWV is very low, dry air turbulence is dominant, especially on scales over a few km.

\subsubsection{Ionospheric errors}
\label{sec-iono}
The ionosphere extends outwards from about 50 km above sea level, as the atmosphere becomes increasingly ionised.    
 At a given location and observing
angle, the ionospheric refraction is directly proportional to the TEC
(total free electron content along the ionospheric path in $10^{16}$
electrons m$^{-2}$). Refraction of radio waves by  free electrons is proportional to $\nu^{-2}$ and is often assumed to be negligible at wavelengths shorter than a few
cm. However, investigations for VERA (\citealt{Jung11};
\citealt{Nagayama20}) showed that -- especially around dawn and dusk --
the difference between the electron density (measured as $\Delta
\mathrm{TEC}$) above well-separated antennas can be enough to cause
very significant delays at 22 GHz. An example  in \citet{Jung11} fig. 3 shows that at 22 GHz, for  $\Delta \mathrm{TEC}=3$ the residual phase error is up to $\sim60^{\circ}$ so $\phi_{\mathrm{rms}\mu} \approx 2200\mu$m (Eq.~\ref{eq-micron-deg}).  Scaling $\phi_{\mathrm{rms}\mu}$ from 22 GHz to 100 GHz by $\nu^{-2}$ 
the delay would be 0.1 mm at 100
GHz, corresponding to 12$^{\circ}$ phase error.  The combined effect of ionospheric delay
($\propto\nu^{-2}$) and refractive delay Eq.~\ref{eq-micron-deg} (phase error
$\propto\nu$) gives a scaling effect on phase $\propto 1/\nu$
for different observing frequencies under the same observing
conditions.

Whilst at present (June 2021) the total TEC across South America at
the latitude of ALMA is 30--40 TEC so the difference over the longest ALMA baselines of 16 km,  during a 
 few minutes scan, is $\Delta \mathrm{TEC}\sim0.1$
(i.e. negligible differential ionospheric refraction at mm wavelengths), TEC values
can exceed 140 during Solar maxima or storms.  If this gives a five-fold increase in $\Delta
\mathrm{TEC}$, this would still only be
a phase error  $\sim$$2^{\circ}$ on a 16 km baseline for ALMA at
Band 3, but would become noticeable if longer baselines are added and
for VLBI.  Whilst these effects are small enough to be solved by
self-calibration, this will be an issue for astrometric measurements
and also if band-to-band calibration \citep{Asaki20a, Asaki20b} is used, since
standard transfer techniques assume that phase errors scale as $\phi_{\epsilon,\mathrm{trop}}\propto\nu$,
without a contribution scaled as $\phi_{\epsilon,\mathrm{ion}}\propto1/\nu$. It is also worth noting that some targets are viewed through 
enough ionised media, such as the centre of the Galaxy towards Sgr A*, to cause strong refraction even at a few mm wavelength.

\subsubsection{Thermal noise}
\label{sec-thermal}
The thermal noise arising from the receiver and other instrumental systems,  atmospheric
emission and other heat sources is  characterised as
$T_{\mathrm{sys}}$.
\begin{equation}
T_{\mathrm{sys}}= \frac{1+g}{\eta e^{\tau_0 \sec z}} \left( T_{\mathrm{Rx}} + \eta T_{\mathrm{sky}} +(1-\eta)  T_{\mathrm{amb}} \right).
  \label{eq-tsys}
\end{equation}
$g=1$ for double sideband receivers (ALMA bands 9 and 10) and 0 otherwise,  $\eta$ is the effective antenna efficiency (see  \citetalias{ALMA-TH} for ALMA values).
$\tau_0$ is the zenith opacity  and $z$ is the zenith angle.   $T_{\mathrm{Rx}}$, $T_{\mathrm{sky}}$ and  $T_{\mathrm{amb}}$ are the receiver system temperature, the sky signal and background and local (e.g. ground) contributions to temperature
(see  \citetalias{ALMA-TH} for more precise details of these terms).
These relationships are used to predict the image noise rms ($\sigma_{\mathrm{rms}}$) for a given observation (see Eq.~\ref{noise}) and by the  ALMA and other sensitivity calculators, see Appendix~\ref{scs}.

The minimum phase error resulting from  noise can be estimated based on \citet{Richards97t}
(derived using \citealt{TMS}) as $\phi_{\mathrm{rms}} =
\sigma_{\mathrm{rms}}/P$ where $\sigma_{\mathrm{rms}}$ and $P$ are the image noise and peak, i.e. the phase noise  in radians
is given by 1/(S/N).

\subsubsection{Antenna position errors}
\label{sec-antposerr}
\citet{Hunter16}   summarises the factors affecting the accuracy
of ALMA antenna position measurements;
since 2016 many of the improvements suggested have been implemented.
If data are taken soon after an antenna relocation, position
updates are supplied within a few days, also see Sec.~\ref{sec-antposerr} for observational symptions and implementing updates.

What is of concern, especially in early data, is unknown
antenna geographical position errors.
 If an inaccurate antenna position
is used during observations, the online corrections for geometric
delay (the difference in signal path for pairs of antennas) will be slightly wrong, causing a phase slope as a function of
frequency or delay error. The geometric delay is zero for a point source at the zenith (see \citetalias{SI99}, Ch. 2 (Thompson) for a fuller explanation). As the Earth rotates, for an E--W baseline, the antennas will point at 4$^{\circ}$ to the zenith after 1 minute. An antenna position error of $\sigma_{\mathrm{antpos}}$ (in mm) will produce an error in the geometric delay correction of  $\sigma_{\mathrm{antpos}}\times \sin (4^{\circ})
\times 1000$ $\mu$m min$^{-1}$, or $\sim$70  $\mu$m min$^{-1}$ for $\sigma_{\mathrm{antpos}}=1$ mm.
Using 
Eq.~\ref{eq-micron-deg}, for $\lambda$ in mm, the phase error of data from that antenna  is given by
\begin{equation}
\sigma_{\phi \mathrm{rate, antpos}}=360 \times \left(\sigma_{\mathrm{antpos}}\times \sin (4^{\circ})
\times 1000\right) \times 10^{-6}/(\lambda \times 10^{-3})
\label{eq-phirateantpos}
\end{equation}
giving a phase error of $\sigma_{\phi \mathrm{rate, antpos}}\sim$25$^{\circ}$ min$^{-1}$ at $\lambda=$ 1mm.  This is a `worst case' as the rate of change of the delay error is lower in other directions and the actual determination of antenna positions is a much more complicated process \citep{Hunter16}.  Errors tend to be larger for antennas with only longer baselines.

There is an associated phase error of
$\phi_{\epsilon,\mathrm{antpos}}$ in the transfer of corrections
to the target:
\begin{equation}
  \phi_{\epsilon,\mathrm{antpos}} \approx 2 \pi \Delta_{\mathrm{pcal-tar}}\frac{\sigma_{\mathrm{antpos}}}{\lambda}
  \label{eq-antpos}
\end{equation}
$\sigma_{\mathrm{antpos}}$ and $\lambda$ are in the same units.  For recent observations using a
phase calibrator -- target angular separation of
$\Delta_{\mathrm{pcal-tar}} = 3^{\circ}$ (for example), a typical antenna position error
$\sigma_{\mathrm{antpos}}=0.1$mm at $\lambda=1$ mm produces a phase
error $\phi_{\epsilon,\mathrm{antpos}}$ of only $\sim2^{\circ}$.  
It is, however,
harder to establish the positions of distant antennas contributing
only long baselines. The same $\Delta_{\mathrm{pcal-tar}}$ and  $\sigma_{\mathrm{antpos}}=2$ mm, but at $\lambda=0.8$
mm, gives $\phi_{\epsilon,\mathrm{antpos}}\sim50^{\circ}$ which is
enough to cause $>30\%$ decorrelation and, more seriously, to shift
the position of the target in the image plane. This can cause severe distortions for
sources with a complex structure but  for a source
with one, central unresolved peak, phase self-calibration will
compensate for small position errors.  Like
$\phi_{\epsilon,\mathrm{trop}}$,
$\phi_{\epsilon,\mathrm{antpos}}\propto \nu$.

\subsection{Amplitude errors}
\label{sec-causeamp}
Water vapour and other tropospheric molecules such as ozone  cause amplitude errors directly as they both absorb
and emit in ALMA bands (as well as corrupting the phase by refracting the incoming radiation).
You can use the ALMA Atmospheric Model (see Appendix~\ref{ap-other}
to check if your observations are particularly badly affected, or overplot the transmission in {\tt plotms}.
The effects of this on the visibility amplitudes vary on timescales of minutes (rather than
seconds, as for phase).
ALMA's $T_{\mathrm{sys}}$
measurements allow correction for absorption and emission
including gain-elevation effects and also compensate for any
differences between antennas or signal paths.  These corrections are
made per scan or every few scans and the residual time-dependent
errors are usually only a few percent. 

Other contributions to amplitude errors include pointing errors, Sec.~\ref{sec-point}.
Nonetheless, after applying $T_{\mathrm{sys}}$ corrections,  the main
contribution to time-dependent amplitude errors is usually
decorrelation due to phase errors, as shown in Eq.~\ref{phnoise}.
The lowest possible level of uncertainty is the thermal noise, Sec.~\ref{sec-thermal}.

\subsubsection{Flux density scale}
\label{sec-flux}
We use `amplitude errors' to refer to fluctuations during an observation, in time or across frequencies (other than target variability/spectral index), distinct from the overall flux scale. The overall flux scale usually remains constant (per spw) during a single EB, or even a series of observations made in the same mode with no instrumental changes. 

Immediately after correlation, ALMA  visibility amplitudes are in units of (correlated signal/system noise),  \citetalias{ALMA-TH} Ch. 10.  Other interferometers use different scalings but in any case a scaling factor is needed to convert the flux scale to Jy or other physical units.  
The flux scale can be derived from $T_{\mathrm{sys}}$ (Appendix~\ref{tsystoJy}) but is normally derived using an astrophysical standard, used to calculate the flux density of other calibrators including the phase calibrator.
Amplitude solutions are derived for the calibrators and then task {\tt fluxscale} is used to produce a scaled amplitude correction table for the phase calibrator.
The derivation usually assumes point sources, hence the target flux density is not determined at this point. Instead, the scaled amplitude corrections derived from the phase calibrator can be applied directly to the target and phase reference. (see Fig.~\ref{VYCont_phref.png} bottom right, where the  orange  circles and (darker) squares show phase calibrator visibilities before applying the flux scale; the darkest, brown diamonds are on a scale 30$\times$ higher, after flux scaling). Alternatively, {\tt setjy} can be used to set the derived phase calibrator flux density and then a fresh amplitude calibration table is derived and applied.

Normally the flux scale can be assumed to be accurate during self-calibration (within uncertainties,  \citetalias{ALMA-TH}  Ch. 10  for ALMA)
but if a continuum source contributes a 
significant fraction
of $T_{\mathrm{sys}}$ (which is several tens K at the lowest bands to a few
hundred at the highest, see  \citetalias{ALMA-TH}), you may need
to correct for its contribution to $T_{\mathrm{sys}}$, see Appendix~\ref{sec-avc}.
You may also  need to
check that the corrections have been derived from $T_{\mathrm{sys}}$
accurately in the case of bright spectral
lines, see Appendix~\ref{ap-other}.
If the target was not phase-referenced, see Sec.~\ref{sec-nophref}.
 
\subsection{Closure relations and the minimum number of antennas for solutions}
\label{sec-closure}

Although CASA and other modern packages do not use closure expressions
directly to self-calibrate, the principles underlie the methods and
provide an intuitive understanding of why there are 3 and 4 degrees of
freedom for phase and amplitude, respectively.  Take antennas 1, 2, 3
where the observed phase on the baseline between the first two
antennas, including the antenna-based errors, is $\phi12 +
(\phi_{\epsilon1} -\phi_{\epsilon2})$, etc.. \citet{Jennison58} realised that if you
combine the phases for 3 baselines between 3 antennas as
\begin{equation}
[\phi12 + (\phi_{\epsilon1} -\phi_{\epsilon2})]+[\phi23 +
  (\phi_{\epsilon2} -\phi_{\epsilon3})]+[\phi31 + (\phi_{\epsilon3}
  -\phi_{\epsilon1})] = \phi12+\phi23+\phi31
\label{eq-phclos}
\end{equation}
the phase errors sum to zero.

Similarly, as amplitude errors are multiplicative, the relationship for baselines between 4 antennas is:
\begin{equation}
\frac{[A12 A_{\epsilon1}A_{\epsilon2}]\times[A34
    A_{\epsilon3}A_{\epsilon4}]}{[A23
    A_{\epsilon2}A_{\epsilon3}]\times[A41 A_{\epsilon4}A_{\epsilon1}]}
= \frac{A12\times A34}{A23\times A41}
\end{equation}
Thus there are 3 degrees of freedom for phase solutions and 4 for
amplitude and in order to obtain unambiguous calibration you need a
minimum of 2 baselines per antenna for phase and 3 for amplitude,
i.e. a minimum of 3 and 4 antennas with all good baselines,
respectively.  In reality, on the one hand a higher number is
desirable to ensure good solutions and on the other, for a source with
very simple structure, self-calibration can be performed with fewer,
even on a single baseline for a point source.

\section{Effects of errors after phase referencing}
\label{sec-afterphref}

Self-calibration starts after applying corrections derived from other
sources or instrumental measurements. For ALMA, this includes
corrections derived from $T_{\mathrm {sys}}$, WVR, antenna positions
and bandpass calibration, and any polarisation calibration,  \citetalias{ALMA-TH}.  Bad data
in the calibration sources should have been flagged and relevant flags
extended to the target.  For convenience we refer to data with these
corrections as `raw', prior to applying time-dependent corrections
derived from the phase calibrator and the flux scale, which should
also be done before target self-calibration.
Section~\ref{sec-errorrecognition} describes more general error
recognition.

We start by assuming that these prior corrections are as good as possible.
After applying phase calibrator solutions, the remaining errors
affecting the target visibilities  are mainly due to the troposphere (as described in
Sec.~\ref{sec-causeeffect}), specifically, atmospheric differences
in time and in direction between the phase calibrator and the target scans.
This means that the target signal amplitude suffers slightly different
sky absorption and emission and the phase suffers slightly different
refraction. Self-calibration provides corrections for such residuals to a greater or lesser extent depending on the S/N and possibly other factors like $uv$ coverage so this may not be perfect.  Antenna position and pointing errors also contribute, and may affect
the starting model, see Secs.~\ref{sec-antposerr} and~\ref{sec-point}.

\subsection{Illustrating phase and amplitude errors and their correction}
\label{sec-VY}

We use ALMA VY CMa Band 7 science verification
data\footnote{\url{https://almascience.org/almadata/sciver/VYCMaBand7/}} to
illustrate residual errors and how to correct them using
self-calibration. The data are summarised in Table~\ref{tab-VYobs}, see \citet{Richards14} for more details. Only one (out of 3) execution blocks are used here, so the
image sensitivity is worse than that obtained from the whole data set
\citep{Richards14} and initially we consider a single spectral window. 
All corrections apart from phase referencing have already been applied.
The observations were made
with alternating phase calibrator and target scans, starting and finishing on the phase calibrator, with a cycle time $\sim$7 min.

\begin{table}
  \begin{tabular}{|c|rll|lc|llll|c|}
    \hline
  Target\,\,\,\,  Phase-cal &$\Delta_{\mathrm{pcal}}$&Cycle &$\Delta t$&$N$ &$B_{\mathrm{min}},B_{\mathrm{max}}$&$N_{\mathrm{pol}}$&$\nu_{\mathrm {cen}}$ &spw&chan&$\Delta{\nu}$\\
           &      $_{\mathrm{-tar}}$ &\multicolumn{2}{c|}{(min)}&& (km)&  &(GHz)&\multicolumn{2}{c|}{(MHz)}&(GHz)\\
\hline
  VY CMa J0648-3044&$\approx2^{\circ}$&1.5:5            &30&20& 0.014, 2.7&2&325& 1875&0.488&1\\
\hline
  \end{tabular}
  \caption{\small Data used for examples, taken from uid\_\_\_A002\_X6cdf83\_X1488, VY CMa Band 7 SV data 2013. $\Delta_{\mathrm{pcal-tar}}$ is the phase calibrator-target angular separation, observed alternately in the cycle time (where $\tau_{\mathrm{scan}}$ is the target scan length, 5 min) for a total time $\Delta t$ on target. $N$ and $B_{\mathrm{min}}, B_{\mathrm{max}}$
are the number of antennas and the minimum, maximum baselines used.  $N_{\mathrm{pol}}$ is the number of receiver polarisations (X and Y).   $\nu_{\mathrm {cen}}$ is the central frequency of the  spw used here and spw and chan are the total spw and individual channel widths. $\Delta{\nu}$ is the line-free continuum bandwidth.}
\label{tab-VYobs}
\end{table}

Note that some figures show visibility data on a
single baseline compared with solutions derived per-antenna, as part
of calibration fitting to data from all antennas, hence there is not
an exact correspondance between data and solutions.  Moreover, the
plots have been assembled to highlight relevant features -- not for
exact science. 

\begin{figure}[t]
\includegraphics[width=8.5cm]{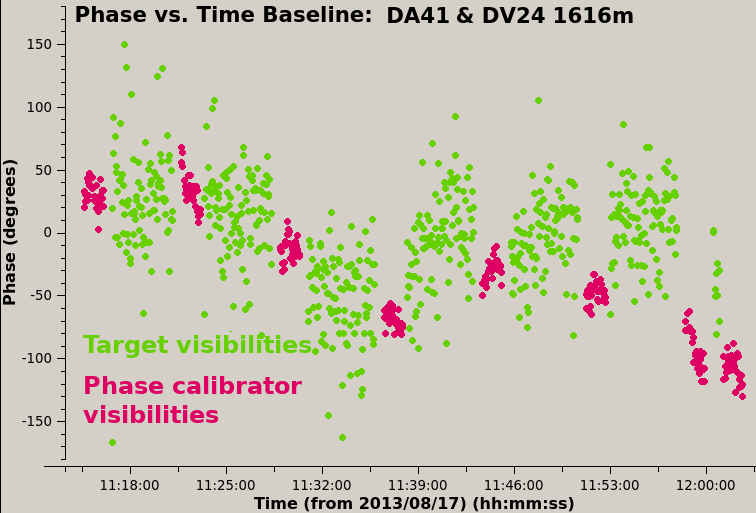}
\includegraphics[width=8.5cm]{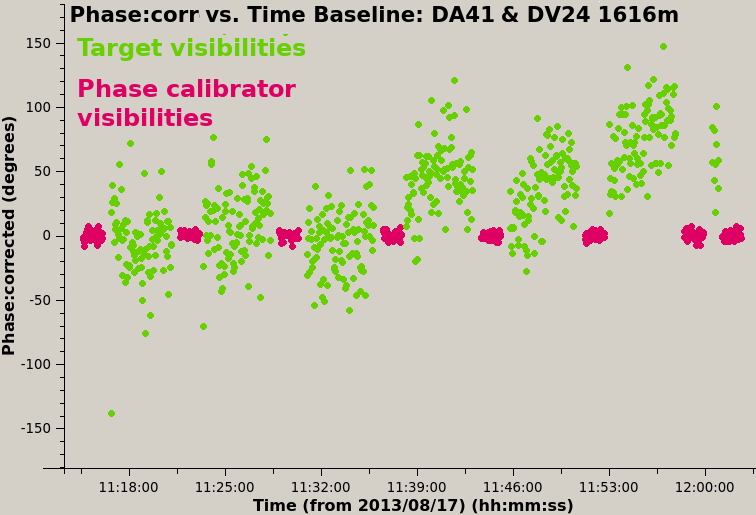}
\includegraphics[width=8.5cm]{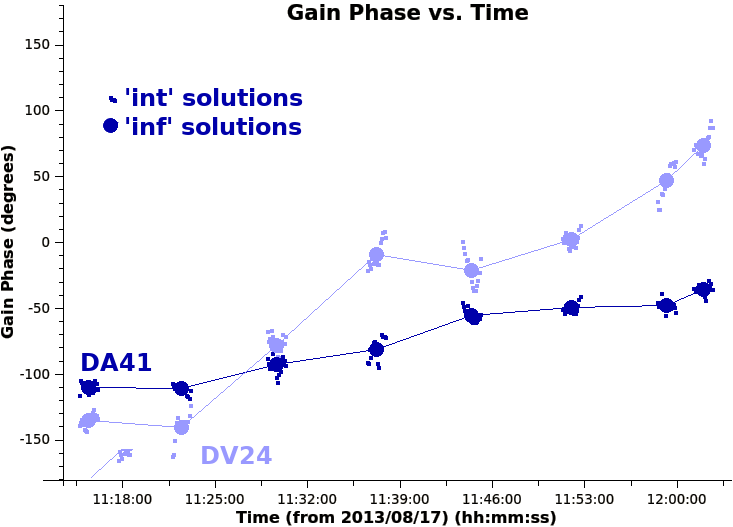}
\includegraphics[width=8.5cm]{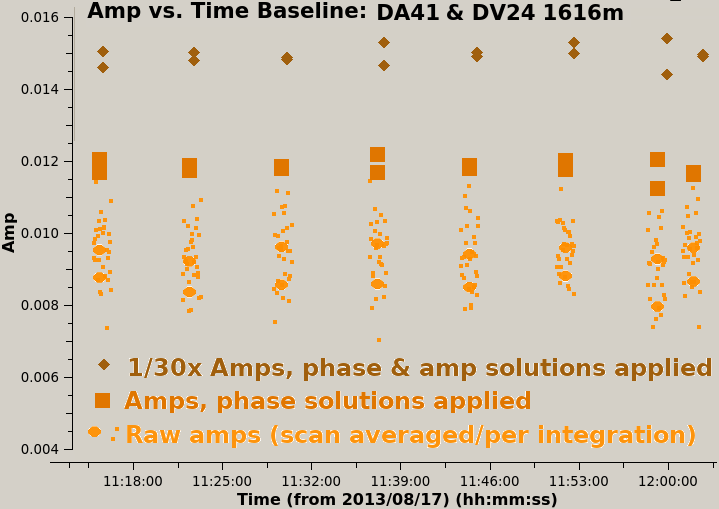}
\caption{\small Illustrations of the use of the phase calibrator to correct the target visibilities. In discussions we ignore the final two very short scans. Top left: Raw visibility phases per integration, for target (continuum) and phase calibrator, in red (dark) and green (light), respectively.
  Bottom left: Phase calibrator phase solutions for X polarisation, derived per-integration (int) and per-scan (inf) (small and large circles), for two antennas, DA41 and DV24 (dark and light blue).
Top right: Phase calibrator phases corrected per integration, target phases
corrected per scan with interpolated phase calibrator solutions.
Bottom right: Phase calibrator visibility amplitudes (both correlations) before and after
    applying phase and amplitude solutions. Raw data, unaveraged and averaged per scan are shown by  small and large orange circles, respectively. Corrected data averaged by scan after applying per-integration phase solutions are shown by darker orange squares. 
    The darkest, brown diamonds show data with the flux scale and amplitude solutions
    also applied and are plotted at 1/30 of the
    actual flux densities, around 0.44 Jy.}
\label{VYCont_phref.png}
\end{figure}

In Fig.~\ref{VYCont_phref.png}  the phase calibrator and target scans are shown in red and in
green, respectively. Plots of visibilities have a grey background and plots of solutions have a white background. The phase calibrator is point-like with an accurate position so
its model has zero phase and a constant amplitude ($\sim$$0.44$ Jy).
DA41 and DV24 are approximately 0.2 and 1.4 km, respectively, from the
reference antenna. Fig.~\ref{VYCont_phref.png} top left shows the raw phase for a single
baseline, single spw, single polarisation.  Fig.~\ref{VYCont_phref.png} bottom left shows
phase solutions for the two antennas for the phase calibrator per-integration 
(small dots) and per-scan  (large dots). The per-integration
solutions are applied to the phase calibrator visibilities, leading to corrected phases
which are very close to zero (see  Fig.~\ref{VYCont_phref.png} top right).
Any remaining offsets are due to the noise limit of the data and,
possibly, small, baseline-dependent errors.

We can estimate the effect of phase errors as a function of time from
the raw phase calibrator data (this is easier than from the target
data themselves, as the target may be weak and may have real phase
variations due to structure, and the basis of phase referencing is that similar errors affect both calibrator and target). Fig.~\ref{VYCont_phref.png} top left,
shows that within a phase calibrator scan the typical raw phase change
is $50^{\circ}$.  Eq.~\ref{phnoise} predicts that this leads to a
reduction of amplitude to $68$\% of the true value.
Fig.~\ref{VYCont_phref.png} bottom right shows that the raw
amplitudes (orange circles) are indeed only  $\sim$$70$\% of the corrected values (orange
squares). That is, the phase errors cause $\sim$$30$\%
decorrelation. Applying the phase calibrator per-integration phase
solutions prior to averaging the visibility amplitudes per-scan
improves the coherence by the expected amount.  

The raw, unaveraged phase calibrator visibility amplitudes (small
circles) have about 20\% random scatter per polarisation and a similar
systematic variation between scans over the whole 45 min
of observation (in contrast to the rapid raw phase changes).  Thus, once phase solutions have been applied, the remaining scatter within a scan is noise-like and the solutions for amplitude calibration can usually be a scan length or longer, since calibration cannot remove noise.
Applying the amplitude solutions flattens the differences between visibility amplitudes per-scan (diamonds).
The final phase calibrator amplitude corrections (diamonds) have also
been scaled so that, when applied, they not only remove small
fluctuations between scans but also contain a constant factor to
convert the target visibilities to the physical flux scale (Sec.~\ref{sec-flux}).

Fig.~\ref{VYCont_phref.png} bottom left shows phase calibrator solutions
averaged per scan (large dots) with lines indicating the interpolation
across the target scans.  This can correct atmospheric effects on the
timescale of a scan cycle ($\sim$7 min) but not errors
within a scan.   Fig.~\ref{VYCont_phref.png} top right  shows that the
target phase deviations within each scan are only slightly
reduced to 
$50-100^{\circ}$.  The remaining overall phase slope and other
deviations are probably due to the target offset
from the pointing centre and its structure.

\subsection{Quantifying target residual phase errors after phase referencing}
\label{sec-qpr}
In order to understand how to correct residual errors, this section provides rough estimates of their origins, timescales and magnitudes.
Note that quantifying phase errors per antenna is a convenient concept, but in practice visibilities are formed per-baseline. Calibration algorithms compare the Fourier transform of the model image with the visibilities. The discrepancies are decomposed using a least squares (or other) minimisation process to find the per-antenna corrections needed to bring the observed visibilities closer to the model.  Thus, when comparing the observed data per-baseline phases and amplitudes (raw or corrected) with per-antenna solutions, as in Fig.~\ref{VYCont_phref.png}, remember that the correspondance is not exact, especially as the visibilities for a resolved or off-centre source will also show patters due to the source structure.

The expressions and typical values derived for errors and the required corrections given here assume that all the antennas in an array have errors of similar magnitude which may not be the case if some antennas contribute  mostly to short baselines and others mostly to long, or if antennas have intrinsically different sensitivities (see Sec.~\ref{sec-weights}).  To start with, we assume that the frequency dependence of errors is
negligible within an observation, and the phase errors described are
mainly tropospheric. The contributions here are listed in approximate order of increasing seriousness for these observations -- this order could be different e.g. in spectral regions where atmospheric noise is high.  Items 1., 2,. and 3.--5. relate to Secs.~\ref{sec-thermal},~\ref{sec-antposerr} and ~\ref{sec-trop} (occasionally \ref{sec-iono}), respectively. 

\begin{enumerate}

\item {\textbf{Noise errors} The errors specifically connected to phase
referencing can in principle be corrected, given high enough target
S/N.
{\textbf {\textit{Thermal noise cannot be
      removed by self-calibration or anything else}}} although
sophisticated modelling algorithms may be used in specific cases to
assign a significant probability to visibility or image features below
the usual detection threshold (e.g. \citealt{Nakazato20}).
 Here, the phase calibrator is $\sim440$ mJy and for a single polarisation, single spw centred on the observing frequency of 325 GHz, in a single 90 s scan, the sensitivity calculator predicts 4.5 mJy rms noise, so the phase error due to thermal noise, given in radians by 1/(S/N), is slightly less than $1^{\circ}$. In transferring solutions from this phase calibrator it is negligible compared with  calibration
errors. More usually, the noise in a final image may be significant in position accuracy for faint  components, see
Sec.~\ref{sec-poserr}}

\item {{\textbf{Antenna position errors:}}
  For a phase reference source with a good position, a smooth slope of phase over many scans suggests an antenna position error (this would show a sinusoidal pattern over 24 hr). Fig.~\ref{VYCont_phref.png} top left shows a total phase change of $\sim-$150$^{\circ}$ in 48 min, or an average of magnitude  $\sim$3$^{\circ}$ min$^{-1}$ (with greater excursions between scans). Using Eq.~\ref{eq-phirateantpos}, $\sim$3$^{\circ}$ min$^{-1}$ at  $\lambda 0.9$ mm corresponds to an antenna position error $\sigma_{\mathrm{antpos}}\sim$ 0.2  mm. This is a very crude estimate but is typical for ALMA observations at that time \citep{Hunter16}. The delay error associated with an antenna position error, and thus the correction, is direction-dependent, and so from Eq.~\ref{eq-antpos} this would lead to an error in transferring solutions to the target of
$\phi_{\epsilon,\mathrm{antpos}}\sim2.8^{\circ}$. }

\item {{\textbf{Short-timescale errors: }}  Fig.~\ref{VYCont_phref.png} top left shows small but systematic
phase deviations  within a phase calibrator scan, in this instance
$\delta\phi \sim 20^{\circ}$ in 0.75 min.  In the worst case, superimposed
on a steep residual transfer error, their effect per 5-min target scan
will be  $\phi_{\epsilon, \mathrm{short}}\sim \delta\phi \times \sqrt(5/0.75)\sim50^{\circ}$.}

\item {{\textbf{Phase errors due to phase calibrator -- target time
      difference:}} The offsets, in both time and angular separation, between
  phase calibrator and target, contribute to the residual phase errors
  in the target visibilities.  An averaged correction
  from the bracketing phase calibrator scans is applied to every
  target scan and the scan cycle is 7 min. We can use the phase reference visibility phase to estimate the phase rate.  The average slope (assumed to be due to antenna position errors) in   Fig.~\ref{VYCont_phref.png} top left is
$\sim-20^{\circ}$ per 7-min cycle, plus additional scan-to-scan raw phase-changes of magnitude $\sim$28$^{\circ}$, giving $d\phi_{\mathrm{atm/min}}\sim4^{\circ}$ min$^{-1}$.  The corrections will thus
  have an inaccuracy due to the offset in time leaving a residual
  error per target scan of $\phi_{\epsilon, \mathrm{time}} =
  d\phi_{\mathrm{atm/min}}\times(7/2) \approx 14^{\circ}$. The factor
  of 2 is is allowing for the maximum discrepancy when interpolating
  linearly the per-scan averages of the phase reference solutions
  across the target scans. 
}
\item {{\textbf{Phase errors due to phase calibrator -- target angular
      separation: }} Following this, the change in phase per unit time
  seen as the telescope tracks sidereal targets can be regarded as
  equivalent to the phase change between different source directions
  on the sky, by converting angular separation to minutes of time,
  equivalent to Right Ascension.  1$^{\circ}$ angular separation is 4
  min of R.A. $\times$ cos(Declination).  In these data VY CMa and its
  phase calibrator have an angular separation of
  $\Delta_{\mathrm{pcal-tar}}\approx2^{\circ}$.  At the target
  Declination of $-26^{\circ}$ the angular separation is equivalent to
  $(\Delta_{\mathrm{pcal-tar}} \times 4 \times \cos(-26)) \sim 7$
  minutes.  This leads to a likely discrepancy of $\phi_{\epsilon,
    \mathrm{angsep}} = d\phi_{\mathrm{atm/min}}\times 7 \approx
  28^{\circ}$ between the phase corrections calculated in the phase
  calibrator direction and the corrections needed for the target.}

\item {\textbf{Accumulated errors after phase referencing} The combined error
affecting target data due to the separation between phase calibrator and
target is
\begin{equation}
\phi_{\epsilon, \mathrm{transfer}}=
\sqrt{\phi_{\epsilon, \mathrm{time}}^2+\phi_{\epsilon,
    \mathrm{angsep}}^2+\phi_{\epsilon, \mathrm{antpos}}^2}
\label{eq-transfer}
\end{equation}
Using the values derived above for
$\phi_{\epsilon,\mathrm{antpos}}\sim2.8^{\circ}$,  $\phi_{\epsilon, \mathrm{time}}\sim14^{\circ}$ and $\phi_{\epsilon,
  \mathrm{angsep}} \sim28^{\circ}$ gives $\phi_{\epsilon,  \mathrm{transfer}}\sim31^{\circ}$.

Adding in the systematic short-time-scale error $\phi_{\epsilon, \mathrm{short}}\sim50^{\circ}$
 leaves a total target phase error per antenna of 
\begin{equation}
\phi_{\epsilon, \mathrm{target,ant}} = \sqrt{\phi_{\epsilon,
      \mathrm{time}}^2 +\phi_{\epsilon,\mathrm{angsep}}^2+\phi_{\epsilon,\mathrm{antpos}}^2 + \phi_{\epsilon, \mathrm{short}}^2     }
 \label{eq-targetpherr}
\end{equation}
 so here, $\phi_{\epsilon, \mathrm{target,ant}} \sim 60^{\circ}$. 

Fig.~\ref{ALMAMemo606Fig1.png} shows phase errors as a function of
  baseline length, 1.6 km in this VY CMa example. 
Crudely approximating the target phase errors for antennas
contributing to this range by taking half $\phi_{\epsilon,
  \mathrm{target,ant}}$ as
$\phi_{\mathrm{rms}}\sim 30^{\circ}$, 
then using Eq.~\ref{eq-micron-deg}, at $\lambda = 0.9$ mm this is equivalent to
$\phi_{\mathrm{rms},\mu}\sim75$
$\mu$m.  Fig.~\ref{ALMAMemo606Fig1.png}
shows that this  falls between the grey and yellow lines for the expected
target phase rms for 1.6 km baselines after applying PWV and phase calibrator
corrections on 2 min and 20 sec timescales, respectively. The VY CMa observations were comparable to the test, but with a smaller target--phase calibrator
angular separation and a longer cycle time than that for the predicted gray line. The PWV was significantly lower, 0.3 mm, contributing to the better performance seen in the VY CMa data. }
\item {{\textbf{Phase errors on many baselines: }}
These estimates and expressions  are imprecise, not least because
generalising from one scan, one baseline is only roughly
representative, and because the effects of short time-scale jitter on images are more visible on sources with complex structure.  However,
assuming that the estimates are typical,
for an observation with
$N$ antennas the overall effect is reduced insofar as each baseline
samples a different combination of atmospheric conditions as illustrated in Fig.~\ref{atmos1.png}.
Different baselines, even of a similar length, will sample different conditions and position angles, so even if
 the magnitude of the effect is similar the sign and
value will change, see \citetalias{SI99} Ch. 5 (Fomalont \& Perley) for a general discussion.
The phase errors for an array of many antennas can be given by:
\begin{equation}
\phi_{\epsilon, \mathrm{target}} = \frac{\phi_{\epsilon,
      \mathrm{target,ant}}}{\sqrt{(N_{\mathrm{indep}})}\sqrt{M}}
\label{eq:target_pherr}
\end{equation}
$M$ is the number of  periods of distinct atmospheric conditions sampled and the change in phase slope from scan to scan in Fig.~\ref{VYCont_phref.png} top left suggests that we can take this as the number of full scan cycles, 6.  This implies a wind speed of $\gtrsim$2.7 km (the maximum baseline) in 7 min, or $\gtrsim$6 m s$^{-1}$, which is a typical wind speed at ALMA (Sec.~\ref{sec-trop}).
If the atmosphere above each antenna is independent, such as for VLBI arrays, the number of independent antennas $N_{\mathrm{indep}}=(N-3)$ (see explanation of Eq.~\ref{noise_ant} in Sec.~\ref{sec-SN}). For the VY CMa example here there will be less  difference between the atmosphere above individual antennas during each scan and Fig.~\ref{ALMAMemo606Fig1.png} suggests that  the atmospheric effects above every antenna are not independent. The 
errors on the longest baselines and the short-term   scatters within each scan may, however, be somewhat independent, and we adopt  $N_{\mathrm{indep}}=2$. 
This leads to 
$\phi_{\epsilon, \mathrm{target}}\sim20^{\circ} \sim 0.35$ rad, leading to a prediction of $\sim6\%$ decorrelation (Eq.~\ref{phnoise}) for the image made  after
applying phase calibrator solutions only. This has a 179 mJy beam$^{-1}$ peak. Comparing this with  the
final VY CMa peak of 199 mJy with  the actual initial
decorrelation is $\sim$10\%.  This is slightly worse than
predicted due to over-simplification in the expressions used and variations in observing conditions, and also because the dynamic range of the first target image is limited by the phase calibrator dynamic range, Sec.~\ref{sec-tderror-DR}.}
  \end{enumerate}

The contributions to phase errors are summarised in Table~\ref{tab-pherr}, numbered as in the above points.
\begin{table}
  \begin{tabular}{|cccccccc|}
        \hline 
Phase noise&$\phi_{\epsilon,\mathrm{antpos}}$&$\phi_{\epsilon,\mathrm{short}}$&$\phi_{\epsilon,\mathrm{time}}$&$\phi_{\epsilon,\mathrm{angsep}}$&$\phi_{\epsilon,\mathrm{transfer}}$&$\phi_{\epsilon,\mathrm{target,ant}}$&$\phi_{\epsilon,\mathrm{target,array}}$   \\
1.         &2.                         &3.                         &4.                        &5.                          &6.                            &6.                              &7.                         \\
\hline
$<$1         &2.8$^{\circ}$               &  50$^{\circ}$                &    14$^{\circ}$            & 28$^{\circ}$                 &31$^{\circ}$                   &60$^{\circ}$                     &20$^{\circ}$                   \\
\hline    
  \end{tabular}
  \caption{\small The various contributions to target phase errors arising from the transfer of solutions from the phase calibrator, in the example of one execution of an observation in one spw, dual polarisation, at 325 GHz, as described  in Sec.~\ref{sec-VY}. The terms are described in Sec.~\ref{sec-qpr}. 1. is the error due to thermal noise within each scan on the 440 mJy phase calibrator. 2. is due to antenna position errors and 3.  is due to the short-timescale phase errors within a phase calibrator scan. 4.--5. arise from the angular and time separations between target and phase calibrator. 6. relates to  the sums in quadrature of the per-antenna errors due to the separation from the phase calibrator and additional short-term errors and 7. is the error when considering all 20 antennas used in the array.  }
  \label{tab-pherr}
\end{table}
  
Image-plane errors and their causes, whether or not they can be corrected by self-calibration, are
summarised in Sec.~\ref{sec-errorrecognition}.

\subsection{Dynamic range}
\label{sec-tderror-DR}
 We take a point
source, flux density $P$, at the phase centre and give errors in radians in this section to simplify expressions (more fully explained in \citetalias{SI99}, Ch. 13 Perley). For a
single scan, an error of $\phi_{\epsilon, \mathrm{baseline}}$ on a
single E-W baseline, length $u_0$, gives rise to a periodic,
antisymmetric image artefact of amplitude $P 2\phi_{\epsilon,
  \mathrm{baseline}}$ and period $1/u_0$.  Analogous reasoning for an
amplitude error of fractional magnitude $A_{\epsilon, \mathrm
  {baseline}}$ shows that this leads to a symmetric image artefact,
see Section~\ref{sec-phase-fidelity}.  If the observation comprises
$M$ scans, then the effect of the errors is reduced by $\sqrt{M}$.

If such an  error dominates over
thermal noise and other contributions to the image rms, the dynamic range $D$ for the image
made from all $N$ antennas is limited to
\begin{eqnarray}
D_{\mathrm {baseline} \phi } &=& \frac{\sqrt{M}
  N(N-1)}{\sqrt{2}\,\phi_{\epsilon, \mathrm {baseline}}} \approx  \frac{\sqrt{M} N^2}{\sqrt{2}\,\phi_{\epsilon, \mathrm {baseline}}}\\ \nonumber
D_{\mathrm {baseline} A}  &=& \frac{\sqrt{M}
  N(N-1)}{\sqrt{2}\,A_{\epsilon, \mathrm {baseline}}} \approx \frac{\sqrt{M} N^2}{\sqrt{2}\,A_{\epsilon, \mathrm {baseline}}}
\label{DR_baseline}
\end{eqnarray}
for large N, for phase ($D_{\mathrm {baseline} \phi }$) and amplitude ($D_{\mathrm {baseline} A}$) errors, respectively. Thus, a
$10^{\circ}$ (0.175 rad) phase error produces the same dynamic range limitation as a
20\% amplitude error. In most ALMA data after phase referencing the amplitude errors  are lower than this whilst the phase errors can be much higher as summarised in Table~\ref{tab-pherr}.
For 20 antennas, a $10^{\circ}$ phase error  on a single
baseline restricts $D_{\mathrm {baseline}}\sim 1600 \sqrt{M}$.  

More commonly all baselines to one antenna are affected by a similar
error. Considering the phase-error
case, this reduces the dynamic range  by a factor of $\sqrt{N}$; if all antennas (and thus all baselines) are affected there is a reduction by another factor of $\sqrt{N}$:
\begin{eqnarray}
D_{\mathrm {antenna} \phi }  &\approx & \frac{\sqrt{M} \sqrt{N}N}{\sqrt{2}\phi_{\mathrm {antenna}}}\\ \nonumber
D_{\operatorname{all-antenna} \phi }&  \approx & \frac{\sqrt{M} N}{\sqrt{2}
  \phi_{\operatorname {all-antenna}}}
\label{DR_all-ant}
\end{eqnarray}

For a $10^{\circ}$ error affecting one of 20 antennas this
corresponds to $D_{\mathrm {antenna}}\sim 360 \sqrt{M}$ or, for all antennas, $D_{\operatorname  {all-antenna}}\sim 80 \sqrt{M}$.

If all the baselines have a different, random error, the dynamic range is
\begin{equation}
D_{\operatorname {all-baseline} \phi }  \approx  \frac{\sqrt{M} N}{\phi_{\operatorname {all-baseline}}}
\label{DR-all-bline}
\end{equation}

For a $10^{\circ}$ error affecting all 190 baselines from 20 antennas, $D_{\operatorname
  {all-baseline} \phi } \sim 115 \sqrt{M}$.  Sparse $uv$ coverage can
further limit the dynamic range, affecting VLBI images, very high
spectral resolution long-baseline ALMA data etc. due to deconvolution
errors.  However close the phase calibrator is, the target dynamic range will be no better than that for the phase calibrator, so if a higher S/N is
anticipated for the target, self-calibration will be needed.

Conversely, even for a very bright target, with 43 antennas of ALMA, a
phase error of 2.5$^{\circ}$ gives a dynamic range $\sim1000$. Whilst
over a whole observation, with a typical continuum
$\sigma_{\mathrm{rms}}$ 0.01 mJy, this only implies a peak of 10 mJy,
achieving such a small phase error requires adequate S/N per solution
interval per antenna, as explained in Sec.~\ref{sec-baseline}.  In
practice, it is hard to exceed a dynamic range of a few 1000 for ALMA,
requiring an excellent target model, bandpass calibration and antenna
position accuracy. If there is extended emission not fully sampled, or other issues, the dynamic range may be only a few 100 even taking account of situations as in Sec.~\ref{sec-sc}. In any case, you cannot reduce the intrinsic thermal noise (although this may be lower than predicted, under better observing conditions). For more discussion of dynamic range, see
\citetalias{SI99} Ch.~13 (Perley) section 3.

\subsection{Flux scale errors}
\label{sec-amp}
Section~\ref{sec-phase-fidelity} introduced decorrelation of
amplitudes due to phase errors. Residual errors
of a few--10\% typically remain after applying $T_{\mathrm{sys}}$
corrections (Sec.~\ref{sec-causeamp}),  \citetalias{ALMA-TH}).  Here, amplitude errors refers to
time-varying relative errors and inconsistencies between antennas, not in the overall flux scale (Sec.~\ref{sec-flux}).
  Amplitude self-calibration cannot correct the overall
flux scale (unless you have an a-priori model for the target) but it
can reduce artefacts by making the amplitudes self-consistent (Sec.~\ref{sec-comb}).

Special problems arise from  sources
which contribute a sizeable fraction of  $T_{\mathrm{sys}}$ (which is
tens -- hundreds Jy
in lower -- higher frequency bands, see  \citetalias{ALMA-TH} and links in Appendix~\ref{sec-avc} and also Appendix~\ref{ap-other} for issues with very bright spectral lines.

If a check source was observed, then comparing its imaged flux density
with that deduced from the visibilities using {\tt fluxscale} gives an
estimate of decorrelation after applying phase calibrator solutions
(allowing for the difference in angular separation from the phase
calibrator).

\subsection{Image resolution and position accuracy}
\label{sec-poserr}

The accuracy of phase calibration determines the distribution of flux
in an image and thus not only the general morphology but also the
accuracy of measurements of the image. This section discusses uncertainties arising from residual errors after calibration, but not those inherent in sparse visibility plane coverage or the accuracy of models. Model goodness of fit, whether in the sky plane or the visibility plane (e.g. \citealt{Marti-Vidal2014}, 
\citealt{Rivi2019}), may be dominated by a genuine discrepancy between the sky distribution and the model such that the  $\chi^2$ value might be worse for well-calibrated data where deviations are more apparent.

\subsubsection{Noise and resolution}
\label{sec-seeing}
As observations generally involve random errors on multiple antennas
over a long enough time for the atmosphere to vary, the image is
smeared by jitters in multiple directions (analogous to optical
`seeing' errors).

In a typical ALMA observation, following \citetalias{SI99} Ch.~28 (Carilli, Carlstrom \& Holdaway) eq 28.8, 
the phase error as given by Eq.~\ref{eq-Kolmogorov}
 can be used to give the baseline
length $B_{1/2}$ where the visibility curve is
reduced to half power, and thus the corresponding phase error:
\begin{eqnarray}
\label{eq-seeing}
 B_{1/2}&=&(69\lambda/K)^{1/\alpha_{\mathrm K}}  \\
\phi_{\epsilon 1/2}&=&KB_{1/2}^{\alpha_{\mathrm K}}/\lambda \nonumber
\end{eqnarray}

This is the seeing limit, for a given set of conditions, such that on longer baselines the
effective resolution will be worse than that predicted. The synthesised beam width is taken as
$\theta_{\mathrm B}\sim \lambda/B_{\mathrm{max}}$ where
$B_{\mathrm{max}}$ is the maximum baseline.  For example,
at
$\lambda$=1.3 mm with an effective $K= 50^{\circ}$ (thus also taking $\phi_{\epsilon 1/2}$ in degrees),  for $\alpha_{\mathrm K}=1/3$, then $B_{1/2} \sim 2$ km.
  As explained in Sec.~\ref{sec-trop}, you may need to iterate with different values of $\alpha_{\mathrm K}$ if
 $B_{1/2}$ appears to be $<$1 or $>$10 km.
Thus,     residual phase errors $\phi_{\epsilon}>65^{\circ}$ will noticeably
degrade the resolution (as well as redistributing the flux, as noted
using Eq.~\ref{phnoise}).

\subsubsection{Relative position accuracy}
\label{sec-gaussfit}
Stochastic (noise-based) position errors are derived from the
noise in the image used for measurement and cover the relative uncertainty between
sources in an image due to phase noise (Sec.~\ref{sec-thermal}). \cite{Condon97} and \cite{Richards99} describe
the effects of phase errors on the accuracy of determining source
parameters by fitting 2-D Gaussian components to an interferometric image (assuming that the
emission is smaller than a few synthesised beams and intrinsically well-described by a 2-D Gaussian).  The errors returned by tasks such as {\tt imfit} should be considered as guidance since the uncertainty is affected not only by random noise but by deconvolution errors, non-Gaussian source structure, nearby sources and, in such situations, by the size of the mask used for the fit.
The
relative position error is given by
\begin{equation}
\sigma_{\mathrm {pos fit}} = C\times\frac{\theta_{\mathrm B}}{(\mathrm{S/N})}
\label{eq-gauss}
\end{equation}
where $C=0.5$ is the theoretical value for a
well-filled array and high phase stability, but can be 1 or even more for sparse visibility
plane coverage such as VLBI.
 \citetalias{ALMA-TH} gives $C=1/0.9$ for ALMA, to allow for decorrelation, with a limit of 0.05$\theta_{\mathrm B}$, appropriate for a target after phase-referencing. However, if self-calibration allows  S/N of a hundred or better,   the relative position of bright, compact
peaks can  be found to an accuracy of a few percent of the
restoring beam size, although residual calibration and deconvolution  errors set an eventual limit.
The uncertainty in the width of the component is 
$\sqrt{2}\sigma_{\mathrm {pos fit}}$.

The same reasoning can also be
used to estimate the uncertainty in a width measured between
$3\sigma_{\mathrm{rms}}$ contour boundaries, as
$\sqrt{2}\times C,\theta_{\mathrm B}/3$.

Uniform weighting e.g. a low value of {\tt robust} in {\tt tclean},
does not necessarily improve position accuracy despite a smaller synthesised beam. It may increase the
beam sidelobes and rms noise near bright peaks.  This means that
although $\theta_{\mathrm B}$ is smaller, the S/N is also reduced and
the chance of clean artefacts is greater.

\subsubsection{Astrometric accuracy}
\label{sec-astrometry}
The apparent position of a target is
determined by the accuracy with which the phase corrections are
transferred from the phase calibrator (i.e. the effects of the time and
angular separation offsets and any antenna position errors), as well
as the accuracy of the phase calibrator position (which is usually better
than 1 mas for calibrators with VLBI positions). 
{\emph {The astrometric position should be
  measured from an image made after applying phase calibrator solutions
  only}}, so the phase error is based on Eq.~\ref{eq-transfer}.
Self-calibration using such an image model will retain the position
but cannot improve it.

For short observations (such that the atmosphere does not change noticeably) with a very compact array,  the signal
entering all antennas experiences the same atmospheric refraction (antennas 1,2 in Fig.~\ref{atmos1.png}), a
bulk shift of apparent position is possible, see \citetalias{SI99},  Ch.~15 (Ekers),
 eqs. 15-4 and 15-5. Such a situation is unlikely for ALMA except possibly for the ACA, but is more likely for compact, low-frequency telescopes and provides an introduction to the origin of position
errors: 
\begin{equation}
\sigma_{\mathrm{pos, compact}} = \theta_{\mathrm B} \times
\phi_{\epsilon, \mathrm{transfer}} \times \pi/180
\label{bigshift}
\end{equation} (including $ \phi_{\epsilon , \mathrm{transfer}}$
conversion to radians). Thus, a $60^{\circ}$ phase error could cause a
displacement of about $\theta_{\mathrm B}$. If the observation is
longer, the noise-error will be randomised and, as with 
longer baselines, the atmospheric turbulence dominates, causing the 
smearing in all directions previously described.

In the more usual case of a longer observation with a more extended array, the mean phase error is half of the difference between
successive phase calibrator scan solutions
$\phi_{\epsilon,   \mathrm{transfer}}$ (assuming the actual phase slope is
linear), reduced by the number of independent measurements. Using Eq.~\ref{bigshift}, the target
absolute position error is
\begin{equation}
\sigma_{\mathrm {pos Target}}=\theta_{\mathrm B} \times
 \frac{ \phi_{\epsilon,\mathrm{transfer}}\times \pi/180.}{2 \sqrt{(N_{\mathrm{indep}})}\sqrt{M}}
\label{poserr}
\end{equation}

Noise-related errors are reduced for many antennas if the atmosphere is independent above each antenna or group of antennas.
$N_{\mathrm{indep}}$ can be $N-3$ for VLBI arrays and as explained in Sec.~\ref{sec-qpr} for Eq.~\ref{eq:target_pherr}; for these VY CMa observations we take $N_{\mathrm{indep}}=2$ and the number of independent time intervals $M=6$.

If the phase calibrator position uncertainty is significant, it must be added in quadrature to $\sigma_{\mathrm {pos Target}}$. For weak targets and short enough observations, the noise errors can be significant,
so $\sigma_{\mathrm {pos fit}}$ should be
added in quadrature to $\sigma_{\mathrm {pos Target}}$ (as in Eq~\ref{eq-targetpherr}).  In the VY CMa
example this is not needed. Using $\phi_{\epsilon,\mathrm{transfer}}
 = 31^{\circ}$ and $\theta_{\mathrm B}$=200 mas gives $\sigma_{\mathrm
   {pos Target}}\sim31$ mas.

 If the phase calibrator has low S/N such that
its observed position error (see Eq.~\ref{eq-gauss}) is significant in
comparison with other errors this should also be included.
If the
target has S/N  much better than the phase reference source, you can improve astrometric accuracy by 
swapping 
the roles (``reverse phase-referencing''). First, complete normal self-calibration to get a good
image model (see Sec.~\ref{sec-nophref} if the phase calibrator is too weak to
use at all).  Then, take the target and phase calibrator data before
applying the phase calibrator solutions to either, and use the target model to
derive phase and amplitude solutions for the target as if it was a phase calibrator, and apply final per-scan
corrections to the phase calibrator as if it was a target.
Comparison of the resulting apparent and catalogued positions of
the phase calibrator provides the offset to be applied to the apparent
target position to derive its astrometric position.  

Dual-polarisation receiver systems on alt-az telescopes rotate during
observations and thus the separate polarisations undergo phase
rotation of the parallactic angle (as well as the amplitude effects,
Sec.~\ref{sec-polarisation}).  The parallactic angle rotation is given
by eq. 3 of \citet{Cotton2012} and the difference for two antennas on
a baseline is, very roughly, similar to their longitude separation  $\delta L$ (and thus the hour angle difference for a target).  This can lead to position errors for arrays observing in circular polarisation  when comparing separate LL and RR positions for circularly polarised sources, such as masers. For example, e-MERLIN has
maximum baselines spanning $\delta L\sim4^{\circ}$ and  synthesized beam
$\theta_{\mathrm B}$ 12--200 mas at $\lambda$ 1.3--21 cm. The offset per polarisation is of order
$\theta_{\mathrm B}$$\times$$\delta L/\pi$ e.g. $\!4/180\theta_{\mathrm B}$ (of opposite sign for each hand).  This is significant if it is comparable to the noise-based position error, i.e. if the  S/N $\gtrapprox \delta L/\pi$. The effect will give a spurious position offset, which can be cured by repeating the appropriate {\tt applycal} applying the parallactic angle correction prior to imaging.  This is less of an issue for ALMA as not only are its baselines shorter but it observes in linear polarisation and separate XX and YY maps are very rarely made. In total intensity the effect averages out and for polarisation observations the parallactic angle correction is customarily applied.

Comparing  apparent and catalogue positions for the check source (if any) gives a separate
estimate of astrometric accuracy, allowing for differences in S/N and angular separation from the phase calibrator.
 \citetalias{ALMA-TH} 10.5.2 outlines the ALMA strategy if high accuracy is needed.

\section{Practical self-calibration}
\label{sec-practical}
This section provides an overview, starting with a  walk-through of the main steps for typical straightforward self-calibration  in Sec.~\ref{sec-overview}. The subsequent sections provide guidance for the main decisions: whether self-calibration is possible in Sec.~\ref{sec-whether}; making the model image in Sec.~\ref{sec-imstart_cont}; parameter settings in the first rounds of phase and amplitude self-calibration in Secs.~\ref{sec-gaincal} and~\ref{sec-ampscal}.  Sec.~\ref{sec-goodbad} illustrates how to tell whether the solutions are good and Sec.~\ref{sec-applycal} covers applying the solutions. How to modify the workflow is introduced in Sec.~\ref{sec-alt} and Sec.~\ref{sec-stop} provides criteria for when the self-calibration is as good as possible.
Examples are taken from the VY CMa Science Verification data introduced in Sec.~\ref{sec-afterphref} using continuum channels. For additional issues when using a spectral line for self-calibration or other special circumstances, see the next Section~\ref{sec-sc}  and  for more practical examples, see the links in Appendices~\ref{ap-schools} and~\ref{ap-casaguides} or the descriptions in \citet{Brogan18} sec. 2.4 covering ALMA continuum, spectral line and mosaic.

Sec.~\ref{sec-overview} builds up corrections incrementally, so the previous solution tables are applied when deriving the next round of calibration. This makes it easy to check the tables have improved (e.g. the second round of phase solutions should show small deviations compared with the first), and if there are initially large errors, applying longer-timescale corrections for this can improve the stability of incremental shorter-scale solutions.  On the other hand, if the first model is not accurate and/or too short a solution interval is used,  bad solutions can lead to artefacts `baked in' to the image and subsequent models, and if many rounds of calibration are used, accumulating a long list of tables is prone to human error.

An alternative strategy, of improving the model in each round but generating each phase calibration table without applying any previous self-calibration solutions is used in the NRAO template imaging script, see
Appendix~\ref{ap-other}. This avoids carrying forward calibration errors due to bad data or an inadequate model, but if large corrections are needed there may be more failed solutions and it is harder to spot bad solutions.  However, for full polarisation, this strategy can make it easier to ensure that the final solutions applied use a consistent reference antenna. This strategy is outlined in Sec.~\ref{sec-alt} and a hybrid approach can be used (with careful book-keeping).
In either strategy, the best previous phase corrections are applied when deriving amplitude solutions.  

\subsection{Quick-start overview}
\label{sec-overview}
This section summarises how to self-calibrate using a typical continuum target in  phase-referenced ALMA data in a single configuration and band, with all standard
calibration applied. If your target is so extended that there is missing flux on short baselines see  Sec.~\ref{sec-configs}; a variety of other situations are also covered in  later sections. 

We assume that all reduction was performed using CASA 4.2.2 or later; for very early cycle data, check the weights (see Appendix~\ref{ap-casaguides} CASA Guides).

At risk of stating the obvious, check the observing proposal if available, the QA2 report, pipeline weblog or any other information about the observational set up and notes on data processing so far.  Also check any
other information about your target to guide you in what field of view
is needed, what sensitivity and resolution is intended, where to
expect emission to appear etc. -- although you should also be prepared
for the unexpected.

Sec.~\ref{sec-whether} explains when self-calibration is possible; in brief, for  ALMA 12-m array with $\sim40$ antennas and $\sim30$ min on-target, this is likely to require image S/N of order 100; for less sensitive (e.g. ACA) observations S/N 50 or less might suffice.

\begin{figure}
 \includegraphics[width=17cm]{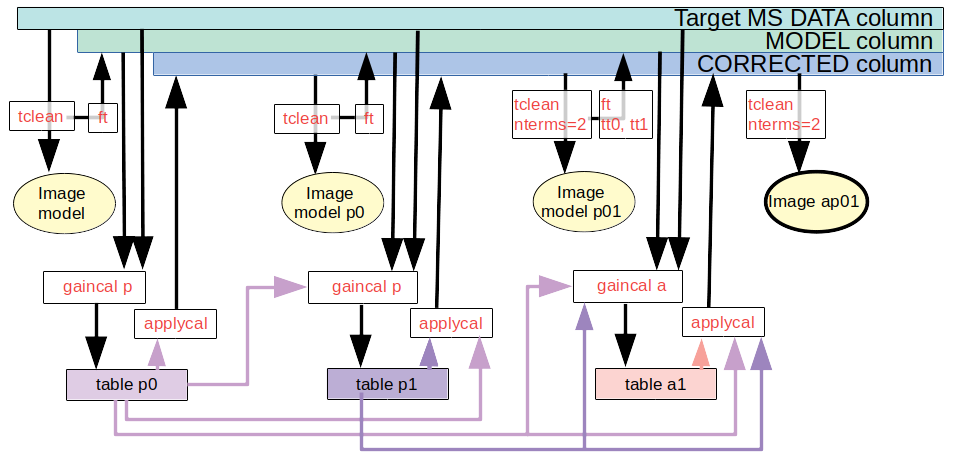}
\caption{\small Flow chart of self-calibration using incremental solution tables. Start  at top left. The black arrows connect products such as the target MS and the initial image via the tasks which are used and the mainly horizontal arrows shown the accumulation of calibration tables.}
\label{CalFlow.png}
\end{figure}

Figure~\ref{CalFlow.png} shows a typical workflow for self-calibration
accumulating gain tables (see Sec.~\ref{sec-alt} for alternative approaches).  The
starting point is a Measurement Set (MS) with phase calibrator
corrections applied. In all cases it is assumed that you are making an
image from a continuum-only channel selection but all spw are
represented (for other situations see Sec.~\ref{sec-transfer}).  As implemented in CASA {\tt gaincal},  the visibilities are compared
with the model per baseline and the corrections needed are estimated
and optimised per-antenna using a least squares (or similar)
minimisation.  Secs.~\ref{sec-gaincal} and  Appendices~\ref{ap-gaincal} and ~\ref{ap-applycal} summarise guidance in setting the main parameters for the CASA tasks to derive and apply calibration.

Each stage below covers making a correction table, following Fig.~\ref{CalFlow.png} in performing two rounds of phase-only self calibration followed by amplitude self-calibration.  
\begin{description}
\item[{\bf Self-cal 1.}] Start with the MS with  phase reference corrections applied.
  \begin{enumerate}
\item    Split out the target data, so
  this now forms the {\sc data} column of a new MS.
\item Make a target continuum image (see Sec.~\ref{sec-imstart_cont}, check that the model will be saved). Mask carefully close to the most believable emission. Using a simple,
but possibly incomplete model is better than starting with a more
complicated model that may contain doubtful features. 
\item The model is used by {\tt gaincal} (Appendix~\ref{ap-gaincal}), which  compares it
  with the MS {\sc data} column. Specify the same spw/channel
  selection as used for the image. Solve for phase only and, for simplicity, assume you can
  solve for each spw separately and use {\tt gaintype='G'}, solving
  for polarisations separately. For most data, the target scan length is a good starting solution interval. Here, the resulting gain table is called p0.
  Check that the solutions are OK (see Sec.~\ref{sec-goodbad} and examples in Fig.~\ref{tab-cals}; (a) shows good solutions after phase referencing).
\item {\tt applycal} p0 to the MS (all spw, no channel selection), forming the {\sc corrected} column (Appendix~\ref{ap-applycal}).
  \item Image again, making sure to use the {\sc corrected} column (the
    default in {\tt tclean}, if it is present).  Check that the image
    $\sigma_{\mathrm{rms}}$ has gone down and the peak is stable or has increased (not decreased significantly), and the structure does not have any artefacts e.g. sidelobe-like.
\end{enumerate}
\item[{\bf Self-cal 2.}] Simply having a image model more accurately representing the target field can allow you to refine phase calibration. A  shorter solint may be
  possible (see Sec.~\ref{sec-SN}),
  \begin{enumerate}
    \item Check that the model is in the MS (Sec.~\ref{sec-ft}).
\item Repeat step 1.(d) but in {\tt gaincal}, apply table p0 as a gain
  table, generate a second table p1 (possibly with a shorter solint).  p1 should show smaller, additional corrections. See Sec.~\ref{sec-goodbad} for guidance, a quick check is that solutions should show a consistent trend for both polarisations; if the discrepancies are noise-like and comparable to the range of time-dependent correction the solint is probably too short (or the previous calibration has not been applied).
\item Repeat step 1.(e) but in {\tt applycal} apply both p0 and p1, which will \emph{overwrite} the previous {\sc
  corrected} column.
\item Image again; use {\tt deconvolver='mtmfs', nterms=2} if there is enough S/N to
  measure the spectral index (Sec.~\ref{sec-mfs}), and check again.
  \end{enumerate}
  \item[{\bf Self-cal 3.}] (and further) If you have not reached the expected noise or dynamic range limit, investigate whether the model can be improved or the solution interval shortened for more rounds of phase
    self-calibration. If enough S/N do amplitude self-calibration:
 \begin{enumerate} 
    \item  Check that the model is in the MS (Sec.~\ref{sec-ft}).
 \item Run {\tt gaincal}, amplitude only, applying p0 and p1, with a
   longer solution interval, typically at least a scan or even longer. For ALMA, normally, use {\tt gaintype='T'}
   to average polarisations. This makes table a1, see Fig.~\ref{tab-acals} for examples of good and bad solutions.
\item In {\tt applycal} apply p0, p1 and a1 (and any additional tables
  if needed).
 \item Image again, check as before. If the model has changed greatly (e.g. more extended, bright flux is included) consider repeating phase calibration cycles.
  \item  For S/N many hundreds or more, you might
  want to do more cycles of self-calibration as the model improves (or even use a better model to start again from the phase-only stages). You can decrease the solution interval but be very careful if the structure is complex. See Sec.~\ref{sec-stop} for stopping criteria; once there is no more improvement, make a perfect final image
    with PB correction!
  \end{enumerate}
    \end{description}

\begin{figure}[t]
  \includegraphics[width=8.5cm]{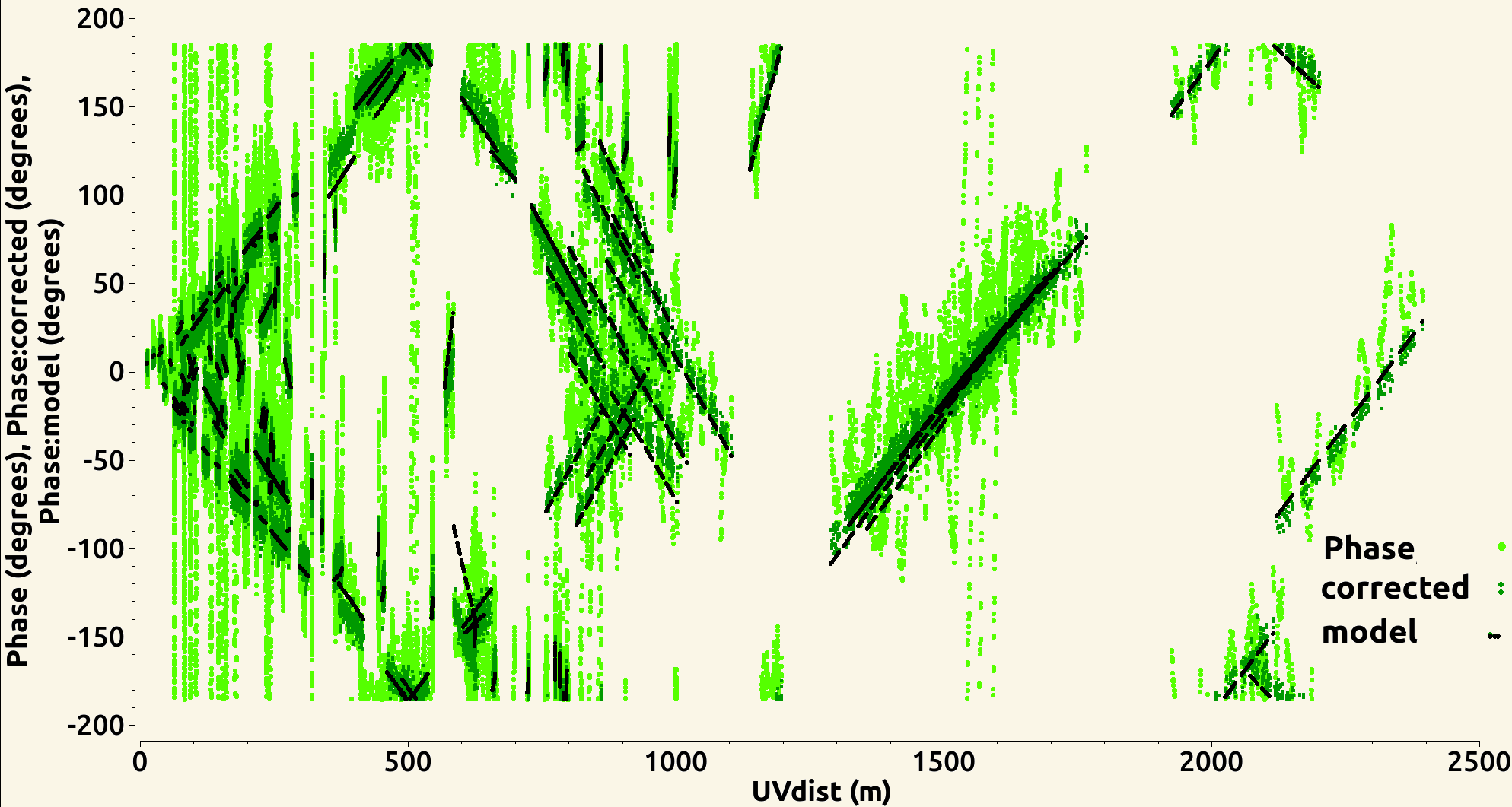}
    \includegraphics[width=8.5cm]{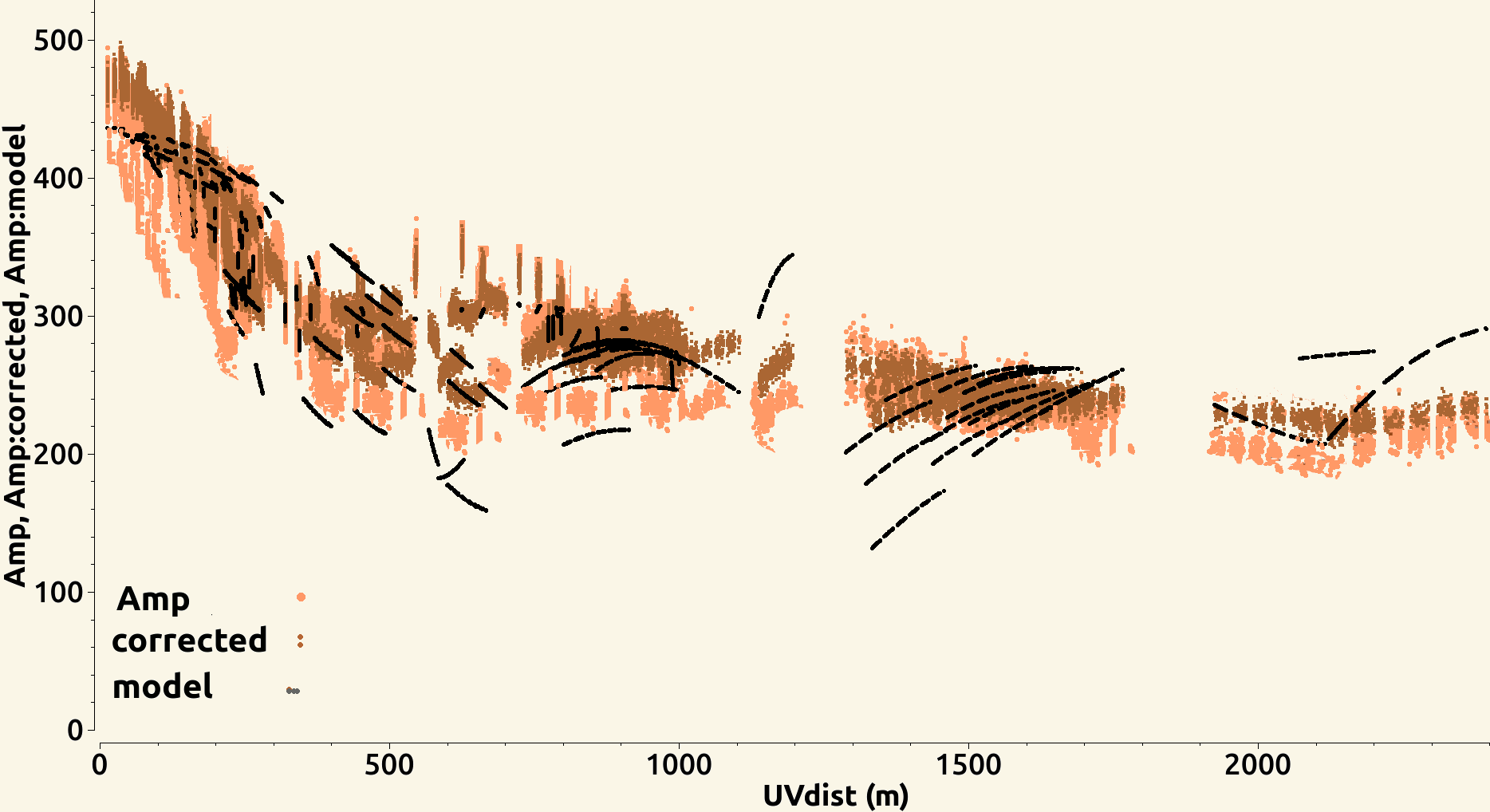}
    \includegraphics[width=8.5cm]{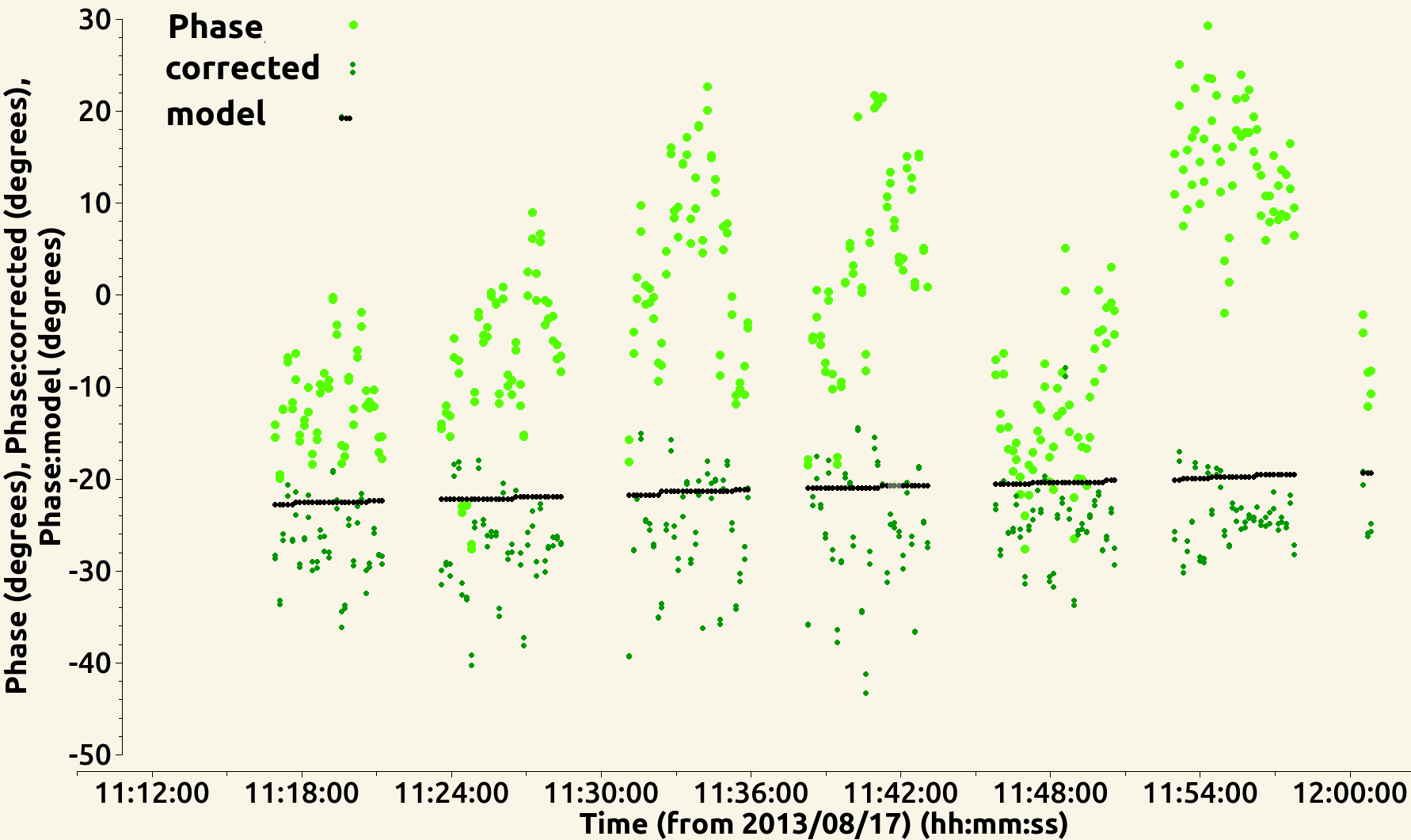}
      \includegraphics[width=8.5cm]{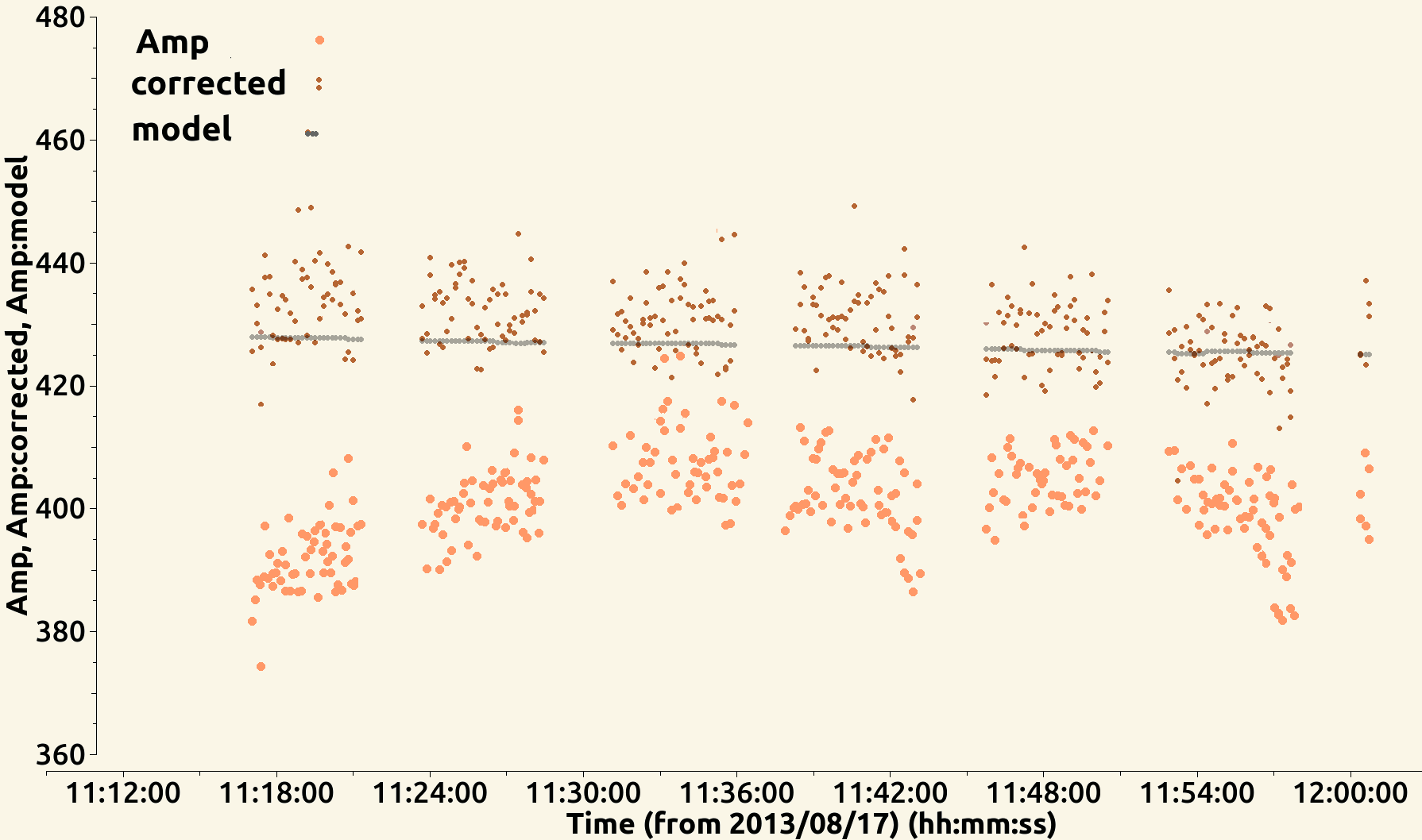}
      \caption{\small VY CMa data as used in Sec.~\ref{sec-afterphref}.
Visibility data for a single channel after applying phase calibrator solutions, before and after (`corrected') self-calibration,  compared with model.
Top left: Target visibility phases v. $uv$ distance, all baselines. 
Top right: Target visibility amplitudes v. $uv$ distance, all baselines. 
Bottom left: Target visibility phases v. time, single baseline.
Bottom right: Target visibility amplitudes v. time, single baseline.
      }
\label{VY_raw-corr-mod.png}
\end{figure}
If you change your mind and want to repeat a
 step, make sure to close any plotting windows.  To repeat an imaging
 step, make sure that the correct table(s) have been applied first; to
 repeat a calibration step make sure that the correct model has been
 Fourier transformed into the MS and the correct tables, if any, are applied in {\tt gaincal} and {\tt applycal}. 
 To go back to the start, use {\tt clearcal} and
 {\tt delmod}.  If you want to undo the effects of a gaintable used to
 flag data by failed solutions, also use {\tt flagmanager} to restore
 the flagged data (assuming that you backed up the flagging state first), see Sec.~\ref{sec-flag}. See Sec.~\ref{sec-alt} for more guidance in modifying the workflow.

Fig.~\ref{VY_raw-corr-mod.png} top left shows the target
phase  before/after self-calibration. The divergence with $uv$ distance indicates a
source offset from the phase centre. Phase
self-calibration has significantly reduced the scatter and the calibrated phase is mostly very close to the model. This can also be seen in phase as a function of time, bottom left.
Fig.~\ref{VY_raw-corr-mod.png} top right shows a similar reduction in scatter from amplitude self-calibration.  The model does not represent the data fully, probably due partly to weak flux not being included in the model (corrected data exceeds the model, more so on some shorter baselines) and partly due to some remaining phase decorrelation (model exceeds corrected data, mostly on some longer baselines). 
Fig.~\ref{VY_raw-corr-mod.png} bottom right shows the increase in amplitude for a shorter baseline, due to correcting decorrelation.
Fig.~\ref{VYcont_images.png}
right and Table~\ref{VYstats} show that the most substantial
improvement (a factor of 3 in S/N) comes from the initial phase
self-calibration.  The final amplitude self-calibration should not
change the flux density by more than a few percent, as this is mainly determined by the input model, but
it does reduce the noise.

The basic principles of simple self-calibration are, {\emph{whatever tables you apply in
{\tt gaincal}, apply these tables plus the new one in {\tt applycal}}}  (there are rare exceptions, e.g. Sec.~\ref{sec-dgc})
and, {\emph{always check that the solutions are not pure noise (Sec.~\ref{sec-goodbad}) and the image
S/N improves}.  The following sections explain how to make the choices
for this.
  
\subsection{How to tell if a data set is worth self-calibrating}
\label{sec-whether}

ALMA science observations are calibrated and imaged using a pipeline or standard scripts and the quality is checked carefully at all stages (known as Quality Assurance stages QA0 to QA2, or QA3 if later improvements are made, see  \citetalias{ALMA-TH}. 
In a typical ALMA data delivery or
retrieval of archive data, there are ready-made sample images,
with phase calibrator and other corrections applied. Products which passed QA2 are usually well-calibrated  but if you
suspect problems, see the pipeline or QA2 logs and refer to
Sec.~\ref{sec-errorrecognition}. You can contact your ARC\footnote{\url{https://almascience.eso.org/help}} if you have queries.

Thus, the standard images can usually be used to start assessing whether they could be improved by self-calibration.

\subsubsection{Measuring the signal to noise ratio}
In order to decide whether it is worth self-calibrating,  measure the S/N for a target image, usually for line-free continuum. The web log may give you values for the peak and
the off-source rms measured before PB correction. Otherwise, the pipeline or QA2 produces an  aggregate continuum image, which should have been made
excluding any lines. If you need to make an image yourself, do not apply PB correction.

To measure S/N (peak/noise rms) using the viewer draw round the peak and use the Maximum in Jy/beam -- it is the peak which counts, not the total flux. The noise $\sigma_{\mathrm{rms}}$ should be measured in as large a region as possible avoiding the target emission and any noisy regions at the edge of a  primary-beam corrected image,  as shown in Fig.~\ref{PBim.png}, left.
You can also use {\tt imstat} in a script; see Appendix~\ref{imstat} for a script fragment to do this.
If
there is very extended emission and it seems impossible to find a
suitable region, you can make an image without PB correction
and measure the off-source noise to the edges, or the $\sigma_{\mathrm{rms}}$ of the
residual image.  You will need to make an image for self-calibration
anyway, see Sec.~\ref{sec-imstart_cont}. 
\begin{figure}
\includegraphics[width=6cm]{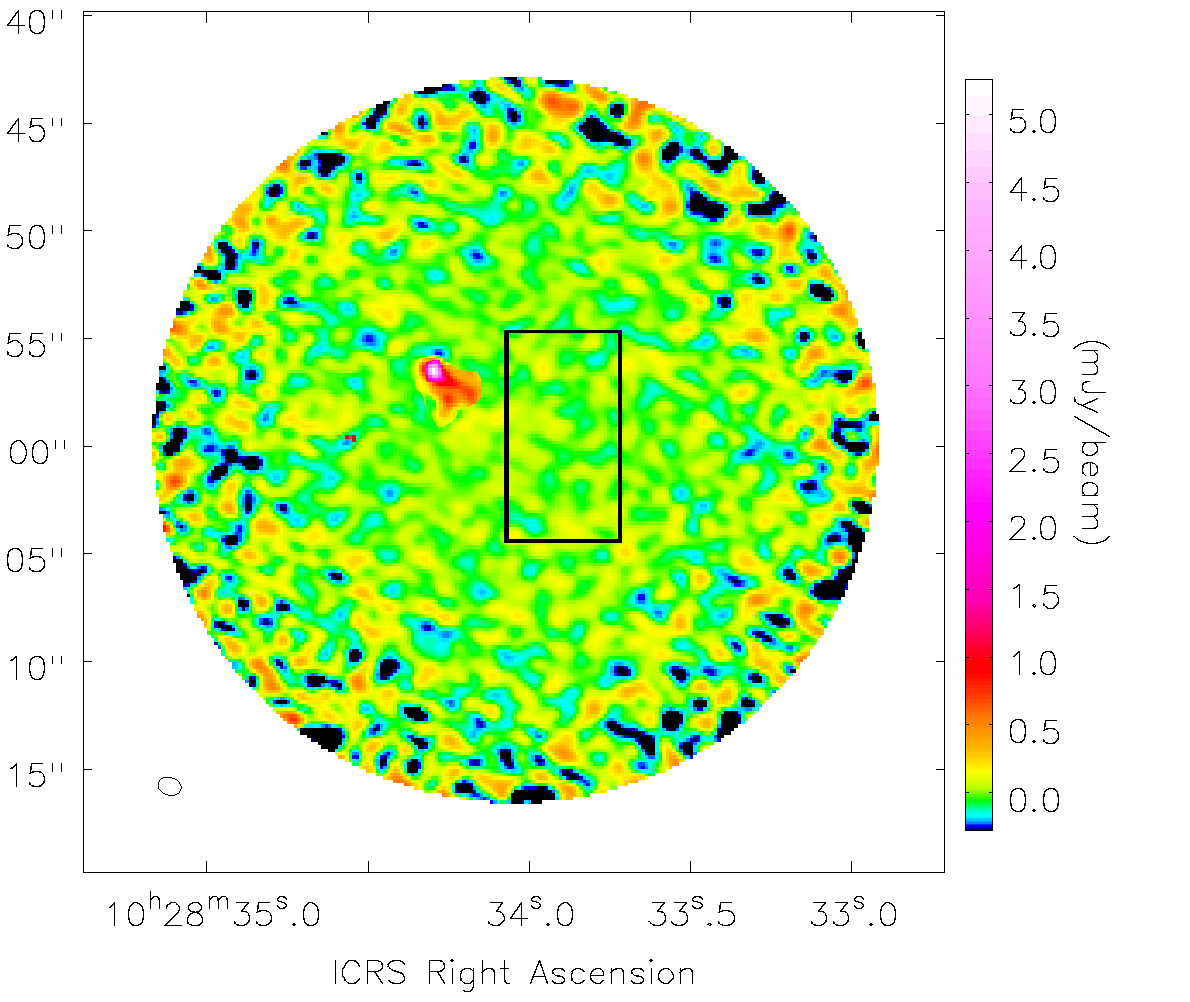}\hspace*{-0.1cm}
\includegraphics[width=5.8cm]{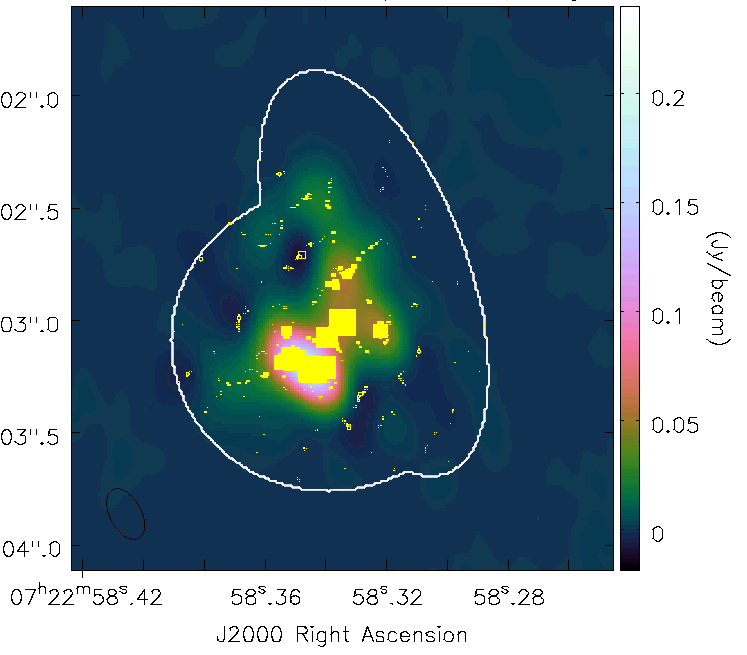}
\includegraphics[width=5cm]{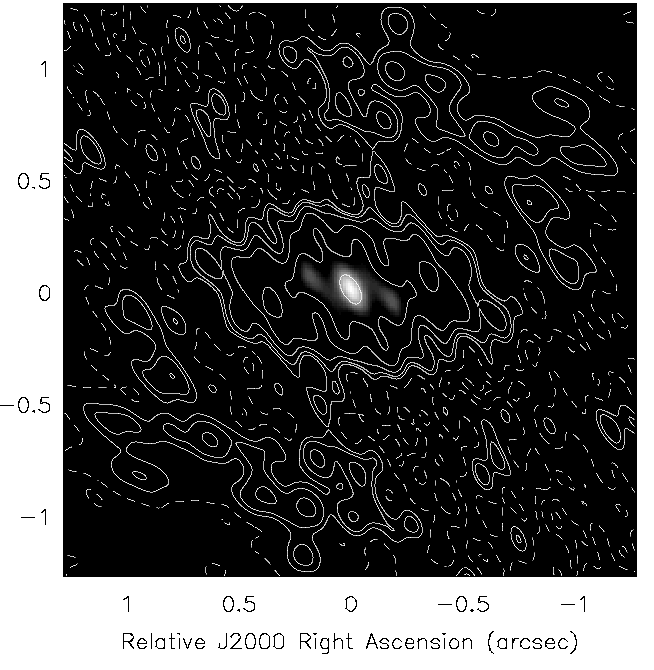}
\caption{\small Left: Example pipeline image with PB correction
  showing possible location of box to measure rms. Centre: VY CMa
  showing Clean Components (yellow) including positive (filled) and
  negative (hollow) spurious components due to over-cleaning, within
  and around the rim of the  mask (white). Right: VY
CMa synthesised beam (psf); contours at (--5 , --2.5, 2.5, 5, 10, 20,
40, 80)\% of the beam peak; the greyscale shows the beam at $>$50\%.}
\label{PBim.png}
\end{figure}

You can also inspect the visibility phases in {\tt plotms}, such as are shown in light
green in Fig.~\ref{VY_raw-corr-mod.png} lower left. For speed, just plot all or some baselines to the reference antenna (covering the longest baselines) and iterate through them.  If you can see a clear pattern, then the target probably can be self-calibrated.
For continuum, average line-free channels. You can only average continuous channel selections.  This may not give enough S/N; if so, you can back up the flags and then flag the line channels, which will allow entire spw (or even all spw) to be averaged. Restore the flagged channels before the actual calibration. If necessary average in time, which will also give an idea of a suitable solution interval. Beware that if some averaging intervals contain less data they will appear noisier, e.g. Fig.~\ref{tab-cals}(h).
If the phase is coherent but wraps very
fast on some or all baselines (as in Fig.~\ref{tab-cals} (g)), see Secs.~\ref{sec-selective} or~\ref{sec-rate}; this may indicate that the target is very offset from the pointing centre. In this case, you can improve plotting the S/N by using the {\tt plotms transform} tab to shift the phase centre used for averaging to the peak position, to reveal residual wiggles due to phase errors.

\subsubsection{Conditions for self-calibration}
\label{sec-cond}

The conditions can be assessed more quantitatively.
It is likely to be worth self-calibrating, at least for phase, if condition (1) below is met; (2) and (3) are typical situations.
\begin{enumerate}
\item The target S/N is high enough (typically 3), per antenna,  in the longest  interval over which useful corrections can be made, typically a scan. For a typical execution of a single EB with the 12-m array in full operations ($\sim$43 antennas),  30 mins of on-target data in 5-min scans using the full available bandwidth of which about half,  $\sim$4 GHz, is continuum, self-calibration should be possible if the image S/N is 100 or more. This is not a hard limit as the initial $\sigma_{\mathrm{rms}}$ is probably higher -- maybe by a factor of a few -- than the potential noise. Moreover, additional averaging over spw and polarisations may be possible.   
\item The image noise $\sigma_{\mathrm{rms}}$ is worse than expected for the actual observing conditions. However, it can be worth self-calibrating even if this is not the case, as small to moderate phase errors can degrade an image without noticeably increasing the noise. 
\item The target dynamic range is hoped to be higher than that of the phase calibrator, so even if the latter's solutions were transferred perfectly, the target solution accuracy can be improved (see also  Sec.~\ref{sec-tderror-DR}). Since the target is typically observed in longer scans, this can be the case even for a brighter phase reference source (although the relative bandwidths used also have to be considered).
\end{enumerate}

Section~\ref{sec-SN} derives expressions for the minimum image signal to noise ratio,  S/N$_{\mathrm{sc}}$, which is  `enough' for self-calibration, based on  the relationship between the whole image S/N and the S/N per antenna, per averaging interval.  The longer the whole observation contributing to the image, and the more antennas, the smaller is the contribution of an individual antenna or a single scan.  For example, for the VY CMa data used here, with only 20 antennas and 1 GHz continuum bandwidth, the minimum S/N required for self-calibration is no more than $\sim${50}.

It is usually worth self-calibrating even in the absence of obvious noise or dynamic range problems, as  moderate phase errors can degrade an image without noticeable increasing the image $\sigma_{\mathrm{rms}}$.  If your data have a lower S/N$_{\mathrm{sc}}$ than is predicted to be required for straightforward self-calibration, see Sec.~\ref{sec-lowsn}, but if the S/N in phase-referenced data is less than about 10, it is very unlikely to be self-calibratable under any conditions. In any case, not all images can be improved by self-calibration -- see Sec.~\ref{sec-errorrecognition}.

\subsection{Making a preparatory image and model}
\label{sec-imstart_cont}

Calibration compares a model with the data and derives the corrections needed to make the latter correspond more closely to the former. The model for an unresolved, extra-solar calibration source may be simply a point of known flux density at the pointing centre, inserted using task {\tt setjy}. 
For target self-calibration, the model is usually composed of the Clean Components (CC) from a previous set of image products for the target.  We describe here how to prepare that, but sometimes you may use a product from a different observation of the same source at comparable resolution (see  Sec.~\ref{sec-ephemeris} for Solar system sources or Appendix~\ref{ap-CC} for using {\sc AIPS} images).

For normal ALMA observations, 
all previous calibration and flagging up to phase referencing should
be applied, e.g. run {\tt scriptForPI.py} on the archive products (see Appendix~\ref{ap-casaguides}, the
pipeline CASA guide). Alternatively, a local ARC may be able to supply a calibrated MS, see e.g. the NRAO Science Ready Data Products service, Appendix~\ref{ap-other}.

Split out the target without continuum subtraction.  Find the
line-free channels -- e.g.  use the pipeline/QA2 selection, but check
for yourself. In ALMA data there may be lines even in TDM spw designated continuum.
At mm and shorter wavelengths, the S/N is often greater in the continuum than for a spectral line, relative to the available bandwidth for each, so we start by describing using a continuum model.  See Sec.~\ref{sec-line} for additional issues in using a spectral line for self-calibration.

If the data are FDM with thousands of channels, it can
save a lot of imaging time to make a second copy with averaging to
TDM
resolution (and select the renumbered continuum channels).  If there has been any channel-selective flagging (e.g. due to  RFI or  arising from failed bandpass or $T_{\mathrm{sys}}$ solutions) then channel-dependent weights are required, see Sec.~\ref{sec-weights}. In this manual, we use channel selection to exclude lines from continuum self-calibration but the lines can alternatively be (temporarily) flagged. This approach is used in the NRAO template imaging script (link in Appendix~\ref{ap-other}) and requires channel-dependent weights. The script also gives guidance on  limits to frequency- and time-averaging to avoid smearing. This will not affect typical ALMA averaging within the primary beam FWHM;  Appendix~\ref{ap-smear} shows approximate limits for channel- and time-averaging which can cause smearing; see Sec.~\ref{sec-errorrecognition} for more references, especially if you are using wide-field cm-wage data.  

The advantages in using all spw assume that they provide roughly equal continuum sensitivity; see 
Sec.~\ref{sec-comb} and Secs.~\ref{sec-transfer} -- \ref{sec-selective} if you are combining multiple executions and/or need to
be more spectrally selective and make an image accordingly.  

This manual assumes the use of {\tt tclean} but in principle any method of generating an image model could be used, as long as the model can be Fourier transformed into the MS and any additional information (e.g. frequency dependence) is recognised.

\subsubsection{Imaging for self-calibration}
\label{sec-imsc}
If you are unfamiliar with interactive imaging, see the relevant CASA Guides (Appendix~\ref{ap-casaguides}).
Difficult/exceptional situations (for ALMA) are covered in Secs.~\ref{sec-nophref} to~\ref{sec-ephemeris}.

For early phase-only calibration, it is better to underclean than
overclean; calibration will not usually remove real emission but
spurious CC may lead to solutions freezing the artefacts into the
data.  You can usually tell if this has happened as the image S/N will
not improve or sidelobes will worsen. The risk is greater for sparse
visibility plane coverage.  Use any a-priori knowledge of the target
field to guide you.

For the first  image model:
\begin{itemize}
  \item If continuum, either make a flag backup and then flag the line
    channels, or set the continuum channel selection in {\tt tclean
      spw} parameter; here we
    assume the latter. If using a spectral line, see Sec.~\ref{sec-line}.
  \item Use {\tt robust=0.5} or greater, usually, as lower values can produce
    artefacts.
\item For homogeneous arrays and many similar baselines (e.g. ALMA 12-m array in all but the most extended configurations) the weights should be proportional to sensitivity, see Sec.~\ref{sec-weights} and \citet{Brogan18}.  If some antennas are very distant or otherwise hard to calibrate, or for heterogeneous arrays e.g. e-MERLIN,
    avoid tapering or other downweighting of particular
    antennas if possible unless vital to avoid artefacts (all spacings should be fully represented to
    optimise the calibration of all antennas).  
\item You do not necessarily need to image the whole field of view but
  all detected emission must be included, whether or not it is of
  scientific interest, leaving an outer signal-free margin of $\sim$$10\%$ to avoid aliasing. At this stage don't do the PB correction (see Sec.~\ref{sec-het} if this would affect your target).
  \item Use {\tt savemodel ='modelcolumn'} so that your final CC are Fourier transformed into the MS, to act as a model for the next round of self-calibration (see Sec.~\ref{sec-ft} to check this).
\item Mask carefully (usually interactively). Take care not to include sidelobes. Clean
  gently.  If you have one or more bright peaks plus possibly weak emission, it is often best to mask just the peaks initially and use a low  {\tt cycleniter} e.g. 10 or 20; as the peaks are subtracted and stored as CC, the residual S/N of weaker emission will improve and you can extend the mask and increase the {\tt cycleniter} as needed, also see e.g. \citet{Brogan18} 2.4.1 Mira example.
\item Stop cleaning when you can no longer be sure that emission structure left in the residual image is definitely part of the source (rather than just noise fluctuations). At this stage there is no need to include faint, extended emission you are not absolutely sure is real, nor to clean down to a predefined level of e.g. $3\sigma_{\mathrm{rms}}$.
\end{itemize}  

 Fig.~\ref{PBim.png}, centre, shows good Clean Components 
  overlaying real emission,  but also artefacts due to
  overcleaning. Fig.~\ref{VYcont_images.png} right shows a safer collection of CC. When masking, be wary of emission 
  which has a positive or negative counterpart in a position
  mirrored about a strong peak. These may be artefacts due to
  amplitude or phase errors (Sec.~\ref{sec-phase-fidelity}).

The peak (not integrated) flux density is what determines whether you
can self-calibrate, since that represents the  signal on baselines to
all antennas.  However, as all emission must be imaged,  if the source is very extended, {\tt tclean
  deconvolver='multiscale'}  processes multiple CC more effectively, equally
usable as models. When you come to making a model for continuum amplitude self-calibration, {\tt 'mtmfs'} also lets you fit for spectral index or higher terms, see Sec.~\ref{sec-mfs}. Sec.~\ref{sec-optimise} discusses further issues
which can make imaging trickier.

You can compare the location of the Clean
Components ({\tt .model} image, display in the casaviewer as a marker map)
with the {\tt .psf} (see Fig.~\ref{PBim.png}, centre and right) to see if the
distribution is too much like sidelobes. It is not easy to edit CC in
CASA, but in any case it is better to optimise cleaning in the first
place by careful masking and setting a realistic threshold (a few
times $\sigma_{\mathrm{rms}}$ to begin with).  The once-common
practice of rejecting components below the worst negative has no real
basis (see \citetalias{SI99} Ch.~13 Perley) since artefacts can be both positive and negative.

If a plot of the target visibility amplitude against $uv$ distance shows a steep rise in emission on the shortest baselines, suggesting  extended emission which is not properly sampled, or there are nulls in the visibilities (e.g. for a planet), see  Sec.~\ref{sec-ephemeris}.
 
\subsubsection{Inserting the model for self-calibration in the MS}
\label{sec-ft}
      {\tt tclean savemodel='modelcolumn'} will Fourier transform the CC into the MS, see Appendix~\ref{ap-casadoc} for link to {\tt tclean} documentation. 
      This allows  {\tt gaincal} to use this as a model in the next round of self-calibration. Each time the model is saved, it overwrites previous models.  Check the casalogger at the end of  {\tt tclean} to ensure the model has been saved, especially if  {\tt tclean} has been interrupted or if running CASA versions $<$ 6.2.1.

The  {\tt tclean} documentation explains how to run the model insertion process without re-cleaning, if necessary. This can be done if {\tt tclean} is interrupted, or even if you want to revert to a previous model as in Sec.~\ref{sec-alt}, as long as you still have the clean products.
Re-run  {\tt tclean} with the parameters and image name used for the desired model, but  with {\tt niter=0, savemodel='modelcolumn', calcpsf=False, calcres=False, restoration=False} (see the {\tt tclean} documentation}).

If this is not possible (e.g. you are using a model from a different observation) you can  use task {\tt ft} to insert the model (similarly, if possible  {\tt usescratch=True}, to write a full modelcolumn). If you  used {\tt tclean mtmfs nterms=2} or higher, set the same value for {\tt nterms} in {\tt ft} and include {\tt model.tt0, model.tt1}.

In all cases, use of the model column is preferred to a virtual model and the latter should never be used for ephemeris objects or imaging using the wide field parameters.  If in doubt use {\tt plotms} to check the model is present and reasonably close to the data, as in e.g. Fig.~\ref{VY_raw-corr-mod.png} top right. In initial rounds of phase-only self-calibration weaker or extended structure may not be well represented, but for amplitude self-calibration the model flux should follow the corrected data well.
If not, use  {\tt gaincal solnorm=True} (see Appendix~\ref{ap-gaincal}).

\begin{figure}
  \includegraphics[width=10.cm]{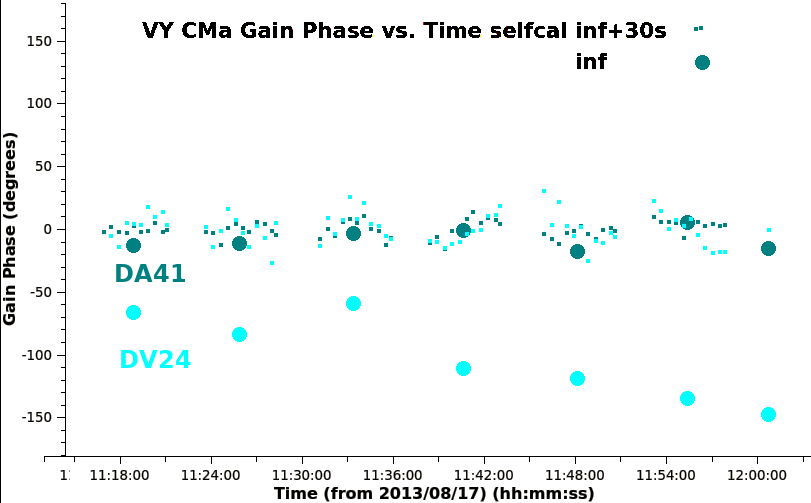}
    \includegraphics[width=7.cm]{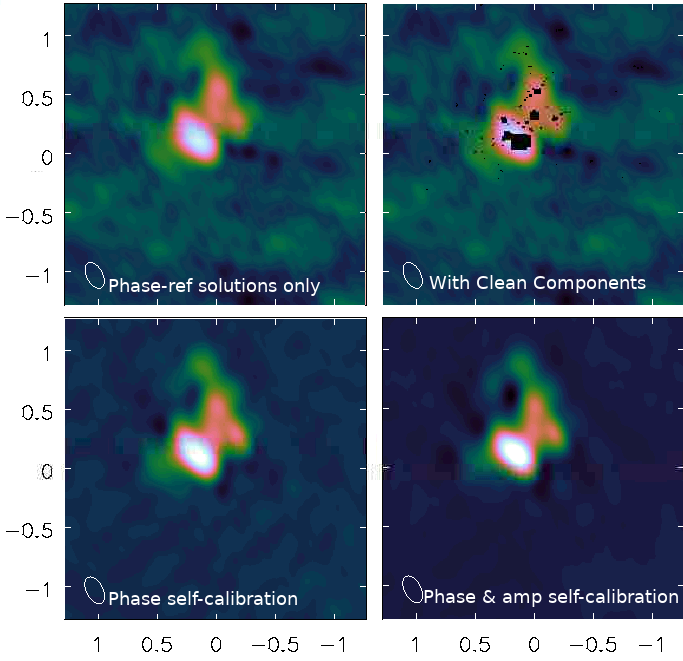}
\caption{\small Left: VY CMa continuum per-scan (`inf') phase solutions for X polarisation, shown
  as large dots, overlaid with incremental 30-s solutions, for two antennas. Right: VY CMa continuum images at the calibration stages labelled, see Table~\ref{VYstats}.}
\label{VYcont_images.png}
\end{figure}

\begin{table}
  \begin{tabular}{|l|ccc|}
    \hline
                               & Peak  (mJy beam$^{-1}$) &
    $\sigma_{\mathrm{rms}}$ (mJy)& S/N\\
\hline
    Phase calibrator solutions only   &179 & 3.0  & 60\\
    Phase self-calibration (solint inf)   &198 & 1.1 & 180\\
        Phase self-calibration (solint 30s)    &199 & 0.8 & 250\\
    Phase \& amp self-calibration&199 & 0.6 &325\\
\hline    
  \end{tabular}
  \caption{\small Improvements in the signal to noise ratio (S/N) during
    incremental continuum self-calibration of VY CMa (see also Fig.~\ref{VYcont_images.png}).}
  \label{VYstats}
\end{table}

\subsection{Deriving the self-calibration solution table ({\tt gaincal})}
\label{sec-gaincal}

The CASA task {\bf{\tt gaincal}} is used to derive solutions by comparing the data and model columns of an MS and calculating per-antenna corrections to apply to remove discrepancies from the visibility data. Each visibility represents a single baseline measurement but the dominant errors are due to the atmosphere or other effects on specific antennas. Thus,  solutions are derived for all baselines to each antenna and (by default) a least squares procedure is used to decompose the baseline solutions to find a representative solution for each antenna which is compatible with good solutions for all antennas.  See Sec.~\ref{sec-goodbad} for assessing the effects of changing parameters on the quality of calibration tables.
Some parameters can usually be left as default (see Appendix~\ref{ap-casadoc}, CASA documentation). The {\tt gaincal} parameters most relevant for deriving and applying ALMA self-calibration and suggested values are given in Appendices~\ref{ap-gaincal} and~\ref{ap-applycal}.

\subsubsection{Choice of reference antenna}
\label{sec-refant}
Having inserted a good model in the MS, the next choice is reference
antenna (known as refant).  The reference antenna phases are set as the origin of phase. The refant(s) should be antenna(s) with as much good data as
possible, near the centre of the array. Phase errors are usually less
on short baselines so a central reference antenna  helps ensure that it contributes to as good solutions as possible and thus an accurate origin of phase\footnote{For
  heterogeneous arrays, use a sensitive antenna,
  e.g. phased ALMA is normally the refant when in mm VLBI}.

Look for what was used  by the pipeline  or other prior bandpass and phase calibrator calibration. The pipeline provides logs for the {\tt hif\_refant} task, or you can check plots of phase solutions, which should be zero for the reference antenna.
For full polarisation,
the same refant should be employed for each set of data using a given polarization calibrator observation, using {\tt gaincal refantmode='strict'}, which will fail all solutions in a given interval if the solution for the reference antenna fails. Similarly, if using multiple mosaic fields to derive the model, a fixed reference antenna should be used.  Otherwise,  {\tt gaincal, refantmode='flex'} allows alternative antennas to be used but it is worth specifying a list of refants, since the defaults may not be the most suitable.

\subsubsection{Phase self-calibration and choosing the solution interval}
\label{sec-SN}
One almost always starts with phase-only self-calibration, since for ALMA and most cm to sub-mm interferometers, phase errors have the biggest impact on image fidelity, as explained in Sec.~\ref{sec-afterphref}.  Phase solutions should be derived separately for the two hands of polarisations ({\tt gaintype 'G'}) and each spw, at least for the first round of self-calibration, since there are often significant phase deviations, unless they have to be combined in order to get enough S/N, see Sec.~\ref{sec-lowsn}. 
The choice of averaging interval for each phase solution depends on the S/N of the data and on how good the model is, both of which should improve in successive rounds of self-calibration.

The usual ALMA approach is to start with the length of a scan, in
order to reduce the effects of the separation from the phase calibrator and
provide an improved model. 
If a bright target is known to have a very simple structure entirely
visible in the first image, you might start with a shorter solution
interval.  In any case, after applying each set of solutions, the
image S/N should improve and shorter solution intervals may be possible. Section~\ref{sec-sols}, Figs.~\ref{tab-cals} provides guidance for inspecting solution tables.
How  short an solution interval to go down to depends on S/N and on how well the image is improving (so a much shorter interval could be used for a much better model)
down to a minimum (see next paragraphs).
If possible, choose solution intervals which are both an exact multiple of the
integration time and factorise into the scan length, e.g. for 6-s
integrations and 300-s scans, ideal intervals would be e.g. 30-s or
150-s; 75-s would be OK but $<6$ sec would not make sense and
96-s would leave a 12-s interval at the end of each scan, see Fig.~\ref{tab-cals}(h).

However you set the solution interval, most solutions should succeed (see logger Fig.~\ref{logger.png}) and it should lead to a better image.
The minimum solution interval can be determined in several ways: \\
\begin{description}
\item {\bf Trial and error} is a typical approach for small data sets, reducing
the solint until many solutions fail (typically, $>$10\%) or the solutions are just random and the image stops
improving.
\item
 {\bf Inspection of the visibility phase} to determine the rate of change of errors with time.
 Plot target phase (channel-averaged) against time for baselines to the reference antenna, as in Sec.~\ref{sec-cond}.
 You are looking for the minimum interval in which phase variations are due to systematic errors.  It can be hard to distinguish atmospheric phase drift
from phase slopes due to target structure.  The phase calibrator phases (after applying scan averaged solutions) can give a clearer indication of the residual phase error rate (since there is usually no structural phase), Sec.~\ref{sec-qpr}.
 If possible, the  solution interval should be such that the
 errors change by  $\le$$45^{\circ}$ (0.785 rad), which should reduce the decorrelation per interval to $\lesssim$$25\%$ (Eq.~\ref{phnoise}).
 One way to test this for intervals shorter than a scan for the target is to apply the per-scan solutions, and then plot the corrected data
 averaged in time by the same interval as the solint you are thinking of using and look at the jump between averaged phase points.  The  phase  should show a systematic change with time, at least over a few averaging intervals; if it looks random then probably it is dominated by noise and you need to use a longer averaging interval, see examples in Fig.~\ref{tab-cals}. NB if you apply calibration tables for tests, you still have to specify whatever you decide to use  in {\tt gaincal}  (and thus the next {\tt applycal}) as  {\tt gaincal} uses the {\tt data} column, not {\tt corrected}.
\item
{\bf Statistical assessment of solution intervals}  A more rigorous
empirical approach is described in the VLA self-calibration CASA guide (see Appendix~\ref{ap-casaguides} link)
testing the S/N of multiple solints to find the interval which gives
the most reliable solutions.
\item
{\bf Analytical estimate of the shortest solution interval}
The shortest feasible solint can also be estimated analytically,
useful for large data sets and non-interactive calibration.
\end{description}
The analytical estimate is derived as follows. The  noise limit of an interferometry observation with $N$ antennas (for a particular
frequency and other parameters), is given by
\begin{equation}
\sigma_{\mathrm{rms}} = \frac{k
  T_{\mathrm{sys}}}{\sqrt{(N(N-1)/2)\Delta t \Delta \nu
    N_{\mathrm{pol}}}}
\label{noise}
\end{equation}
where   $T_{\mathrm{sys}}$ is
system temperature (Eq.~\ref{eq-tsys}) multiplied by a constant $k$  covering conversion from K to Jy (for ALMA this term includes estimating the atmospheric contribution for a given elevation and PWV and antenna efficiency see  \citetalias{ALMA-TH}, sec. 9.2.1 in 2021); there can be
different considerations for other arrays.  The noise is inversely
proportional to the square root of the number of baselines $N(N-1)/2$, the time
on source $\Delta t$, bandwidth contributing to image (e.g. continuum-only) $\Delta \nu$ and the number of
independent polarisations correlated $N_{\mathrm{pol}}$. This is the
predicted ideal image noise.

You can  use a 
sensitivity calculator (Appendix~\ref{scs}) to find the predicted $\sigma_{\mathrm{rms}}$ for your whole observation, or indeed for a single scan. 

 Since atmospheric, and most instrumental, errors are antenna-based,
 the S/N for calibration is usually maximised by deriving solutions
 per antenna.  If all antennas have the same sensitivity then the
 noise per baseline over the whole observation is thus
 $\sigma_{\mathrm{rms}} \times \sqrt{(N(N-1)/2)}$.  Eq.~\ref{noise} can be rearranged to give the rms in  a solution interval of
 $\delta t$, per antenna, per polarisation, per frequency calibration interval
 $\delta \nu$, which would typically be  continuum-only channels for each separate spw:
\begin{equation}
\sigma_{\mathrm{solint, ant}} = \sigma_{\mathrm{rms}} \times
\sqrt{\frac{(N(N-1)/2)}{N-3}} \times \sqrt{\frac{\Delta t}{\delta t}
  \frac{\Delta \nu}{\delta \nu} N_{\mathrm{pol}}}
\label{noise_ant}
\end{equation}
where $N-3$ is number of antennas -- degrees of freedom.
There are 3 degrees of freedom for phase: $N-1$ baselines
per antenna; the origin of phase set to zero and the reference antenna phase (see \citet{Cornwell81} and Sec.~\ref{sec-closure}).
A more formal derivation is presented in \citet{Sob2021} (although we incorporate an additional degree of freedom).  

Typically a S/N of 3 per antenna solution is wanted (see Appendix~\ref{ap-gaincal} parameter {\tt minsnr} but also Sec~\ref{sec-goodbad}),
i.e. $\sigma_{\mathrm{solint, ant}} = P/3$ where $P$ is the image peak,
leading to a minimum solint of:
\begin{equation}
  \delta t_0 = \left(\frac{3\sigma_{\mathrm{rms}}}{P}\right)^{2}
  \frac{(N(N-1)/2)}{N-3} \Delta t \frac{\Delta \nu}{\delta \nu} N_{\mathrm{pol}}
\label{min_solint}
\end{equation}

For the VY CMa example, there is only one spw so $\delta{\nu} = \Delta{\nu}$. From
Table~\ref{tab-VYobs} and Eq.~\ref{min_solint}, the minimum solint
$\delta t_0$ at the start of self-calibration is 1.7 min or $\sim100$
sec. This fits well as it is close to 1/3 of the scan length $\tau_{\mathrm{scan}}$. During self-calibration, the S/N should improve, allowing a shorter solution interval.  The theoretical minimum solint is $\delta
t_0\sim3$ sec but, as in this case, it is not always desirable to go to
the minimum solint. Not only is it less than an integration time but the shortest-timescale corrigible errors are over a few tens seconds,
below that is noise.  Moreover, the rms level of 0.6 mJy obtained
after a 30-sec phase solint and 60-sec amplitude solint is already the
theoretical sensitivity for this frequency and observing parameters,
see also Sec.~\ref{sec-stop}.

Given that  $\frac{\sigma_{\mathrm{rms}}}{P}$ = 1/(S/N), Eq.~\ref{min_solint}  can also be used to estimate the minimum S/N needed to perform per-scan, per-spw self-calibration: 
\begin{equation}
\mathrm{S/N}_{\mathrm{sc}} = 3\times\sqrt{\frac{(N(N-1)/2)}{N-3} \frac{\Delta t}{\tau_{\mathrm{scan}}} \frac{\Delta \nu}{\delta \nu} N_{\mathrm{pol}}}
\label{eq-minsnr}
\end{equation}
For the observing parameters given it Table~\ref{tab-VYobs}, S/N$_{\mathrm{sc}}$ needs to be at least 35, and we can see from Table~\ref{VYstats} that the starting S/N is well over this. Note that the example data have only one spw; the commonest real-life difference is that if there were 4 spw each contributing 1 GHz to the image, for example, S/N$_{\mathrm{sc}}$ would need to be doubled.

The minimum useful solution interval is the larger of
$\delta t_0$ (Eq.~\ref{min_solint}) and the shortest interval which shows error structure. Do not use so short an interval that the visibility scatter within it
looks entirely like random noise. If the solutions show a noise-like scatter,  usually this means that you need more averaging to
increase S/N, but if you know that the S/N should be high, it could be
that you have averaged too much, see Sec.~\ref{sec-sols}, Fig.~\ref{tab-cals}(f) and (g).
Even if the minimum solint with sufficient S/N is a scan or more,
you may be able to correct systematic but more rapid phase errors by fitting a
gradient or spline, see Sec.~\ref{sec-rate}.

\subsection{Amplitude self-calibration}
\label{sec-ampscal}
Amplitude corrections require a higher S/N than for phase.
As explained in Sec.~\ref{sec-phase-fidelity}, averaging data with
phase errors decorrelates 
amplitudes and the
effect of even a few  bad (high) solutions can add a lot of noise,
so usually the  phase is calibrated first and corrections are applied in
{\tt gaincal} and {\tt applycal} during amplitude
self-calibration. Section~\ref{sec-sols}, Fig.~\ref{tab-acals} gives examples of good and bad solutions. Amplitude-only self-calibration {\tt gaincal calmode='a'},
is normally used since the phase should already be well-calibrated and the amplitude self-calibration often uses longer solution intervals.  If {\tt calmode='ap'} is used, residual phase solutions should be small; if many solutions fail and/or the phase solutions have  slowly-changing structure (Figs.\ref{tab-cals}(b)~\ref{tab-acals}(m)) this probably means that the model is not a good match to the data. Isolated, wild amplitude solutions (Fig.~\ref{tab-acals}(l)) imply bad data.
Sometimes {\tt calmode='ap'} is used for one-step calibration of simple, bright sources (often the case for the `old' VLA).

It is usual to start amplitude calibration with a solint of a scan.
Formally, you can use Eq.~\ref{min_solint} to estimate the minimum solution interval by replacing
$(N-3)$ by $(N-4)$ (as there is an additional degree of freedom in the flux scale, also see Sec.~\ref{sec-closure}). This  increases the minimum
solution interval by roughly 10\% compared with that for phase, but usually,
correctable amplitude errors change much more slowly than phase errors so even for high S/N it is rare to go to the shortest possible interval.
If you are considering a solution interval less than a scan,
inspect  suitably-averaged phase-corrected visibility
amplitudes against time. If the target is well-resolved, the amplitude variations due to structure will be faster on long baselines but errors will vary on similar scales at a given time on most baselines for ALMA (although for arrays with hundred+ km baselines the weather may be consistently worse at some antennas).
If your target peak is not at the pointing centre, use the {\tt Transform} tab in {\tt mstransform} to enter the offsets in R.A. and Dec., which will adjust the phase in plotting and avoid amplitude decorrelation simply due to the phase offset.  As in assessing phase solution intervals, you need to identify the timescale on which errors are systematic, not just noise.  In Fig.~\ref{VYCont_phref.png} bottom right, the scatter within a scan (small dots) is noise-like so amplitude solution intervals less than a scan are unlikely to be of any benefit for these data (see also Fig.~\ref{tab-acals} (k)). In some cases where the error is the same throughout an observation
 (as in Fig.~\ref{tab-acals} (n)), many scans can be averaged if necessary.
 
For linear feeds like ALMA, amplitude calibration normally averages X and Y polarisations,  {\tt gaincal gaintype='T'} (see Appendix~\ref{ap-gaincal}) to avoid the effects of parallactic angle rotation. Special care is needed with circular feeds for circularly polarised targets, see Sec.~\ref{sec-polarisation}. This assumes that at least one phase
 solution table previously applied e.g. from the phase calibrator did use {\tt
   gaintype='G'} to remove phase offsets between X and Y. This can be checked from the corrected phase calibrator visibilities.
 If no polarisation nor parallactic angle correction has been applied, then for an unpolarised target there should be no offset between XX and YY visibility amplitudes; if it is polarised, typically a few percent or less, the offset should appear no more than this and change smoothly in opposite senses over hours as the parallactic angle of the feeds rotates with respect to the sky.  If there is an excessive or abrupt/irregular offset (more common for e.g. VLBI arrays than ALMA) then the polarisations can be calibrated separately ({\tt gaintype='G'}) if you are only interested in total intensity and the actual target polarisation is small.

 It is important that the model used for amplitude self-calibration is
 as faithful as possible a representation of the target and includes all the target flux (unlike
initial
 imaging for early phase-calibration when``if in doubt leave it out'').  As seen in Fig.~\ref{n_allerrors.png},
 after phase calibration,  amplitude errors give symmetric artefacts --
 so if you see a peak with one other peak beside it and no mirrored
 positive or negative, both peaks are probably real (see Sec.~\ref{sec-spur} for the hazards of over-calibrating on an incorrect, single-peaked model). Check the residual image when making the model, and if there is remaining emission which looks like it is part of the target field, adjust the mask if necessary and clean more deeply.
 
 As explained in Appendix~\ref{ap-gaincal},  {\tt gaincal solnorm=True}, is normally used if the flux scale is thought to be accurate, especially if your model may not
contain all the target flux, e.g. due to much weak, extended
emission. See Fig.~\ref{tab-acals} (m) for the effects of a poor model. On the other hand, if there are some data which are initially
very badly scaled (e.g. an antenna for which $T_{\mathrm{sys}}$ or
phase referencing failed for some scans, see Sec.~\ref{sec-selective})
then you should not use normalisation.  If amplitude
self-calibration with {\tt solnorm=False} produces a significant decrease in flux density of the subsequent image you
should check the model and {\tt gaincal} parameters, but if nothing can be improved, repeat {\tt gaincal} with {\tt solnorm=True}.  If you do not
normalise the solutions and the flux density increases by more than a
few percent, you should work out why. It is often because you have
over-cleaned and piled too much flux into a small area of model, so the image will have a higher peak but worse sidelobes and/or missing extended structure.  If
so you should re-make the input model (and check it has been inserted into the MS). Alternatively, it may be a
sign that the target is simply too weak for amplitude
self-calibration, at least without more averaging.

See Sec.~\ref{sec-assessamps} for more on assessing amplitude calibration tables.

\subsubsection{Multi-frequency synthesis for amplitude self-calibration}
\label{sec-mfs}
Multi-frequency synthesis (MFS) in making a multi-channel continuum image involves fitting for the variation of flux density with frequency (e.g. spectral index) when deriving Clean Components.  You do not usually need this before phase-only self-calibration of an ALMA observation,  as the position of continuum emission in the field of view is not normally frequency dependent.  
If the continuum S/N is high enough to consider amplitude self-calibration,  however, it is often worth fitting
for spectral terms in {\tt tclean} when preparing the prior model, after applying the phase self-calibration. 
Even if the fitted spectral index
({\tt .alpha} image) has large uncertainties the total intensity image will
be improved and the model ({\tt .tt0} and {\tt .tt1}) will make sure that the
flux density slope across spectral windows is correctly calibrated.
For spectral index $\alpha$, the flux density $S_1$ at frequency
$\nu_1$ is related to that at $\nu_0$ by:
\begin{equation}
  S_1=S_0 \left(\frac{\nu_1}{\nu_0}\right)^{\alpha}
\end{equation}  

You can thus estimate the flux density in the lowest and highest
frequency spw present.  The full observation VY CMa observation used 4 spw centred at 314.0, 315.8, 326.0
and 327.8 GHz, the nominal central frequency is 320.9 GHz.  The total
continuum flux density is 200 mJy and from the properties of a red supergiant like VY CMa we expect $\alpha\sim1.5$. So,
the difference in flux density between the highest and lowest spectral
windows is (206.5-193.6) = 6.1 mJy. The sensitivity calculator gives
rms per spectral window $\lessapprox 1$ mJy for each spw (for 1 GHz line-free continuum per spw and 30 min on
target), so it is worth using {\tt tclean deconvolver='mtmfs',
  nterms=2}.  (The VY CMa example used here has just one spw so MFS was not used).

More generally, for a spectral index $\alpha=\pm0.7$, this can make a
10\% difference across the maximum single-execution bandwidth at Band
3.  At Band 7 the difference is 3.3\% which means that it is still
significant for S/N$\ge30$, and even at higher bands, $\alpha=\pm2$
provides a similar significance.  This provides a smooth change in amplitude with frequency and better image quality.  However, when the fractional frequency span of all data used is $\lessapprox$10\% the precise slope and thus the $\alpha$ map itself is often
unreliable.  Occasionally (very bright sources/combined observations
over a large fractional bandwidth) it may be possible to use {\tt nterms
  $>$ 2}. However, if it is difficult to get a good model, or if the S/N is close to the lower limit,
compare imaging with and without MFS to check that it makes an improvement and does not produce amplitude artefacts, as it may be safer to make a simple single-term image.

Continuum amplitude calibration, with a spectral slope if S/N allows,
not only improves the solution accuracy but it ensures that the flux
scale is consistent across spectral windows.

\subsection{Judging the quality of solution tables, failed solutions and flagging}
\label{sec-goodbad}

This section provides guidance in how to assess solution tables and
what to do in case of problems.  This should be done alongside
assessing their effects on imaging (see error recognition from images
Sec.~\ref{sec-errorrecognition}) and on the visibilities.
When checking visibility data, apply all previous good calibration and if necessary experiment with different averaging, 
Sec.~\ref{sec-SN} on {\bf Inspection of the visibility phase} (and the
same goes for amplitude calibration).
If all the data for some antennas are  noisier, but not pure noise, this may be inevitable if they have mostly long baselines (many km for ALMA) or are less sensitive (e.g. smaller diameters in VLBI); this usually should not be flagged as
  downweighting may compensate without removing their contribution altogether, Sec.~\ref{sec-weights}.

\begin{figure}
  \includegraphics[width=17.cm]{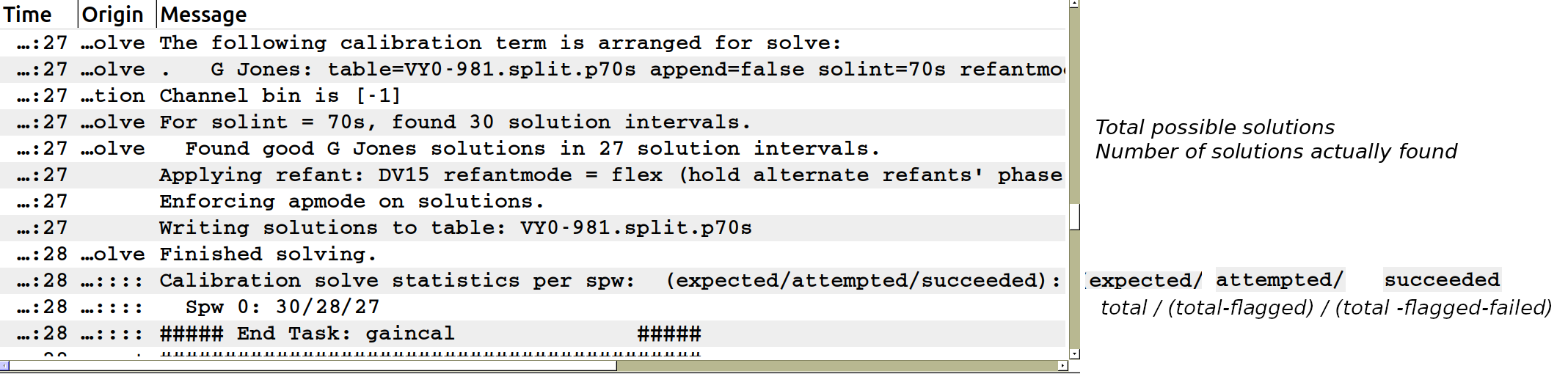}
  \caption{\small CASA logger messages from {\tt gaincal}. The total, {\tt 'expected'}, is the sum of (scan length)/(solution interval), rounded up per scan,  per spw and per polarisation unless these are averaged (but irrespective of the number of antennas). {\tt 'attempted'} is the number after subtracting intervals flagged (previously, or by existing calibration tables applied as  {\tt gaincal gaintable}s). {\tt succeeded} is the number of solutions which converged using the input parameters; this does not guarantee that they are suitable to apply to the data.
  }
\label{logger.png}
\end{figure}

In more common ALMA cases,  first check the CASA logger, see Fig.~\ref{logger.png}.
As explained in Appendix~\ref{ap-applycal} it is safest not to use failed solutions for flagging, i.e. set {\tt applycal applymode='calonly', calwt=False } at least until you are sure that all the parameters and the model used to derive solutions are optimum.  If the 
minimum solution interval from  Eq.~\ref{min_solint} is more than a scan, or some or all antennas/times are hard to calibrate,  refer to see Sec.~\ref{sec-lowsn}. Additional tips are given in \citet{Brogan18} sec. 2.3.5.
Distinguish between:

\begin{description}
\item[{\bf{Improvable}}] data {especially at the start of self-calibration. Even if many solutions fail or show a large scatter, check the visibilities. If the data are not obviously bad,  try improving the model or parameters in {\tt gaincal}.  You can
discard the guilty image and/or gain table (see towards the end of
Sec.~\ref{sec-overview}) and try again with different imaging or
calibration parameters. Or, if phase referencing may not have been very good,  pass the failed solutions  ({\tt applycal applymode='calonly'}) for the initial rounds and hope that fewer fail as the model improves.}

\item[{\bf{Unselfcalibratable}}] {data which has had good phase calibrator corrections applied. The target may be much too faint, so there is no point trying. However, if it is borderline, or some antennas have lower S/N (e.g. extended source/long baselines) you may find many solutions (say, 20--50\%) fail in self-calibration, but the rest are fairly good.  The data giving failed solutions have had the phase calibrator corrections applied, so if they are not obviously bad, usually it is best to let them pass.  You can test by making an image without flagging, then back up the flagging state and repeat  {\tt applycal} with {\tt calmode='calflag'} and re-image; if this image is worse, restore the flags. Also compare with no self-calibration.}

\item[{\bf{Useless}}] {data, e.g. pure noise, zero amplitudes. If the data cannot be salvaged, see Sec.~\ref{sec-flag}.}
\end{description}  

\subsubsection{Assessing phase  solutions}
\label{sec-sols}

This section gives examples of what to look out for in phase solution tables. 
Fig.~\ref{VYCont_phref.png} bottom left shows per-scan (large circles) phase solutions for a phase calibrator, which follow a mostly consistent trend with time, demonstrating that they can be interpolated reasonably accurately across the target data, such that the residual self-calibration solutions, Fig.~\ref{VYcont_images.png} left, are within $(\pm60^{\circ})$, approximately the phase changes between  phase calibrator scans. 
More examples of gain tables are given in Fig.~\ref{tab-cals}. These are mostly plotted using fixed ranges on the axes. When looking at other tables, make sure to check the scales before jumping to conclusions.

\begin{figure}
  \begin{tabular}[t]{p{7cm}r}
    {\vspace*{-2.4cm}(a) DV15 is the reference antenna with zero phase solutions. DV07 and DV19 solutions change smoothly, within the scan to scan  differences between the phase calibrator solutions previously applied. Small
      jumps in the centre of some scans are due to having used {\tt applycal interp='nearest'} (not {\tt linear}). All antennas have a similar number of solutions. }
&\parbox[c]{7cm}{\includegraphics[width=9cm]{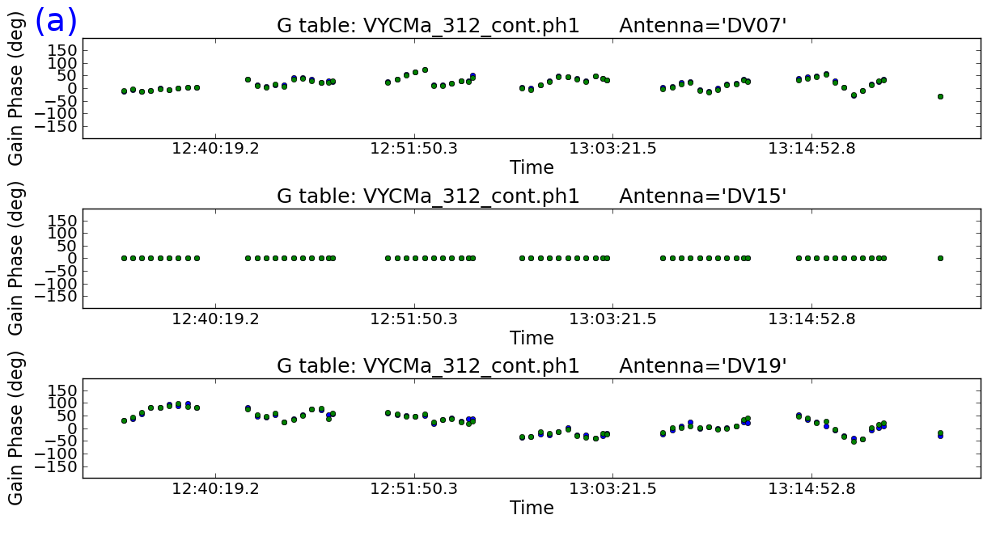}}\\

    \vspace*{-2.2cm}(b) If phase calibrator solutions have not been applied, a larger solution range, even wrapping through 180$^{\circ}$, may be seen. If the data were phase referenced, check that the solutions were applied and the model is correct. Even in this case, a high phase solution rate may be reasonable if the phase changes rapidly (which would also be seen for the phase calibrator). 
&\parbox[c]{7cm}{\includegraphics[width=9cm]{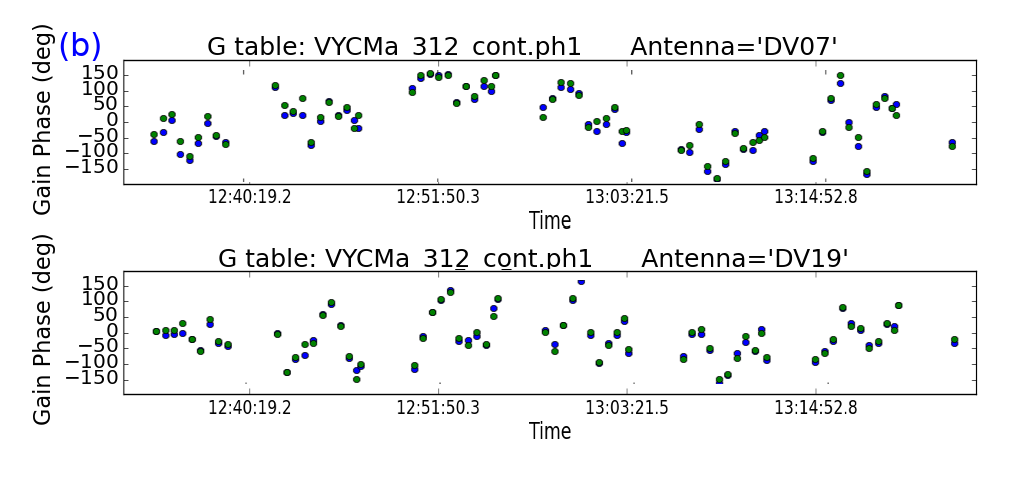}}\\

    \vspace*{-1.7cm}(c) Here, the 30-s solutions for DV07 look quite noisy on short time scales but the average change per scan is just a few tens degrees. DV19 is worse with some periods appearing random and many failed solutions.    The polarisations should be aligned (as in (a)) after applying initial phase solutions; the random offsets between X and Y (blue and green) solutions seen here are due to noise. It would be advisable to use a longer solution interval of 150s or  300s. 
&\parbox[c]{7cm}{\includegraphics[width=9cm]{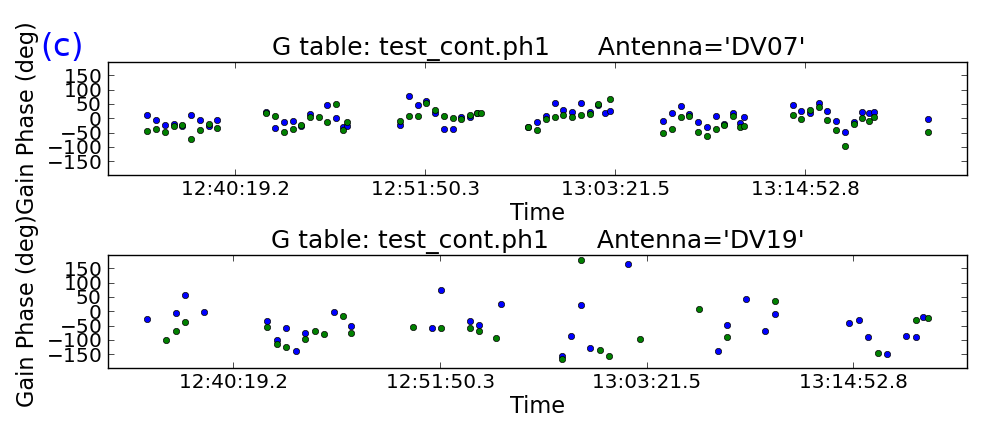}}\\

\vspace*{-1.5cm}(d) Many failed solutions but little scatter, as on DV19, suggests a lower {\tt minsnr} and/or more averaging might be better, or possibly data have been lost due to deliberate flagging or previous failed solutions.
&\parbox[c]{7cm}{\includegraphics[width=9cm]{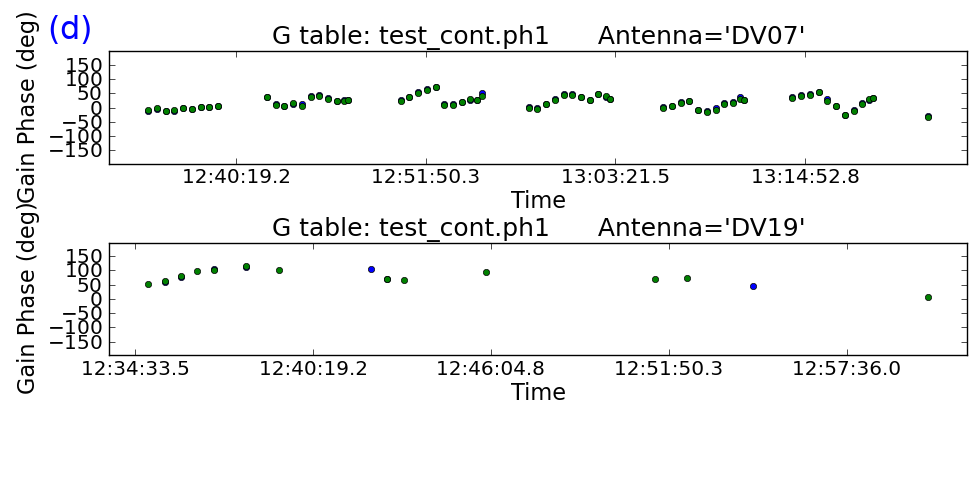}}\\

\vspace*{-2.3cm}(e) If solutions have 180$^{\circ}$ noise scatter on all antennas, try a longer solution interval or other averaging. Check that the correct previous calibration has been applied and the right model is present. If S/N is high,  check the reference antenna/try a different one or see example (f), (g). Alternatively, maybe S/N is too low to self-calibrate. &\parbox[c]{7cm}{\includegraphics[width=9cm]{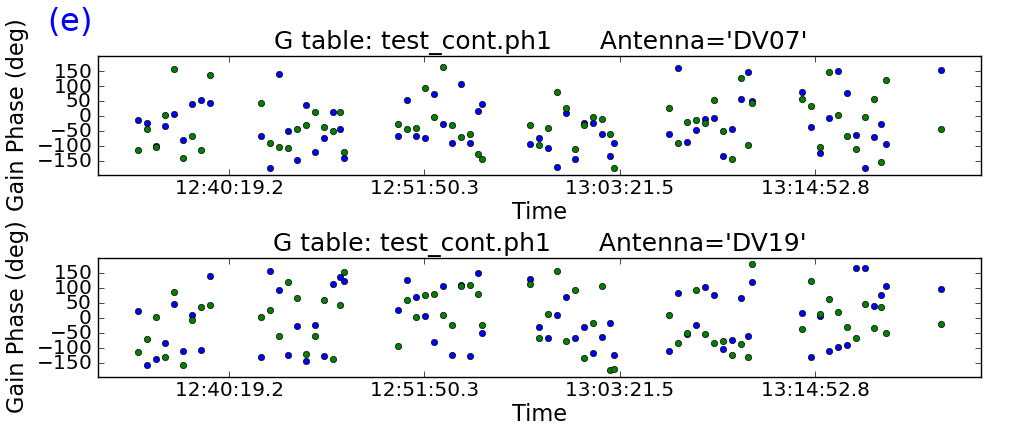}}\\
  \end{tabular}
  \caption{\small Target self-calibration phase solution examples. (a) and (b) are probably good solutions; (c) and (d) need better inputs  (or flagging) and (e) may be all bad. In all plots with both blue and green spots they represent solutions for the X and Y polarisations.  Continued next page.
  }
  \label{tab-cals}
\end{figure}

\begin{figure}
  \addtocounter{figure}{-1}
  \begin{tabular}[t]{p{7cm}r}

    {\vspace*{-2.cm}(f) Most of these solutions look like noise but the first few DV07 scans show a high phase rate. Possibly prior calibration has been forgotten, or there is no phase-referencing. In such a case see Sec.~\ref{sec-nophref};      if the target is complex or peaks many beams from the phase centre it could be that the genuine corrections are undersampled. }
&\parbox[c]{7cm}{\includegraphics[width=9cm]{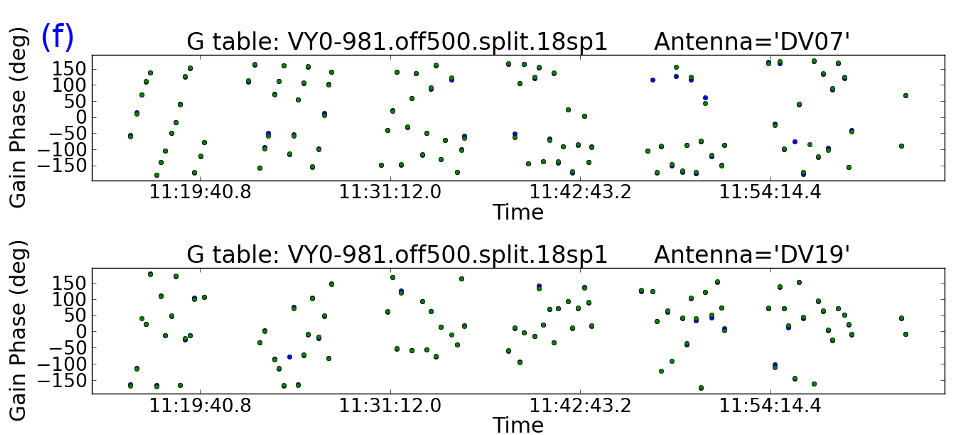}}\\

    {\vspace*{-2.3cm}(g) Using  {\tt solint = 'int'} for the (f) situation shows coherent solutions (it may be necessary to zoom; the lines are to guide the eye). If S/N is too low for short enough solution intervals, see Sec.~\ref{sec-rate} for fitting to the phase slope.
    }
&\parbox[c]{7cm}{\includegraphics[width=9cm]{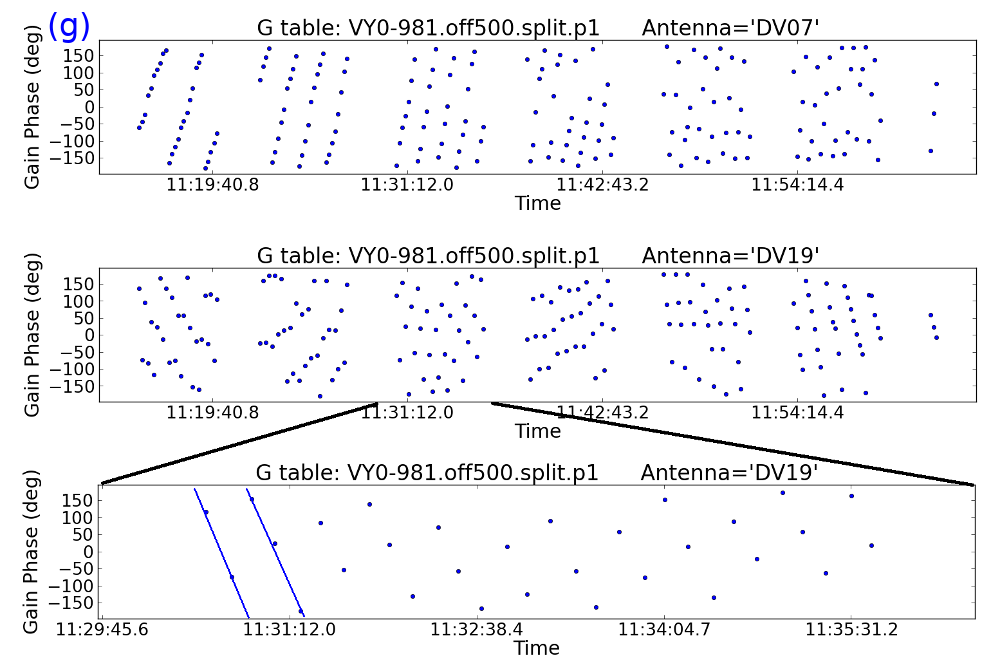}}\\

    {\vspace*{-2.2cm}(h) In each scan, the final solution has failed or is inconsistent relative to the general trend.  A 96-s solution interval was used for 300-s scans so the last interval is only 12 s, with worse  S/N. If you need just a few solutions per scan, use an exact divisor (e.g.  75 s in 300-s scans).  Alternatively, bad solutions at the start or end of scans may be due to slewing, see Sec.~\ref{sec-flag}.}
&\parbox[c]{7cm}{\includegraphics[width=9cm]{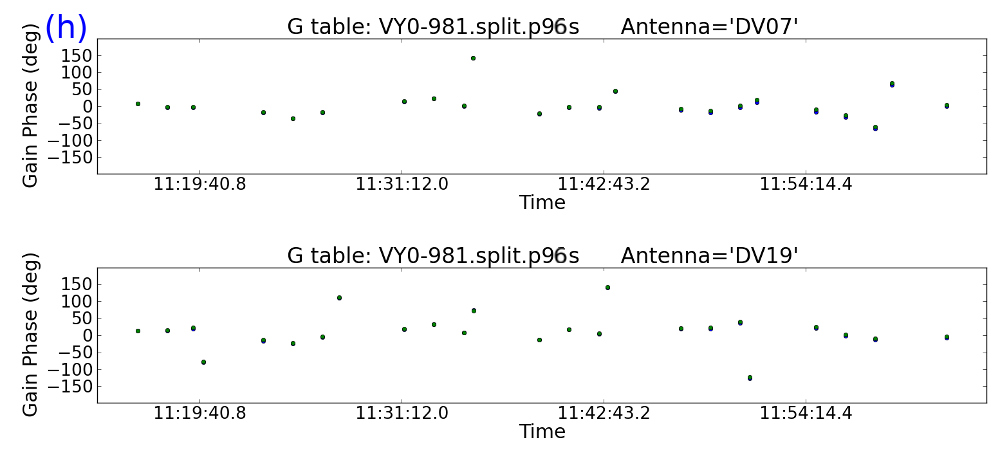}}\\

    \vspace*{-.7cm}(i) When solutions are close to zero the data match the model within the noise; time to stop self-calibration.
&\parbox[c]{7cm}{\includegraphics[width=9cm]{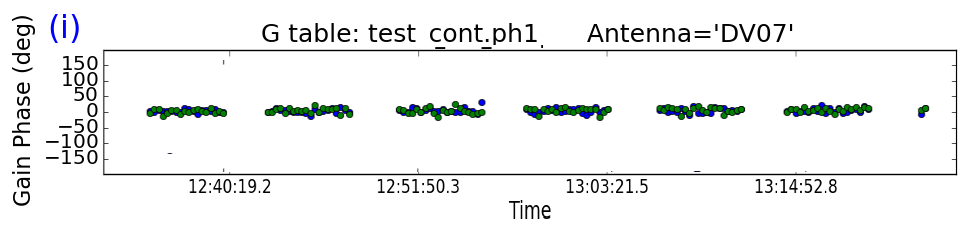}}\\
  \end{tabular}
  \caption{\small (continued) Target self-calibration phase solution examples. (f) needs improved parameters, leading to good solutions in (g). (h) needs better inputs or flagging and (i) shows no more phase calibration is needed.}
\end{figure}
\clearpage
\subsubsection{Assessing amplitude solutions}
\label{sec-assessamps}
 
\begin{figure}
  \begin{tabular}[t]{p{7cm}r}
    
    {\vspace*{-2.1cm}(j) Amplitude self-calibration solutions should be close to 1, as for DV07, if previous calibration was good. Investigate scatter $\gtrapprox10\%$.  The large scatter on DV19  may be due to poor preceding phase solutions, a bad model or too low S/N.  If S/N is enough, many cycles of self-calibration may be needed to improve the model, with shorter amplitude solution intervals. 
    }
&\parbox[c]{7cm}{\includegraphics[width=9cm]{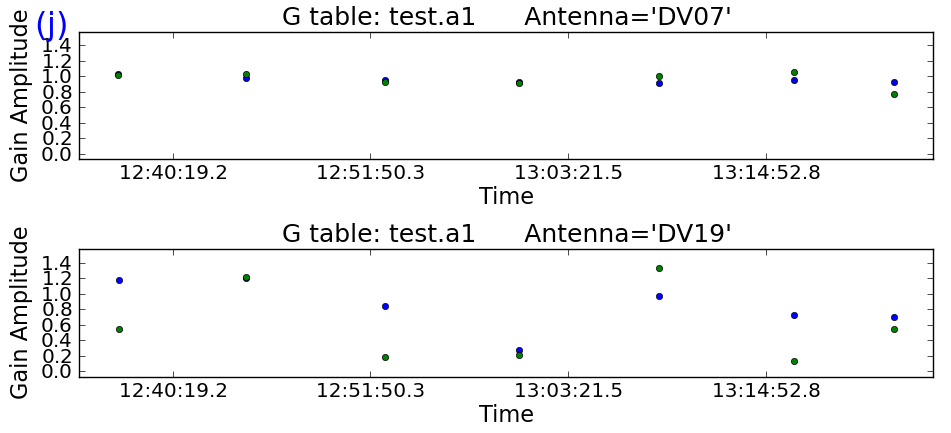}}\\

     {\vspace*{-1.7cm}(k) Using a shorter solution interval for DV07 shows good results with a small variation close to 1, suggesting that the calibration is complete. There is a higher, noise-like scatter for DV19; if there are still amplitude errors a longer solution interval is needed.}
&\parbox[c]{7cm}{\includegraphics[width=9cm]{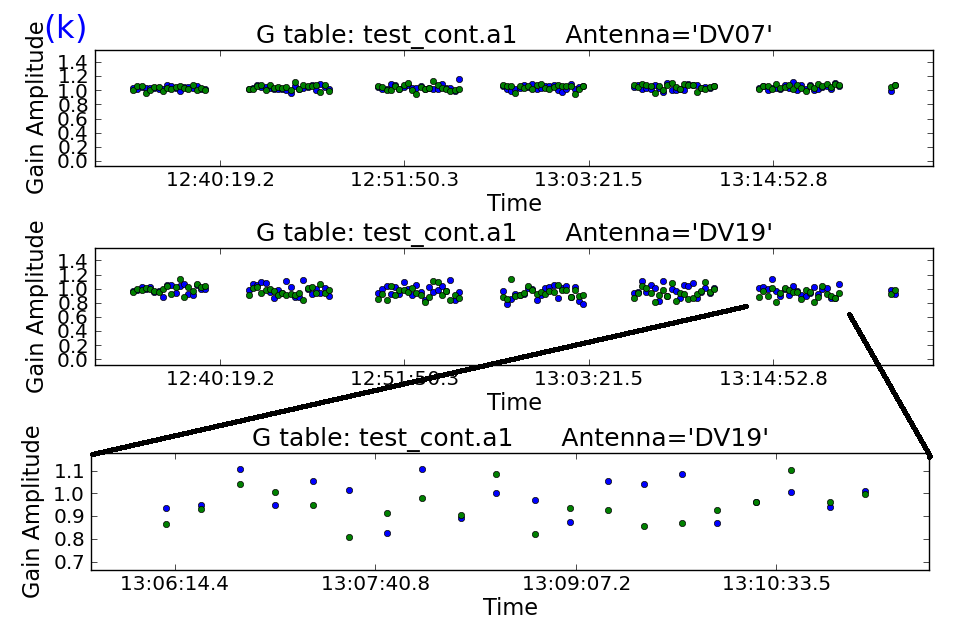}}\\

     {\vspace*{-2.2cm}(l) DV07 single solutions $>$20\% discrepant are probably due to bad data (but check previous phase solutions). Identical, low solutions on one antenna, DV19, show entirely bad data. Identical 1's (0's) on all antennas may mean that the table contains only phase (amplitude) solutions but amplitude (phase) is being plotted.}
&\parbox[c]{7cm}{\includegraphics[width=9cm]{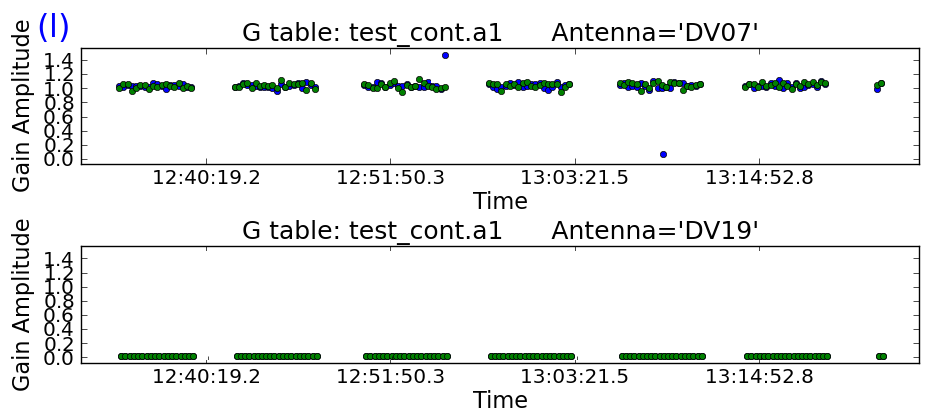}}\\

     {\vspace*{-2.cm}(m) Amplitude solutions $\ll$1 are expected if the flux scale has not been corrected, e.g. no phase referencing. Otherwise it indicates a poor model or decorrelation (see Fig.~\ref{tab-cals}(f)). Structure in good solutions, changing faster for antennas with more long baselines, DV19, shows a data-model discrepancy, e.g. due to a point model used for a complex source (see Sec.~\ref{sec-nophref}).}
&\parbox[c]{7cm}{\includegraphics[width=9cm]{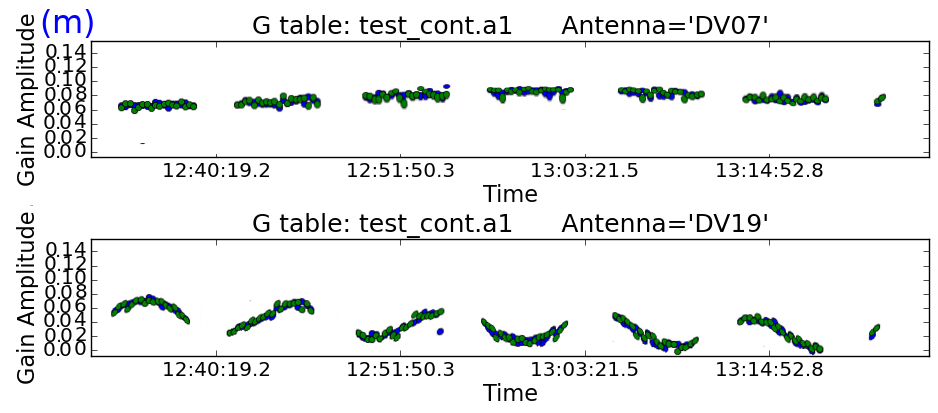}}\\

  {\vspace*{-2.cm}(n) Although DV07 has lower solution amplitudes than DV19, the former also has smaller scatter, maybe due to a $T_{\mathrm{sys}}$ mis-scaling, not bad visibilities. Try making a model excluding DV07 and using this to derive amplitude solutions for all antennas (see Sec.~\ref{sec-selective}). The variation within a scan is just noise so a longer solution interval is needed.
    }
&\parbox[c]{7cm}{\includegraphics[width=9cm]{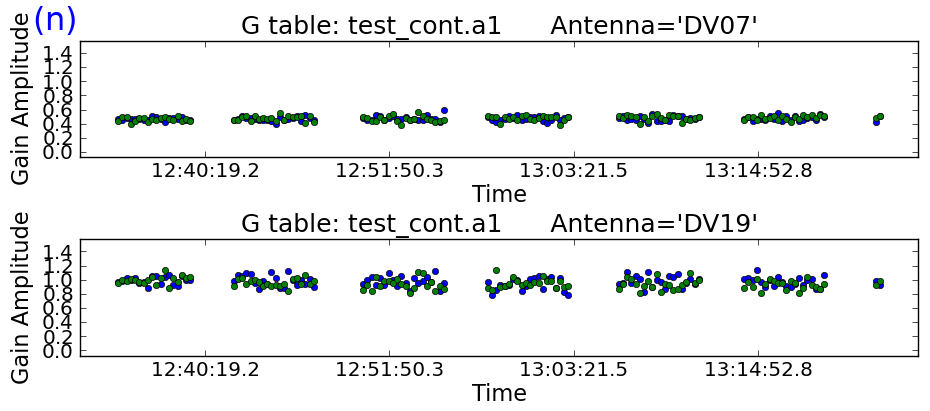}}\\

  \end{tabular}
    \caption{\small Target amplitude self-calibration solution examples. (j), (k) and (l) need some improved inputs. (m) may be good solutions if self-calibrating without prior flux scale/amplitude calibration. (n) shows probably-good solutions but DV07 needs re-scaling and a longer solution interval is needed.}

  \label{tab-acals}

\end{figure}

 For already well-calibrated data, amplitude solutions are likely to be within $\lessapprox$20\% of 1.  Fig.~\ref{tab-acals} gives example solution tables.   It is important to check for low
 points\footnote{for AIPS check high points as the data are multiplied by solutions} as in CASA, the solutions are divided into the data, so a low solution  signifies weak visibility amplitudes and when that is applied the associated noise will also be amplified.  Check the phase solutions but if these or the averaging interval can't be improved, such data should be flagged (Sec.~\ref{sec-flag}).

The diagnostics are analogous to those for phase solutions with these considerations:
  \begin{description}
  \item[{\bf{Scan-length}}] solution intervals are often best if previous, good calibration has been applied. For complex, high S/N sources try shorter intervals but you cannot calibrate away noise.
    However, short solution interval amplitude gain tables can be a quick way to identify bad data, which can then be flagged separately. Discard the diagnostic table and repeat with a more suitable solution interval.
    
    \item[{\bf{Decorrelation}}] giving low amplitude solutions can occur due to previous bad phase solutions (e.g. an unsuitable averaging interval) or intrinsic high phase noise -- the unaveraged amplitudes may look OK but the more averaging, the worse the solutions.  If the phase cannot be corrected, flag the data.
    \item[{\bf{Contamination}}] from spectral lines which have not been excluded, but should be, can
     give bad continuum self-calibration solutions (high and/or noisy). 
\item[{\bf{Target flux}}] should stay about the same, i.e. solutions close to 1 (if flux scale corrections have been applied, Sec.~\ref{sec-flux}). The image flux density often improves due to phase calibration (see Table~\ref{VYstats}) but the main effect of amplitude self-calibration should be to reduce the noise rather than change the image flux, see discussion of normalisation in Sec.~\ref{sec-ampscal}. Using {\tt gaincal solnorm=True} is safest but {\tt solnorm=False} is a good way to check for model problems (gains very different from 1)  See Fig.~\ref{tab-acals}(n) for an example without prior flux scaling. 
  \end{description}

\subsubsection{Flagging}
\label{sec-flag}

To check the cause of bad solutions, and remove bad data, the best practice is to
\begin{itemize}
\item{Back up the current flagging state using {\tt flagmanager} to make it easy to restore or repeat flags if necessary.}
  \item{Check the visibilities in {\tt plotms}. Apply all good calibration and try different averaging intervals if necessary and make a note of any bad data (see the help for the {\tt locate} function).}
  \item{Write a script for {\tt flagdata}.}
\end{itemize}    
Alternatively, you can use failed
solutions to flag data  in {\tt applycal calmode='calflag'}, especially if only a few percent of solutions are affected and it would not mean losing all the long baselines (or other vital data scales), but make sure to back up the previous flagging state.
For more details see the CASA documentation for these tasks and, especially for cm-wave data, which often needs far more flagging than ALMA data, the CASA Guide to Flagging (VLA) (Appendix~\ref{ap-casaguides})

    Bad solutions at the start (or, occasionally, the end) of all or most scans can be due to data recorded whilst the telescopes are slewing. Plot the data with the shortest time-averaging needed to show any issues and this will show up as low visibility amplitudes and phase noise. These can  be flagged using {\tt flagdata mode='quack'}.

\subsection{Applying the calibration ({\tt applycal})}
\label{sec-applycal}
The task {\bf {\tt applycal}} is used to apply the tables derived using calibration tasks such as {\tt gaincal}.
Some parameters can usually be left as default: see  Appendix~\ref{ap-casadoc} for a link to the CASA documentation for applycal,
 for all options and Appendix~\ref{ap-applycal} for some parameters and suggested values most relevant to ALMA self-calibration.

\subsubsection{Weights}
\label{sec-weights}

Each visibility sample has amplitude and phase (as in Eq.~\ref{eq_V0})
and also a weight.  The weights for ALMA (and most other telescopes)
are initialised  in proportion to integration time, channel
width etc., at the very start of processing ({\tt importasdm}) if CASA
4.2.2 or later was used by the standard pipeline/QA2 scripts. The initial weights should be automatically
modified correctly in any visibility data averaging (proportional to
the antenna sensitivity and the time and frequency span of the data).
However, by default, all channels in each spw have the same weight.  If channel-averaging is used and channels have been selectively flagged, per-channel weights should be derived. This can be achieved by using {\tt mstransform usewtspectrum=True} to do the averaging, or by running {\tt initweights}, see the NRAO template script (link in Appendix~\ref{ap-other}) for an example.

The weights can also be modified when calibration is applied ({\tt applycal calwt=True}).
The pipeline/QA2 applies weights derived during initial calibration
e.g. from $T_{\mathrm{sys}}$.  The weights can occasionally be mis-scaled or
inconsistently scaled between observations; this is rare but if you
find otherwise inexplicable problems, the CASA Guide to Weights (see
Appendix~\ref{ap-casaguides}) describes what sensitivity-based weights
should look like. Weights can be plotted in {\tt plotms}.

These weights will be retained when the target data are split out for
self-calibration.
Weights can be further modified when applying self-calibration gain tables, to downweight the data contributing noisier solutions. This will improve the image S/N but reduce the contribution of target structure on the scales of the downweighted spacings.  This is a price worth paying for arrays with good $uv$ coverage 
and/or if the source does not have important structure on the corresponding scales.  However, if, for example, most of the antennas contributing the longest baselines have the greatest errors initially, the smallest scales will not be represented properly in the image model and thus the longest baselines will not be calibrated effectively, 
see Sec.~\ref{sec-lowsn}. The synthesized beam will also become larger, i.e. worse angular resolution. In such cases it is safer to use {\tt
  calwt=False} at least for the first rounds of self-calibration.  This is also the case for calibration of arrays of telescopes of different sensitivities (e.g. VLBI, e-MERLIN).

As an alternative to {\tt calwt} you can, after all calibration, use the separate task {\tt statwt} to modify the weights based directly on the scatter in the visibilities.  This requires selection of data where time-baseline fluctuations are noise-dominated, so it will not work for targets with bright extended structure and any spectral lines should be excluded.
A more sophisticated scheme for connecting the calibration and imaging weights is described by \citet{Bonnassieux2018}.

Calibration weights become part of the visibilities and can only be re-initialised or modified using the CASA toolkit.
Additional weights can be applied to the uv data
during the imaging process  (see help for {\tt tclean}). These imaging weights do not affect the
intrinsic data weights.
In order to ensure all available scales are represented in making self-calibration images, 
 it is best to use close to natural weighting (e.g. for ALMA, typically {\tt robust $\geq$0.5}) when preparing self-calibration models (see also Sec.~\ref{sec-optimise}). In situations where some baselines are initially hard to calibrate, the first rounds of  calibration and imaging can use a restricted {\tt uvrange} and {\tt uvtaper} (see Sec.~\ref{sec-nophref}). The ranges are expanded as the model improves in successive rounds of calibration.

\subsection{Modified workflows}
\label{sec-alt}
Sec.~\ref{sec-overview} described incremental self-calibration, where the
previous solutions are applied in subsequent rounds of self-cal.
This can be modified in two separate ways, described in more detail by \citealt{Brogan18}.

If, after one or a few rounds of self-calibration, you get an image
and model which are much better than the starting model (e.g. the
initial model included CC which turned out to be artefacts), you can
remove the previous self-calibration and model (tasks {\tt clearcal, delmod}) and start phase
self-calibration afresh with
the better model without applying previous
solution tables.
A similar strategy is used in the NRAO template imaging script, see Appendix~\ref{ap-other}.
If the model you want to use exists, but did not result from the immediately previous {\tt tclean}, see the {\tt tclean} documentation or Sec.~\ref{sec-ft} for how to insert it in the MS.
If the S/N permits, in this case you could start with
a solution interval shorter than a scan.
In each  {\tt applycal}, be sure to apply any solution tables  you applied in {\tt gaincal} plus the new table. It helps to make your own flow chart.

Alternatively, instead of updating the model and calibration in the
same MS at each stage (as in Fig.~\ref{CalFlow.png}), you can split
out the calibrated (corrected column) target data after each {\tt applycal} step. The
{\sc data} column of the new MS contains the previous corrections. Use
that for the next round of imaging  and {\tt gaincal}.  This
means that you are only ever applying the table just generated in {\tt
  applycal}.  Some users find this simpler, although it generates a
greater data volume and can be more complicated if you are transferring
solutions between parts of a data set.

\subsection{When to stop self-calibrating}
\label{sec-stop}

The three rounds of self-calibration (Fig.~\ref{CalFlow.png}) are typical but for low S/N you may only be able to do the first phase-only round.  
Conversely, additional rounds may be needed to improve a model of a complex target, or if high dynamic range is needed 
you may be able to go to successively shorter solution intervals as the model improves.  In some cases you may find that the image structure becomes clearer after amplitude self-calibration, and so you can use the better model to improve phase calibration., see note at the end of the list in Sec.~\ref{sec-overview} and Sec.~\ref{sec-alt}.

In all self-calibration, there should be some change in each cycle,
either in the solution parameters or the model, or flagging of the
data, and the image S/N should improve whilst the rms does not
increase (except in the case of a deliberate re-scale). When this is no longer the case stop self-calibrating.  The noise level should be estimated taking into account the
spectral ranges actually used and the duration and number of antennas
after any flagging, and the PWV during observations; see Appendix~\ref{ap-casaguides}, CASA Guides, Analysis
Utilities
for scripts to estimate the amount of useful data, and the web log,
QA2 report or SNOOPi for the WVR measurements.

The obvious time to stop is if you
reach the predicted noise level and inspection of the visibilities
suggests that the solution interval already used is sufficient to
remove all but random noise errors, or an additional round of
calibration produces no further improvement (Figs~\ref{tab-cals}(i) and~\ref{tab-acals}(j) or (k), DV07).

If the noise remains more than a few tens percent higher than
expected, Sec.~\ref{sec-sc} suggests some other approaches which may
be applicable. Be careful not to reduce the solution interval to shorter than the timescale of systematic errors (or a single integration); if the solutions are dominated by random scatter you are probably just moving noise around which is not useful.
The shorter the solution interval the more tightly the data are constrained so the model must be very accurate. If the image peak increases significantly, check that the rms has not also increased and that real, extended flux has not been reduced by being forced to conform to an incorrect model.  Use of a bad model or too short a solution interval can also make  negative or positive artefacts worse near the target even if the noise off-source falls.
If you are using the continuum for self-calibration, image
a line channel to check that it has also improved (or v.v.).

Several situations limit the efficacy of self-calibration. See
Sec.~\ref{sec-errorrecognition} for how to recognise confusion,
missing spacings and issues which require special treatment  and/or cannot be cured by self-calibration.
Previous calibration may not have been optimum e.g. a weak bandpass
calibrator or too-distant phase calibrator, when the target S/N does
not allow short enough solution intervals to compensate
completely. Self-calibration cannot correct fully for antenna position
errors and tracking errors, Sec.~\ref{sec-antpos}, especially for
sources far from the pointing centre.  You may reach a dynamic range
limit, see Sec.~\ref{sec-tderror-DR}; for ALMA it can be difficult to
get dynamic ranges better than a thousand, but S/N $>5000$ has been
achieved (e.g. \citealt{Fenech2018}). At high dynamic range,
especially for lower-frequency arrays, baseline-dependent errors which
cannot be factorised per-antenna may emerge, see
Sec.~\ref{sec-baseline}.

\section{Special Circumstances}
\label{sec-sc}
In this section we cover a number of situations more complex than well-behaved continuum self-calibration where additional procedures are needed.

\subsection{Transferring solutions between spectral windows/across frequency differences}
\label{sec-transfer}

Observations may be made in a spectral configuration where all or some
spw are very narrow or  fall in very noisy parts of the
atmosphere (transmission can be plotted in {\tt plotms}) which nonetheless contain bright lines (e.g. a 100 Jy maser in a region of 50\% transmission) exclude all the affected channels during continuum self-calibration as if they were a spectral line or transfer solutions from another spw.
In such cases, check in case prior calibration has flagged the narrow (or noisy) spw.  The phase calibrator may not provide enough S/N; the ALMA pipeline usually recognises this situation and averages in time or across spw as required, or the phase calibrator may have been observed in wider spw.  However, if this has not been done (e.g. early cycle data)
many solutions may fail for narrow spw, and thus flag the target. In such a case the phase calibrator corrections (or those for any other calibration source with many failed solutions) should be re-derived with more averaging (and a lower {\tt minsnr} if necessary).  This can also be the case during self-calibration.
The principle is the same in both cases.

One method is similar to that used by the pipeline and standard QA2
scripts for the phase calibration source in narrow spw, which seeks to
increase the phase calibrator's S/N per solution interval by using a
longer solution interval (e.g. {\tt 'inf'} instead of {\tt 'int'}) and/or
averaging all spw. Such averaging (e.g. all continuum channels) can
also be used for self-calibration, as summarised in
Sec.~\ref{sec-avg}. 

In this section we outline an alternative method in the case that
you have some spw wide enough to self-calibrate individually, the
solutions are usually less noisy if very narrow spw are left out and
the wide spw solutions are applied to all.  Similarly, if a  line provides the
best S/N for self-calibration, its solutions can be applied to all channels.

\begin{figure}[t]
  \includegraphics[width=8.cm]{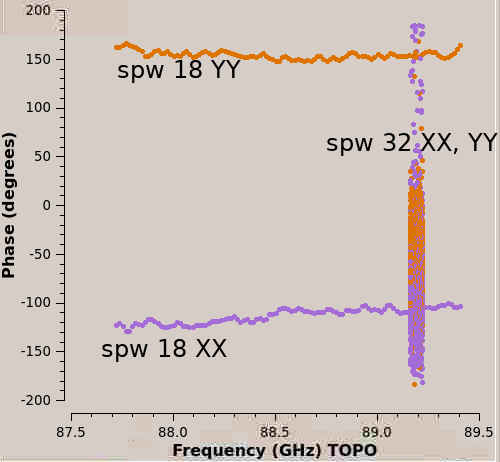}
  \includegraphics[width=7.1cm]{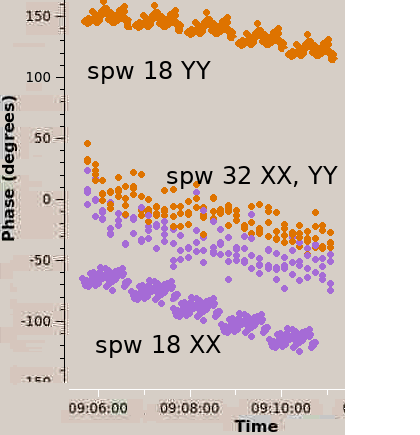}  
  \includegraphics[width=4.9cm]{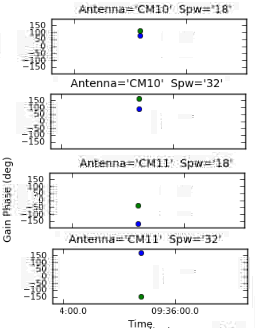}
  \hspace*{1cm}
  \includegraphics[width=10.cm]{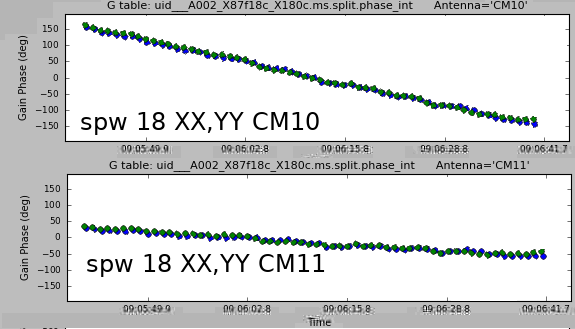}
  \caption{\small Exaggerated example of correcting for offsets between spectral windows and  correlations. Top left: DGC (bandpass calibrator)
    raw phases CM10\&CM11
    baseline, XX (purple) and YY (orange) correlations, for wide spw 18 (87.65--89.45
    GHz) and narrow spw 32 (around 89.2 GHz). Top right: the same spw
    raw phases for the target as a function of time; the narrow spw is
    noisier. Bottom left:
    time-averaged phase solutions for the DGC for CM10 and CM11, polarisations X and
    Y. Bottom right, the target self-calibration phase solutions for
    spw 18 (showing both polarisations now aligned) after
    applying the DGC solutions during gaincal.}
  \label{DGC.png}
\end{figure}

Normally, after applying all corrections including the phase calibrator
solutions to the target, there should be no remaining offsets between
the continuum phases for different spws in an EB. The correction for
spw offsets also removes instrumental offsets between polarisations
(but not the effects of source polarisation or parallactic angle
rotation).  Check that this is the case by plotting phase v. time for
each spw -- you may have to average per scan, even all baselines, for
narrow spw.  If the phase calibrator was observed in the same
configuration it may be easier to examine its calibrated
visibilities. For each correlation there should be no systematic
offset between spw (as in the top row of Fig~\ref{DGC.png}); if there is, see Sec.~\ref{sec-dgc} for use of a `diff gain cal'.

Assuming you are using continuum and have identified line-free channels, and have a measurement of the continuum peak, you can calculate the continuum bandwidth for each spw and thus what the S/N would be. You may already have per-spw continuum images from the pipeline, or make images. Otherwise, if you have an image made from all spw, aggregate bandwidth $\Delta\nu$, the noise per spectral window with line free channels summing to $\delta\nu_{\mathrm{spw}}$ 
is proportional to $\sqrt{\Delta\nu/\delta\nu_{\mathrm{rms}}}$ so you can estimate whether you can use any individual spw for self-calibration or, perhaps, combine two in the same baseband. You may need trial and error to decide whether combining more spw gives more good solutions (Sec.~\ref{sec-goodbad}) and a better image due to better S/N per solution, or whether it makes things worse because the errors are different in different spw.

Using  the chosen spw or
combination,
make a model image.  In {\tt ft} (if used) and in {\tt gaincal} set {\tt spw} to spw/channel selection used for the model.
By default, {\tt gaincal} and {\tt applycal} apply the gaintable
solutions for a given spw only to that spw and flag spw with no
solutions.  In order to apply solutions across boundaries, you have to
use the {\tt spwmap} parameter (see Appendices~\ref{ap-gaincal} and~\ref{ap-applycal}). In this example, the  wide-band spw is number {\tt [18]} and the narrow spw are {\tt [20, 22, 24, 26, 28, 30, 32]}. spw 1--17 are present in the data (e.g. as listed by {\tt listobs} and although they are not being used for the target they must be given in the spwmap list.
A 
`diff gain cal' table {\tt 'DGC.p'} is being used and you have derived a wide-band correction table 'spw18.p0'.  You  need one spwmap list
per table (an empty list applies each set of corrections to the spw
it was derived from), giving: \\
{\tt gaintable=['DGC.p', 'spw18.p0']}\\
{\tt
  spwmap=[[],[1,2,3,4,5,6,7,8,9,10,11,12,13,14,15,16,17,18,18,18,\\
      18,18,18,18,18,18,18,18,18,,18,18,18,18]]}\\
to apply spw18.p to spw {\tt 18} and all higher-number spw.\\
If you want to average all spw, the resulting gaintable will be
labelled for the lowest-number spw in the average, so, here, if you
averaged all science spw between 18 and 32 in {\tt gaincal}, you would use the same spwmap as above in {\tt applycal} (and in the next round of {\tt gaincal}, if you were applying these tables when deriving a new one).

Tropospheric refraction is proportional to frequency (Eq.~\ref{eq-linear}), so a
phase correction of $45^{\circ}$ at 220 GHz is $45 \times 270/220 \sim
55^{\circ}$ at 270 GHz.
If you are transferring solutions across a large frequency interval, gain tables are scaled appropriately in {\tt applycal} by
using the interpolation suffixes 'PD', e.g., for two gaintables, {\tt applycal
  interp=['linearPD','linearPD']} -- note also that
in such cases you are usually extrapolating not interpolating.  Within a single EB this is less serious;
the maximum scaling required, in Band 3,
is 16\%, but is worth using across a larger interval e.g. using a line for self-calibration in a  spectral scan.
For  band-to-band phase referencing, see
\citet{Asaki20a, Asaki20b}.

\subsubsection{Transferring solutions between MS}
\label{sec-cross-cal}
It can be necessary to transfer solutions between separate visibility
data sets.  For example, if self-calibrating using a spectral line, it
may be quicker to split out  the line channel(s) into a
separate, small MS for self-calibration. When you have achieved the best set of gain
tables, these can be applied together to the full data set.  However,
you might want to apply the tables after each round of calibration to
the full as well as to the line MS, and make a test image of another line or
the continuum to ensure that the solutions are suitable for all data.

If you split out a line for self-calibration from spw 2 (for instance),  using
{\tt mstransform reindex=False} then  the MS  spw used to derive the self-calibration gain table will be numbered
2.  If there are 4 spw to be corrected in the MS you are applying it to, in {\tt applycal} specify {\tt spwmap=[2,2,2,2]} (see also example in Sec.~\ref{sec-transfer}).

\subsubsection{Use of Diff Gain Cal}
\label{sec-dgc}
If one or more spw are very narrow, all spw may 
have been averaged for the phase calibrator, or it may have been observed
in a different, full-bandwidth configuration. Before solution transfer
between spw, or spw averaging, the instrumental phase offsets between
spw must be removed  by applying corrections from a source used
as a `diff gain cal', DGC. This is usually the bandpass calibrator but
can be any continuum source where good phase solutions can be derived
per spw by averaging over a long time period (usually all data).

In this example, the phase calibrator wide-band spw are {\tt [18, 20, 22, 24]} and the narrow target spw are {\tt [26, 28, 30, 32]}, the bandpass calibrator being observed in both sets of spw.
Figure~\ref{DGC.png} top right shows the phase offsets between the
wide- and narrow-band (noisier) raw phases for a single baseline. The
bottom right (DGC.p) shows a single, averaged solution per spw per correlation
per antenna, representing the spw and polarisation offsets.  The
solutions in DGC.p are {\emph{not}} designed to correct
phases in the manner of imaging corrections, but, when applied to any
other data in the same EB, they allow the next set of phase
corrections to be derived such that they can be applied regardless of
spw or polarisation offset as long as the DGC solutions are again
applied. Thus, the bottom right panel shows that the target solutions
for spw 18 (spw18.p0), derived whilst applying the DGC solutions, have the
polarisations aligned and are symmetric about zero and in fact can be
applied to spw 32.

You may be able to use the target as its own DGC, see Sec.~\ref{sec-avg}.

\subsection{Spectral line self-calibration} 
\label{sec-line}

\subsubsection{Aligning data in rest frame of target}
\label{sec-align}

Before aligning in frequency check that each separate data set has used the correct position for the target, allowing for proper motions. The imaged target peak should be at the same angular separation (or exactly at) from the pointing centre in each case. In case of alignment problems see Sec.~\ref{sec-comb} for this and other issues when combining executions.

For each execution of an ALMA EB, the requested spw centre frequency is adjusted for the $V_{\mathrm{LSR}}$ of the target but then held fixed for the duration of the observations (except for ephemeris objects, for which see Sec.~\ref{sec-ephemeris}). The rotation and bulk motions of the Earth amount to about 1--2 km s$^{-1}$ per day so the shift during an execution is $<0.1$ km s$^{-1}$ but is more for multiple executions and can be tens km s$^{-1}$ for observations months apart, or if required by observational tuning (which is limited in precision to $\sim$20 MHz).  For simple imaging, {\tt tclean} will adjust to a chosen reference frame correctly but for self-calibration the data must usually be combined
into a single MS adjusted to  a constant $V_{\mathrm{LSR}}$, i.e. a constant frequency
in the target frame. For a single execution or for very broad channels this may not be necessary but as you usually would split out the target anyway this can be done in the process.

Use the {\tt listobs} output to compare the frequency ranges and  channel widths for each spw in multiple executions.  Work out, for each
spectral window, the overlapping frequency range and note the largest
channel width in frequency, and split each spw out separately using the appropriate
values in {\tt mstransform, regridms=True}, to ensure aligned frequencies and a uniform frequency width. If you only have one execution but channels $<0.1$ km s$^{-1}$ then it is also best to regrid the data.
Then, concatenate the regridded MS with
a small frequency tolerance (about 1 channel). You can also use a position tolerance for targets with proper motion, to align the combined data by pointing position (assuming this is correct, see above) rather than absolute position.  If each of e.g. 3 input MS had e.g. 4 spw and the same, single target, then the single output MS should also have 4 spw with  frequency coverage similar but probably not identical to the original and one target field. If the concatenated data set has $>$4 spw (or more than the expected number of spw) then the regridding did not use the correct frequency ranges/widths and/or the frequency tolerance is too small. If the  target is listed more than once, you probably need a larger position tolerance in {\tt concat}.

\subsubsection{Line self-calibration and the presence of continuum emission}

If you have made a concatenated data set, use this from now on, after applying continuum self-calibration if relevant (see below). 

ALMA targets usually exhibit both line and continuum and 
if the S/N per spw is high enough (see Sec.~\ref{sec-SN}
including Eq.~\ref{eq-minsnr}), it is best to self-calibrate the
continuum first. Measure the S/N
for the line-free continuum -- you can use the pipeline/QA2 image if
you are sure it is line-free, or make an image from the data combined
as above -- note that the numbering of line-free channels may have
changed. Making a channel-averaged copy of the data with averaging will speed up continuum self-calibration, but see precautions in Sec.~\ref{sec-imstart_cont}. 

If at all possible do amplitude self-calibration to make sure that the spectral slope is consistent
across all spw, see
Secs.\ref{sec-imstart_cont} to~\ref{sec-mfs}.   Apply the calibration
to the whole target data set.

\begin{figure}
  \includegraphics[width=15cm]{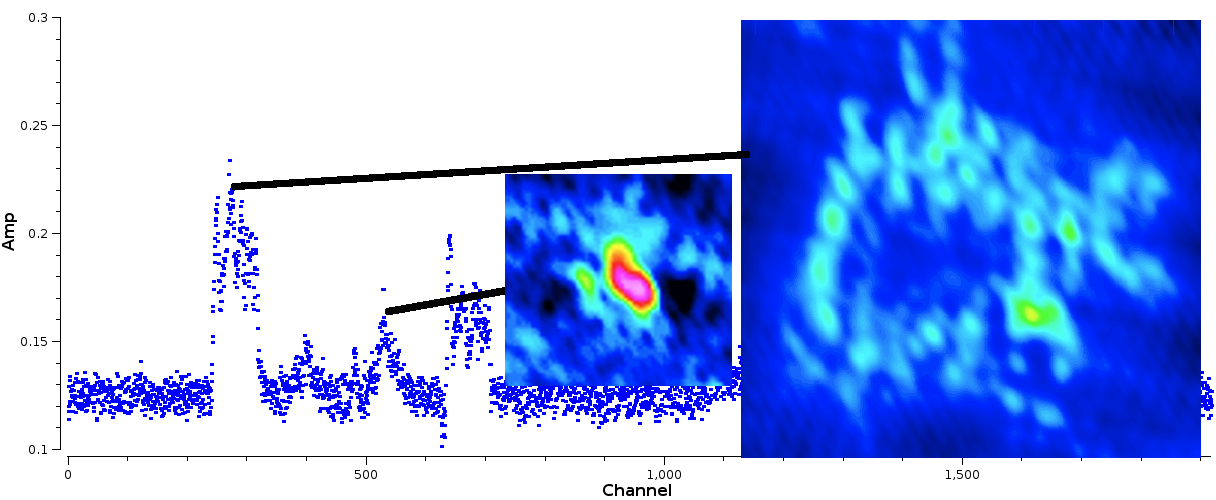}
  \caption{\small Example spectral lines with images of peak channels (on same
    intensity and angular scales); the left-hand image peak of channels around 515 is three
    times brighter than the peak {\emph compact} emission in the more extended line, despite the higher spectral peak, around channel 290, of the latter.}
  \label{Lineshapes_ims.png}
\end{figure}

For self-calibration using a line, choose the line with the brightest compact peak, which may not be the
 brightest spectral peak on all
 baselines. Fig.~\ref{Lineshapes_ims.png} shows that the line around
 channel 520 is most suitable (left-hand image), and in fact lines with a `triangular'
 profile usually provide the most compact emission. You can also plot a $uv$ range confined to the longer baselines to identify compact lines. If possible,
 self-calibrate using a single channel or a few channels, taking care
 not to include weaker emission from the line wings which will reduce
 the averaged S/N.  If the line emission is resolved and the structure
 changes from channel to channel, or becomes more extended away from
 the peak, then channel averaging should be minimised.
 
Whether to subtract the continuum before or after line self-calibration depends on which route will enable the most accurate model to be developed.
Estimate the S/N for the channel(s) to be used for self-calibration for the chosen line.  If the line is much brighter than the continuum you can estimate the rms in the combined channel width to get a rough estimate, or make an image of the chosen channels. If the rms in the combined channels is such that the continuum would be $\lessapprox$3$\sigma_{\mathrm{rms}}$ in those channels, or the line and continuum have very similar structure but the line has much higher S/N,  self-calibrate on the line before subtracting the continuum
(as this will be more accurate for better-calibrated data).  In other cases, subtract the continuum before line self-calibration.
However,
if the continuum is bright enough to be seen in the channel(s) used
for line self-calibration, but has weak, hard-to-image extended
emission, then you can self-calibrate using the data set after
continuum subtraction but apply the corrections to the original
unsubtracted data (see Sec.~\ref{sec-cross-cal}) and then re-subtract before making the final images.  

Line self-calibration then follows the same principles as for continuum (Secs.~\ref{sec-imstart_cont} to \ref{sec-ampscal}). Make sure that you use the same channel selection in imaging and in
calibration as you used in making the model, but use the spwmap
parameter in {\tt applycal} to apply the solutions to all spw
(Sec.~\ref{sec-transfer}). 
 If you have previously used continuum self-calibration then the flux scale and positions of all spw should be consistent, but otherwise, in order to 
transfer corrections from a line in one spw to all spw, see also
Sec.~\ref{sec-dgc} to check that there are not phase offsets and, if the frequency difference is more than 10--20\% (depending on the probable accuracy of phase corrections), see the last paragraph of Sec.~\ref{sec-transfer} for scaling phase corrections by frequency.

\subsection{Extending self-calibration from a time selection to all target data}
\label{sec-selective}
\begin{figure}
  \includegraphics[width=8.6cm]{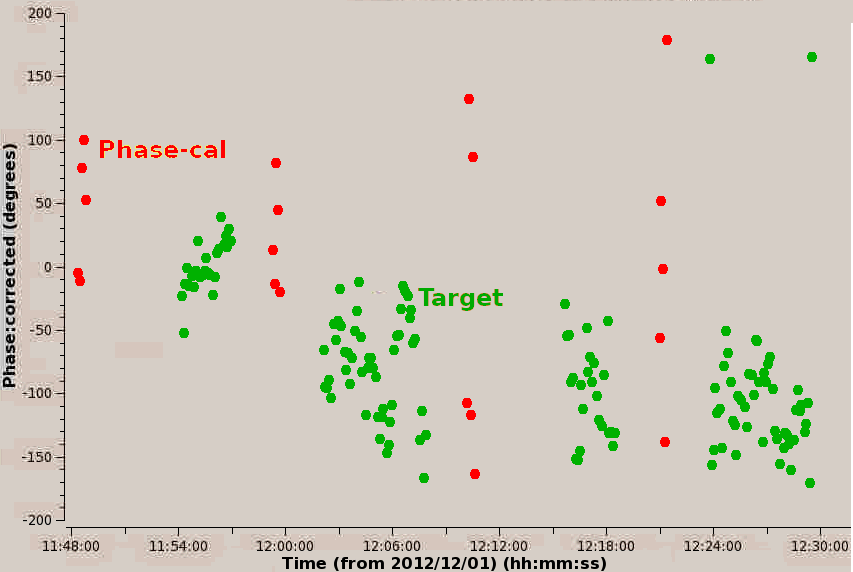}
  \includegraphics[width=8.5cm]{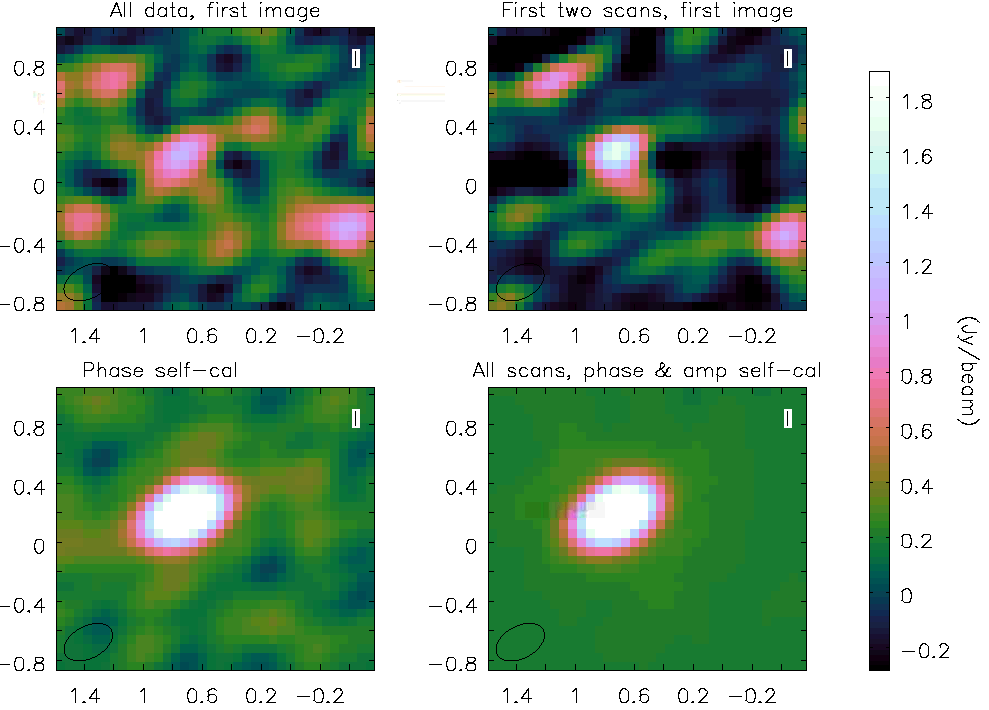}
  \caption{\small Left: phase calibrator and target scans for CW Leo (ALMA Band 9, continuum), one
    baseline, after applying phase calibrator solutions. The phase calibrator was at a higher declination than the target and is seen at low elevation in the last scans.   Right: Images
    of CW Leo during self-calibration.}
\label{CWLeo.png}    
\end{figure}

Sometimes, phase-referencing is less successful for part of an EB, for
example if the phase calibrator is at a lower Declination than the
target and suffers low elevation effects for the last scan(s), or if
an execution is terminated early and the last scan is on the target.
In such cases, an image made from all the target data including
scans without good phase referencing is likely to have serious
artefacts. Fig.~\ref{CWLeo.png} left (data from \citealt{Decin15}) shows that the
last phase calibrator scan is missing and the previous scans are
progressively noisier as the observation progresses.  The amplitude for
the last two target scans is also poorly calibrated.
Fig.~\ref{CWLeo.png} in the top left of the right-hand panel, shows that making an image,
after applying the phase calibrator corrections only, showed more than
one apparent source of similar brightness.  Imaging just the first two
scans gave a noisier image but the star could be identified clearly.
In a case like this:
\begin{itemize}
  \item Make an image model from the good target data, making sure that it is    saved in the MS.
  \item Perform and apply phase self-calibration for the good scans
    (iteratively if necessary).
  \item Re-image the good scans naking sure that the model is saved in the MS and
    perform amplitude self-calibration as usual for the good scans.  
\item Applying this calibration and re-imaging the good scans now
  provides a model M0  for starting to self-calibrate {\emph{all}} the data.
\item {\emph{Don't use the previous gain tables}}; phase self-calibrate all
  scans with the shortest solution interval which was suitable for the
  first two scans (although if that fails many solutions, try longer),
  table p0all.
\item Don't use applycal yet, but apply p0all in {\tt gaincal} for
  amplitude self-calibration of all scans, still using model M0, table a0all.
\item Image all scans and if necessary use this as a model to continue
  self-calibrating all data; however, at every {\tt gaincal} and {\tt
    applycal}, apply tables p0all and a0all as this will ensure that
  all the scans have reasonable starting calibration.
\end{itemize}
This uses the first two scans to provide a model to self-calibrate all
the data; the thing to remember is not to image all scans until you
have applied both phase and amplitude solutions to all scans.  The
first set of solutions for all data will look quite different for the
different scans.  

\begin{figure}
  \begin{centering}
  \includegraphics[width=8.3cm]{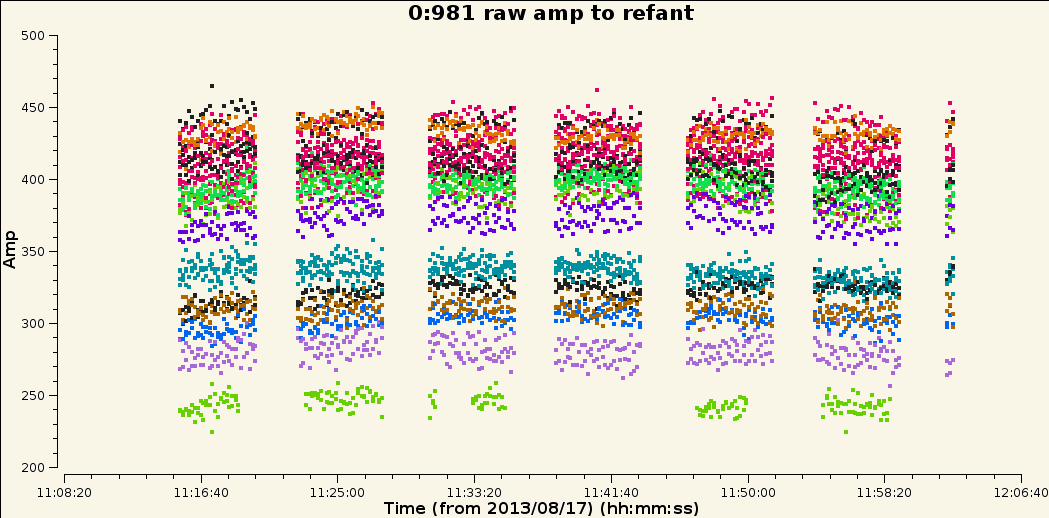}
      \includegraphics[width=8.3cm]{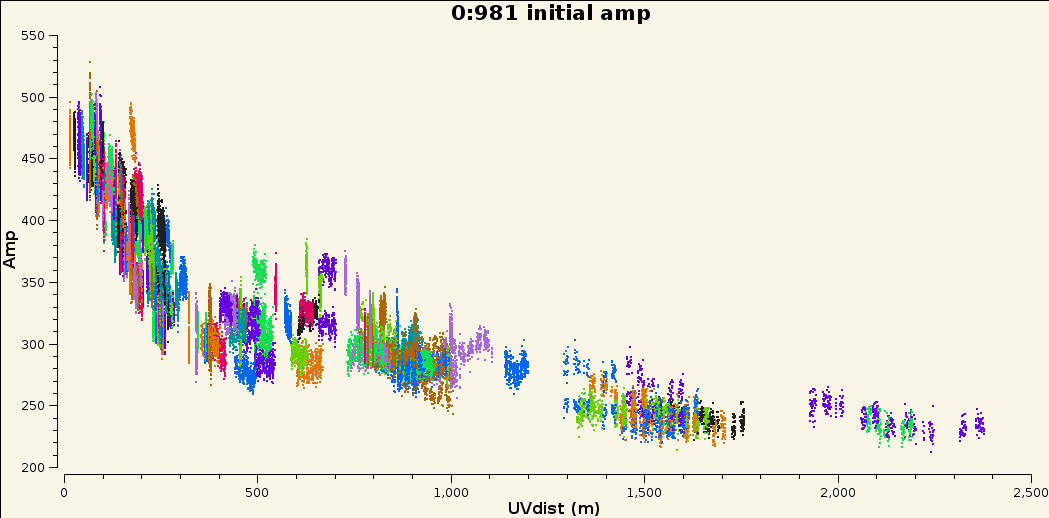}
      \includegraphics[width=8.3cm]{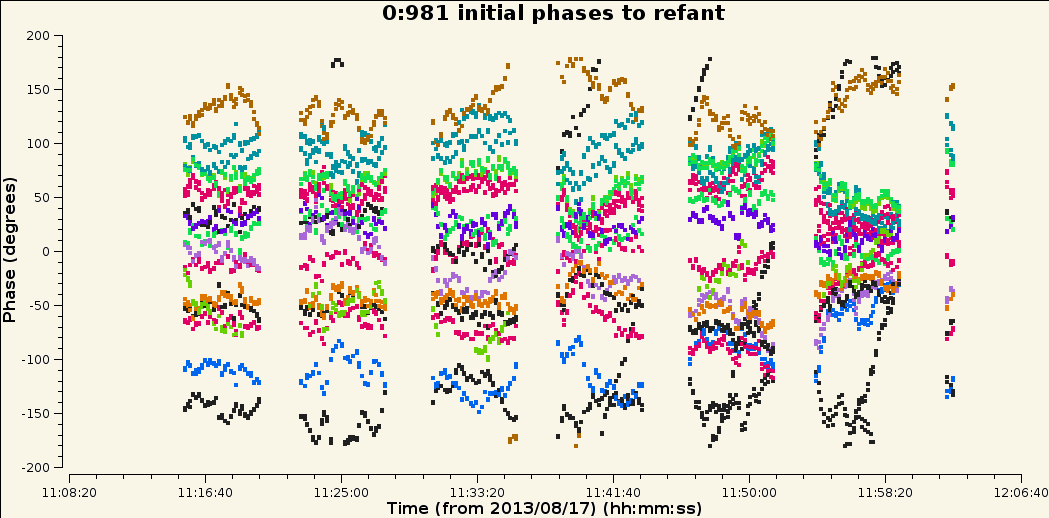}
  \includegraphics[width=8.3cm]{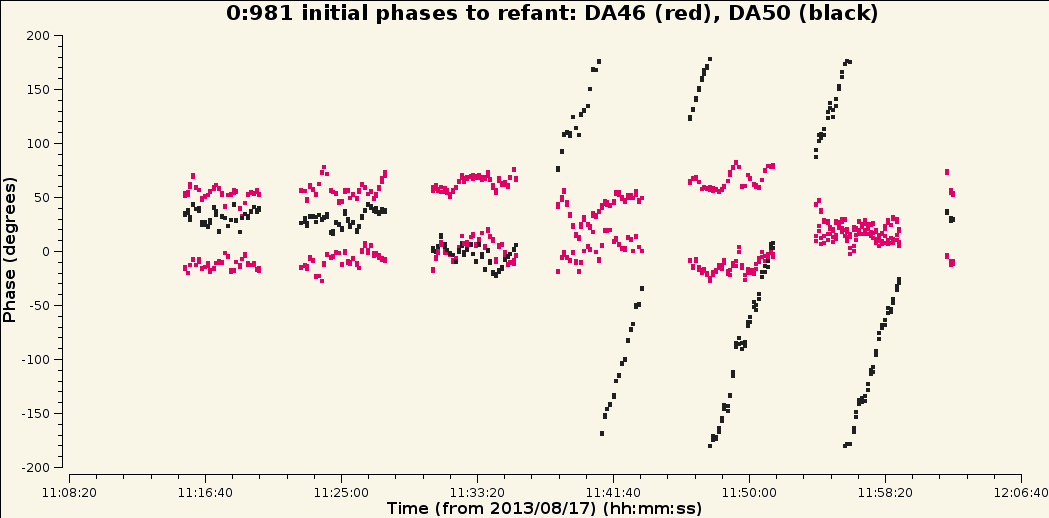}
  \caption{\small Top, left and right: VY CMa amplitudes for a spectral peak after applying
    phase calibrator solutions, all reasonably consistent and indicating
    slight resolution. Bottom left: VY CMa phases
    for this peak after applying
    phase calibrator solutions. This are mostly fairly flat (offset as
    the peak is not at the phase centre) but the last 3 full scans
    show a steep phase slope for one baseline. Bottom right: This is due to
    DA50, contrasted with a better-behaved antenna.}
  \label{0-981.png}
  \end{centering}
\end{figure}

A similar strategy was used in self-calibrating a maser peak in the VY
CMa data, Fig.~\ref{0-981.png}, but in this case although
phase-referencing worked properly, there was an additional error (of
unknown origin)
affecting target phases for one antenna. The amplitudes after applying the phase calibrator solutions
were as expected, showing slight resolution (top row). However, one
antenna, DA50, showed an anomalously high phase rate
(bottom row).  In this case, DA50 was excluded from imaging and
calibration ({\tt antenna='!DA50'}) in the initial phase-only
self-calibration (table p0) using {\tt solint='inf'}, which 
would have averaged over the $360^{\circ}$ wrap. DA50 was then
included for phase calibration using a solution interval of 36 sec,
and, applying all phase solutions, for amplitude self-calibration.
If the model has the right amplitude scale and you want to correct mis-scaled data, in {\tt gaincal} use {\tt solnorm=False} so the solutions for the good antennas (or spw etc.) should come out close to 1, but the problem antenna (or etc.) will have the required solutions, as in Fig.~\ref{tab-acals} (n).

Transferring solutions between spw assumes that the corrections obey  Eq.~\ref{eq-linear}, i.e. the atmosphere is non-dispersive and the phase errors scale linearly with frequency (negligible within a single ALMA tuning).  However, some water masers e.g. 325 GHz, are bright enough to self-calibrate but the data are affected by a deep telluric water line which may produce non-linear effects.  Transferring solutions to other parts of the observed spectrum e.g. the line-free continuum may leave higher than expected residuals and astrometric errors.  If the continuum or any other line is bright enough to self-calibrate, you could derive and apply its solutions first, and then self-calibrate the maser.  This will almost certainly improve the maser channels but compare before and after applying its solutions to other channels to see which gives the best S/N and test for position shifts (which will also give an idea of any additional astrometric uncertainty for the maser).

\subsection{Resolved phase calibrator}
\label{sec-resolved}
Self-calibration with the default point or an external model can also
be used to build up a model for the phase reference source if it is
unexpectedly resolved, similar to the procedure for a target, Sec.~\ref{sec-uvrange}. If using QA2 scripts, make sure that the best model is inserted  into the MS, either from your final {\tt tclean} or with {\tt ft} if you are using an external model (see Sec.~\ref{sec-ft}), before
running the stages calibrating the phase reference. If there is a {\tt
  delmod} in that part of the script, remove it.  If you have no external model, normalise the amplitude solutions while developing a model.
The simplest ways to set the target and phase calibrator flux densities is to use the bandpass calibrator or $T_{\mathrm{sys}}$ as outlined in Sec.~\ref{sec-nophref}.
Or, measure the apparent total flux density of the phase calibrator from an image made after applying corrections derived for all baselines using normalised amplitude solutions, which will be in correlator units. 
Make an alternative amplitude solution table in the usual QA2 fashion but in the {\tt gaincal} leading to {\tt fluxscale} restrict the baseline lengths to those where the phase calibrator is unresolved in order to derive the total phase calibrator flux density.  Then, derive a scaling table to apply alongside the normalised all-baseline amplitude calibration table (and the phase corrections) to the target and phase-reference, see Appendix~\ref{tsystoJy}.  

This is not usually necessary for ephemeris objects used for phase referencing, as the ephemerides used in CASA provide models, see Sec.~\ref{sec-ephemeris}.

\subsection{Observations without phase referencing or problem phase calibrators}
\label{sec-nophref}
Some very early ALMA data, and observations from other arrays, are
observed without phase referencing, or the phase reference source may
be so weak it does more harm than good, so self-calibration is worth
trying. VLBI data are often not phase-referenced and the application
of prior calibration will differ from the description below but the
actual self-calibration principles are the same. See the CASA VLBI
School (linked in Appendix~\ref{ap-schools}) for more details such as the calibration
of delay and rate. The use of ALMA in VLBI requires pre-calibration of the phased array data and special attention to  polarisation, see  \citetalias{ALMA-TH} and \citet{Goddi19}.

Occasionally, a phase (or other) calibrator is bright enough but the model is not correct -- typically, a calibrator is assumed to be a point source but is actually resolved (see~Sec.~\ref{sec-resolved}), or there are other sources in its field.  This leads to an overestimate of amplitude corrections for longer baselines so a target plot of corrected amplitude against $uv$ distance has an unphysical positive gradient, and also  to phase errors, an increase in noise and degredation of astrometry.
Very weak extensions (below the required dynamic range) may not matter, but in other cases the resolved calibrator should be self-calibrated itself,  starting as described below for a target, until a good image model is obtained.  This model can then be provided in the QA2 script or pipeline and the stages from the first use of the calibrator in question should be repeated (taking care that no subsequent steps remove the good model).  If this is successful you can then process the target as normal.

If the you tried applying
phase calibrator solutions to the target unsuccessfully use {\tt clearcal} before recalibration/self-calibration; you may
also have to restore any data flagged by bad solutions.

If the phase calibrator is unusable, you need alternative techniques for both the functions of
phase-referencing, i.e. setting the flux scale (for correlator output units
to Jy) as well as time-dependent calibration.
  Make sure
that the bandpass and earlier calibration are as good as possible.
If you have separate flux scale
and bandpass calibrators, {\tt fluxscale} produces a gain table with
the bandpass calibrator amplitude corrections scaled.  If the
solutions for the bandpass calibrator are pretty smooth and
consistent, including the solution closest in time to the target,
apply that table using {\tt interp='nearest'}.

If the solutions show
more than a few percent scatter, or if you have an a-priori value for the flux density of the bandpass calibrator (or whatever you are using as flux scale calibrator for the target) you can, first, use {\tt setjy} to set its flux density. Derive time-dependent phase and then normalised amplitude corrections for this calibrator.  Apply these to the calibrator and derive another amplitude solution table without normalisation with a long time-averaging interval; this table can be applied to the target also.
The amplitude scaling factor should be the same for all sources unless they contribute a significant
fraction of $T_{\mathrm{sys}}$, in which case see Appendix~\ref{sec-avc}. Don't apply phase solutions from other sources to the target if you don't have phase referencing as these are direction- and time-dependent.

Alternatively, in the absence of any astrophysical flux scale calibrator, you can use
$T_{\mathrm{sys}}$ to scale the data from correlator counts to Jy, see
Appendix~\ref{tsystoJy}.

Estimate S/N from the observing parameters and whatever prior information you have about the target. 
If the S/N is so low  that you need to average all spw and
polarisations, see Sec.~\ref{sec-dgc} for using a bright calibrator to
remove the offsets. 
It is best to apply the fluxscale (and DGC, if needed)
tables along with the bandpass and other prior calibration, and split
out the target; we assume that, however you do it, these tables are
all applied.

\subsubsection{First round of self-calibration without phase referencing}
\label{sec-uvrange}
You will have to do one round of phase self-calibration
{\emph{before}} imaging the target.  See Appendices~\ref{ap-gaincal} and~\ref{ap-applycal} as well as the online help for more explanation of parameters mentioned here.

If a model is not specified in {\tt gaincal}, the default is a 1 Jy
point source at the phase centre.  If you expect
the target to be dominated by a point source this is adequate for the
first round of phase-only self-calibration.  
If you have access to a separate observation of the same target at a
similar resolution (even e.g. a more compact array but a higher
frequency), and can image this to produce a model, this can be used as
long as the structure is similar to what you expect.  You may need to
modify the model header to persuade {\tt ft} to insert it correctly.
An image in units of Jy per pixel can be used.
The CC table attached to an AIPS image (exported from AIPS as FITS) can be converted
to a CASA image, see Appendix~\ref{ap-CC}.  

Be very cautious in the first {\tt gaincal}, phase only (so the model
amplitude does not matter), possibly reduce {\tt minsnr} to avoid
failing solutions. For arrays with few antennas, you might also need to reduce {\tt minblperant}. Use {\tt applycal
  applymode='calonly'} as failed solutions may be due to an inadequate model (not bad data). 
Each self-calibration product (tables, images, MS
{\sc corrected} column) should be inspected carefully (Sec.~\ref{sec-goodbad}), especially in the
early stages.  As well as checking the S/N, make sure that you are not
freezing artefacts into the data or forcing extended emission into a
central point. 

If many solutions fail or look like noise, or
the subsequent image is very noisy or distorted, it may help to
restrict the $uv$ range in the first round of calibration. For example, if you
think that the target is a few synthesised beams in area with an
unknown distribution of peaks, you could restrict the range to less than
about half the longest baseline length (so, effectively,
the point model will be larger, and complex substructure is below the
resolution limit, so the model represents the data better).  If all
antennas have at least some baselines in this range the resulting gain
table will correct all antennas and you can then make an image with
all data as usual.  If some antennas do not have good solutions after
the first round, increase the $uv$ range slightly (enough to include
at least one more antenna) in imaging and use a similar value in the
subsequent {\tt gaincal}, continuing phase self-calibration with small
increases, until all the data are included.

If you are using a spectral line, check the effect of
the calibration on the continuum or a different line (at the end, or after each {\tt applycal} if not too time-consuming).  Conversely, if
you are using compact continuum but you expect an extended line, image a line channel after applying the calibration to check it is as expected.

\subsubsection{Developing the image and model structure}
\label{sec-spur}
Starting with a point model and building up the actual target
structure, as in Sec.~\ref{sec-uvrange}, usually works very well if done cautiously.  There is a
particular hazard if your target is a point double and a point source
is used as the model for amplitude self calibration. This can produce
an apparently triple source -- `spurious symmetrization'.  If you see
such a structure, and you are expecting a double not a triple, you may
want to try re-imaging just masking two components to provide an image
model of a double.   For this, or any, structure, if your best model is
very different from the starting model, you might want to go back to
the beginning and insert the best model (so far) before the first
phase-only self-calibration (as in Sec.~\ref{sec-alt}). If necessary re-apply just the
pre-self-calibration gaintables and use {\tt delmod} and restore any
flagging due to failed solutions.  No or fewer $uv$ restrictions and
fewer rounds of calibration would be needed the second time round.

\subsubsection{Implications of no phase-referencing}
The target position
is fixed at the position of the first model so you cannot use
self-calibration without phase referencing to improve astrometry.
If you need to adjust the position to match that of other
observations, it is easiest to use {\tt puthhead} on the final image,
but you can also change the visibility data position (Appendix~\ref{fixvis}).
See  Sec.~\ref{sec-astrometry} for estimating the
target position using `reverse phase referencing' if you do have a detectable phase calibrator (or check
source) but it is too faint to use for normal phase referencing.

\subsection{Primary beam correction, heterogeneous arrays and smearing}
\label{sec-het}
The model used in self-calibration should represent the sky brightness distribution detected by the array, including field of view limitations. Thus, as noted in Sec.~\ref{sec-imsc}, you should not apply a PB correction until after self-calibration even if there are sources affected by this.   Similarly, if sources in the field are affected by bandwidth- or time-smearing, they can still be included in the self-calibration model (although self-calibration cannot remove the smearing); links on this issue are given at the start of Sec.~\ref{sec-errorrecognition}.

Occasionally (mostly test data) ALMA 12-m and 7-m data are correlated.  Other arrays, such as e-MERLIN, the EVN and other VLBI, customarily use different-diameter antennas. The different PB corrections mean that sources outside the inner parts (1/3 or less) of the PB will be affected differently. This should not affect position, so phase self-calibration is unaffected, but amplitudes are.  If there is a bright source affected differently by PB corrections, and you need to try amplitude self-calibration, one strategy is to image it using the smaller antennas of similar size, to get the least inaccurate model, and use this to self-calibrate all data.  The gains for the larger antennas can then be scaled by the ratio of their PB correction to that of the smaller antennas, at the source position. 

\subsection{Mosaics}
\label{sec-mosaic}
It is usually best to chose a single field, containing the most suitable source for  self-calibration.
Since each pointing is observed separately, any pointing with weak or no signal will have very low S/N and solutions probably will fail.  Within a single set of pointings, if there are two or more bright sources
such that each is also detectable within the field of view of the
other,  use both (or more) such pointings for imaging and calibration, in order to ensure that confusion is mapped out.
The solutions for the best source(s) can then be applied to all pointings, on a principle similar to phase-referencing.  

Usually, ALMA observes all pointings over a short enough time interval that the inaccuracies due to a small time offset are small.  However, for very large mosaics, if only a subset of pointings is observed between each pair of phase-reference scans, you may need to handle each subset separately.

Use
the {\tt listobs} output to identify the scan numbers or source names and specify
these in imaging to make the model, and in {\tt gaincal}.  In {\tt applycal}, apply the solutions to all target scans. If there are more than one subset of target scans with separate calibrator fields use {\tt field} and {\tt gainfield}.

\subsection{Wide-field (anisoplanatic) calibration}
\label{sec-wide}
In the case of arrays with a wide field of view, such as at low frequencies (below a few GHz), or arrays using antennas with apertures which are  small with respect to the maximum baseline, $B_{\mathrm{max}}$ at a given wavelength, the corrections derived in one direction may not be appropriate for the whole field of view (hence the distinction in the first row of Eq.~\ref{eq-me}). This occurs when $B_{\mathrm{max}} \lambda \gtrapprox A_{\mathrm{eff}}^2$, where $A_{\mathrm{eff}}$ is the antenna diameter \citep{Cornwell2005}.

This is not a problem for ALMA but affects  metre-wave, wide-field (anisoplanatic)
arrays, where there are usually multiple, detectable sources throughout the field of view, for example see Appendix~\ref{ap-schools} for the LOFAR cookbook. This can also  be an issue for e.g. the VLA and e-MERLIN at $\lessapprox2$ GHz. It is often compounded by a large fractional bandwidth, requiring simultaneous inclusion of multi-frequency terms (Sec.~\ref{sec-mfs}).
Wide-field, wide-band imaging is available in CASA {\tt tclean} but is also often achieved using
{\tt WSClean} (\citealt{Offringa2014}, \citealt{Offringa2017}).
The technique of deriving solutions for one source in the field, subtracting it and moving on to the next,
is known as peeling, as implemented e.g. in {\tt DDFacet} \citep{Tasse2018}. These packages are available on github. 
Recent overviews of facet-based techniques are given in \citet{vanWeeren2016} and \citet{Tasse2021}.


A similar approach is needed to mitigate the effects of antenna pointing errors (Sec.~\ref{sec-point}) across wide fields, if there are multiple
bright sources extending outside the inner $\sim$third of the
PB, described for the VLA by \citet{Bhatnagar17}.

\subsection{Polarisation}
\label{sec-polarisation}
 ALMA observes in linear
polarisation (X and Y) and, usually, only the total intensity (Stokes I) will be
bright enough to self-calibrate, and the location of the different
hands of polarisation, i.e. the phase, should be the same.
  Thus,
phase-reference and polarisation calibration is used to correct any instrumental or other
$X$-$Y$ offset errors prior to self-calibration. For 
amplitude self-calibration,
make a total intensity image and use {\tt gaincal gaintype='T'} to
derive average solutions for X and Y, thus ensuring that differences
between the two hands are preserved. When the corrections are applied,
all the correlations will be scaled proportionally per solution interval.
For polarisation calibration of ALMA-VLBI data, see \citet{Goddi19}.

In the case of arrays with circularly polarised feeds and targets with strong Zeeman splitting, such as OH masers,
the peak in one hand, e.g. LL, may be offset by many channels from the RR peak, with negligible RR emission in the LL channels and visa versa. In such a situation it is necessary to derive solutions from the brightest single hand peak channel and apply these to all data.  This relies on prior calibration (phase-referencing, parallactic angle, and polarisation leakage where possible) to have removed any instrumental L-R offset. If this has not been possible, a procedure similar to that used to correct for wide-narrow band offsets can be used as in Sec.~\ref{sec-dgc}. 

\subsection{Low S/N and calibration problems}
\label{sec-lowsn}

If the first image is much noisier than expected, check first that there are not phase referencing problems and that all prior calibration has been applied correctly to the target.  You could also inspect images of a check source, if present, or the phase calibrator; if the latter is resolved see Sec.~\ref{sec-resolved}.  
If the problems only affect a particular spectral window, antenna or time segment, see Secs.~\ref{sec-transfer} -- \ref{sec-selective}. If the phase calibrator is entirely too weak or absent see Sec.~\ref{sec-nophref} and
Sec.~\ref{sec-previous} also for general prior calibration problems.  See Sec.~\ref{sec-goodbad} for diagnostics if you have tried self-calibration but many solutions fail and also check the target visibilities in case there are bad data.

This Section covers intrinsically faint targets such as where the S/N of the first target image is
so low that the minimum solution interval $\delta t$ from Eq.~\ref{min_solint} is
longer than a scan, or you try the usual recipe of solving for phase
per scan, per polarisation, per spw but lots of solutions fail.  
  There are several potential strategies to allow calibration solutions to be derived with a high enough S/N to be worthwhile, for targets which are borderline bright enough --  see Sec.~\ref{sec-whether} and inspect the target visibilities with different averaging intervals to see whether self-calibration is likely to succeed.

\subsubsection{Calibration parameters to make the best of low S/N data}
\label{sec-lowsn_params}

Appendices~\ref{ap-gaincal} and~\ref{ap-applycal} outline the use of the parameters which can be changed from default ALMA values for low S/N calibration. In summary, {\tt gaincal minsnr} can be less than 3 (especially if previous application of phase calibrator solutions means that all data have had some corrections already). If at first the model is likely to be only an approximate representation of the target, use {\tt applycal calwt=False, applymode='calonly'} to avoid downweighting or flagging data in early cycles of self-calibration. More averaging and alternatives are discussed in the next two Sections.

\subsubsection{More averaging}
\label{sec-avg}
If Sec.~\ref{sec-whether} and Eqs.~\ref{min_solint} and~\ref{eq-minsnr} suggest that the S/N is too low for self-calibration solutions per spw, per scan, per polarisation, try averaging in frequency or time in  {\tt plotms} to find the intervals which allow you to see some structure in the visibilities. For example, averaging both polarisations and
4 spw with similar continuum bandwidths could reduce
$\sigma_{\mathrm{rms}}$ by $\sqrt{8}$, so S/N 20 without averaging
becomes S/N 56. 
Initial averaging over scans etc. may also be needed even if the optimum solution interval should be a scan or less but many solutions fail or  the image after self-calibration is noisier than you expect. Examine any solutions you have derived
(Sec.~\ref{sec-goodbad}) to help diagnose issues.

See Appendix~\ref{ap-gaincal} for averaging using parameters {\tt gaintype} and {\tt combine}). Averaging for calibration will only work if the errors (on scales larger than thermal noise) are similar across the combination interval. For example in Fig.~\ref{DGC.png} top right, the phase slope is similar for both spw and both polarisations plotted and so the polarisations could be combined per spw, if the constant offset between X and Y was removed. In most `real' cases the offset (for a phase-referenced target) will be much less. Similarly, the spw could be combined; see Sec.~\ref{sec-cross-cal} for use of {\tt spwmap} to allow a single set of solutions to be applied to multiple spw.

Be careful not to average over a such a long time period that the phase changes by more than you are trying to correct for; see Fig.~\ref{tab-cals} (f), (g)
and Fig.~\ref{DGC.png} lower right, where antenna CM10 has a phase slope of $\sim$50$^{\circ}$ in 12 min. Averaging over this interval would leave a residual error of up to $\sim$25$^{\circ}$; this sort of solution accuracy will probably provide improvements in imaging but averaging over the whole time shown with $>$300$^{\circ}$ change would not work (see Eq.~\ref{phnoise}). 

In reality, in a low S/N data set it is hard to distinguish errors from the intrinsically complicated phase structure of a resolved source. You may be able to get a better idea of the error phase rate by looking at the phase calibrator visibilities (before applying its time-dependent phase and amplitude corrections) since offsets between polarisations and spw are instrumental and usually constant within an EB.  However, you may have to experiment and find what combinations (if any) produce genuine improvements from how many solutions pass, whether they look reasonable (Sec~\ref{sec-goodbad}) and whether the image improves.  Assuming that you do get a better model, you may be able to average over shorter intervals in successive calibration rounds.

See Sec.~\ref{sec-dgc} for how to
check that the spw/polarisations are aligned and apply averaged
corrections. Usually the bandpass calibrator is used to derive corrections for offsets between spw or polarizations.
However, you can also use the target
itself as a `diff gain cal'. Before the first self-calibration round, set {\tt solint} to  a long time interval (or all times) of the best target data
and average in time ({\tt gaincal combine='scan'})
but not the spw nor polarisations. As long as this gives good
solutions for all antennas, spw, polarisations, applying this table as
a DGC table (in all {\tt gaincal} and {\tt applycal} allows you to
average spw and polarisations for subsequent rounds.  {\emph{Don't}}
apply the DGC by itself to make an image model; stick with the first
model until you have per-scan solutions as well.  

If, even with averaging spw and polarisations (in
Eq.~\ref{min_solint}, $\delta \nu=\Delta \nu; N_{\mathrm{pol}}=2$),
 $\delta t$ is greater than a scan length, there is usually no point
in  averaging multiple scans for phase solutions, since
this is unlikely to improve on phase calibrator corrections, but it may
help if there are antenna position errors or to remove phase offsets. On the other hand, long-time averaging is often useful for if
one or more antennas are systematically miscaled as in Fig.~\ref{tab-acals}(n); if spw or polarizations are affected differently, keep them separate. In this case amplitude self-calibration may be worth-while even without phase self-calibration.

\subsubsection{Solving for phase or rate by gradient or spline fitting}
\label{sec-rate}

Figure~\ref{0-981.png}, bottom right, shows a steep phase error slope
for DA50.  If the S/N is too low to use a short enough solution interval to sample  the rate of change of phase adequately  (unlike in Sec.~\ref{sec-selective}), one can alternatively
 fit to the phase gradient, where this is stable over a scan (or more), so a longer solution interval can be used.

The task {\tt fringefit} is
designed for VLBI calibration.  It performs a fit to the first
derivative of phase as a function of time, known as as `rate'
(sometimes referred to as `delay-rate').  This can be done
simultaneously with fitting to the gradient of phase with frequency
(`delay') but that is not usually needed for ALMA alone; phase and
rate fitting only is done using {\tt fringefit
  paramactive=[False,True,False]}.  The best available model should be used. If successful, applying rate solutions should reduce the phase rate so that a long (e.g. per-scan) {\tt gaincal} solution can be used for phase calibration.  A number of rounds of calibration are likely to be needed to improve the model and solutions iteratively.

Even if the phase errors do not have a consistent gradient, 
a similar technique to fit to noisy phase variations with time uses
{\tt gaincal gaintype='GSPLINE'}.  This  can be used to interpolate
solutions from a weak phase calibrator over the target, as well as for
target self-calibration, and can be used for amplitudes as well as
phases, see the CASA docs for
gaincal (link in Appendix~\ref{ap-casadoc}).

In both cases, the astrometric accuracy will be compromised.

\subsection{Combining interferometric observations}
\label{sec-comb}

This section covers multiple executions of the same EB or closely related ones (e.g. a spectral scan) where the same array is used,  and also combining different arrays. 
When combining multiple observations, the first issue is to make sure
that they have a consistent target position and flux scale (allowing
for spectral index but assuming no target variability).  See Sec.~\ref{sec-align} for additional considerations for spectral line observations. 
If you are deriving solutions from one EB to apply to another
 you will need to use {\tt applycal
  interp='nearest'} (not linear) as linear interpolates but does not
extrapolate in time.  If you are deriving solutions from more than one EB, you have to concatenate them prior to {\tt gaincal}, see Sec.~\ref{sec-align} if you are using a spectral line.

In the case of extra-solar Galactic objects with proper motions (see
Sec.~\ref{sec-ephemeris} for ephemeris objects), if accurate proper motions were available for each observation, 
{\tt concat} can be
used to align MS
by the target pointing centres with the position at
the earliest time present. It is all right if the peak is offset, as long as the position relative to the pointing centre is the same
at each epoch. If not, due to  proper motion uncertainties or errors in
phase transfer from the phase calibrator, discrepancies can be corrected by self-calibration using a single-epoch model (Sec.~\ref{sec-posalign}). However,  centring the peak at each epoch before combination may provide a better starting-model, using {\tt fixvis} or {\tt phaseshift} and {\tt fixplanets}, see  Appendix~\ref{fixvis}.

Multiple executions of the same EB should give the same target flux
density, or, for different configurations, plotting amp against $uv$
distance should show similar amplitudes for a similar $uv$ distance and
frequency. There may be some variations due to $uv$ angle, and if the
longest baselines appear to have a generally higher scatter of
amplitudes than short ones then probably this may be just noise.
However, any discrepancy greater than the noise needs to be removed.

The ALMA accuracy goal is 5--10\%
depending on band, worse in the early cycles, but if different
observations are made separated in time and/or with different
calibrators these errors accumulate.  

Whether additional self-calibration can be used to reduce position or flux scale
errors, and improve combined data images, depends on what is being combined.

\subsubsection{Same configuration, different executions or spectral scans}
\label{sec-posalign}

For a single array configuration, if the target position (allowing for any proper motion)
is expected to be the same, as is the flux density (allowing for
spectral index), combined self-calibration can be useful to align EBs.
More data allows more accurate image models to be made (although the S/N
in each solution interval is not changed).  Since the structure must be in common
for all spectral ranges observed at different times, only continuum  can be used for joint self-calibration of
a spectral scan.  If the tunings are closely interleaved in time (within phase referencing intervals) you could use a bright line, see Sec.~\ref{sec-line} for issues like
aligning the velocities.

If the position or flux discrepancies are significant, and one EB
gives a significantly better quality image or is more accurate based
on a-priori knowledge, you might want to use its image model as a
starting model to phase self-calibrate all executions.
See Sec.~\ref{sec-refant} for choosing the refant.  Alternatively you
can concatenate the data and make an image model from all the data, if
any discrepancies don't cause noticeable artefacts, and perform phase
self-calibration.  If amplitude self-calibration is possible it is
best to use all data and to solve for the target continuum spectral index if the S/N permits,
especially for a spectral scan, see Sec.~\ref{sec-mfs}.

\subsubsection{Combining array configurations}
\label{sec-configs}
There are many considerations when combining arrays (or interferometry plus single dish total power data); this subsection only covers self-calibration aspects.
For a fuller overview, see the CASA Guide for M100 (ALMA) (link in Appendix~\ref{ap-casaguides}).  In this
situation, e.g. combining ACA and 12-m array data observed separately,
at any given time, each antenna will only have baselines within one
array or configuration, so calibration tasks will handle the data for
each array separately and the S/N per antenna cannot normally be
improved by self-calibrating the data after array combination.  The
separate configurations should be self-calibrated (where possible)
before combination as an image model made from one configuration is
probably not representative of the signal on the spatial scales of
another configuration. One exception is if there is a bright source
which is completely unresolved in any configuration, but otherwise it
is usually better to deal with any offsets in position or flux scale
by other means, see Appendices~\ref{fixvis} and~\ref{tsystoJy}.

However, if there is a good range of overlapping baseline lengths,
image models made from a different or multiple configurations may be useful in
self-calibrating individual configurations, if one or more are hard to self-calibrate due to missing spacings or low S/N. Self-calibrate individual configurations first as long as this leads to improvements and then make a combined image model from all calibrated target data.   For instance, if you have MS for extended and compact 12-m array configurations, (generally referred to as TM1 and TM2), 
you can using {\tt concat} to combine the MS, and then image TM1+TM2 together to provide starting model to self-calibrate TM1.  Use the {\tt listobs} output for the concatenated MS to select the scan numbers corresponding to TM1.

Alternatively, you can make the first combined image without concatenation (as {\tt tclean} will take a list of MS), and then use {\tt ft} to  Fourier transform the model  into TM1. It will only be inserted on
scales corresponding to baselines actually present in the data. Using this for self-calibrating TM1 won't lead to sensitivity to larger scales,  but it will avoid model errors giving rise to spurious calibration solutions. If the starting model does not represent the smallest scales properly you might need to exclude the longest baselines in the first rounds of TM1 self-calibration.  

Similar situations include using the model from well-calibrated data to be combined with another configuration which is not phase-referenced, see Secs.~\ref{sec-selective} and~\ref{sec-nophref}.

Total Power (single dish) observations cannot be conventionally self-calibrated, even if you have created pseudo-visibilities, since there is no observational phase information, although they can be used in combination with array data to constrain large scales in a model for array self-calibration.  Time, frequency or amplitude/flux scale errors should be corrected using specialised single dish tasks before combination. 

\subsection{Extended emission, Solar System Ephemeris Objects and the Sun}
\label{sec-ephemeris}

\begin{figure}
  \begin{centering}
      \includegraphics[width=7.8cm]{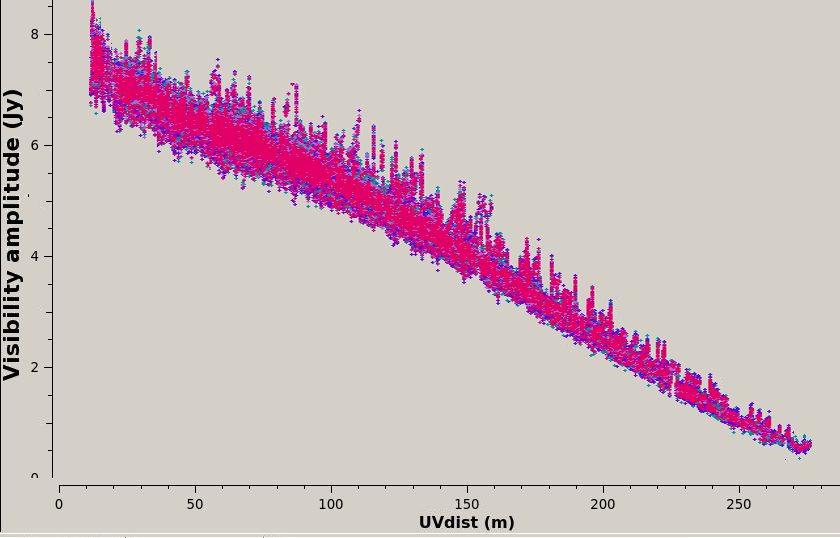}
  \includegraphics[width=8.6cm]{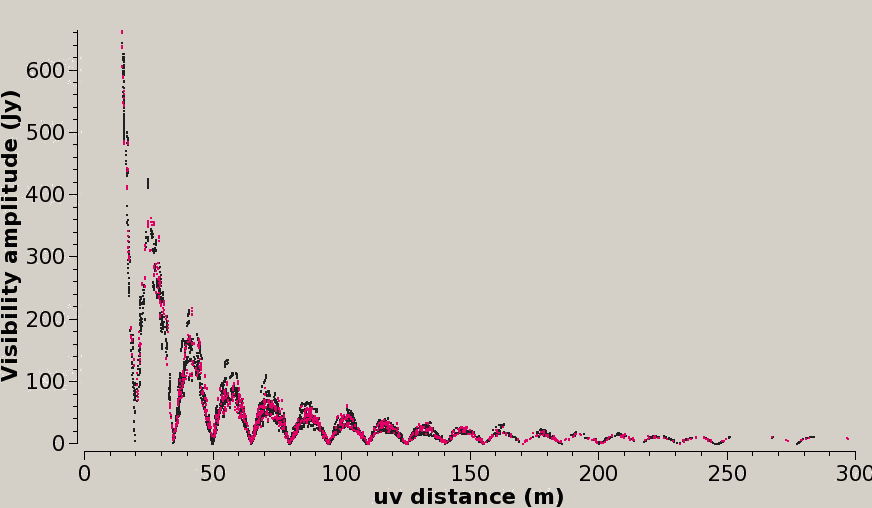}
  \caption{\small Left, a source likely to have some missing flux on baselines $<$20m.
    Right, Venus, showing extreme missing flux and many nulls, suggesting that it comes close to filling the imageable field of view.}
  \label{missing.png}
  \end{centering}
\end{figure}

Many Solar system objects -- planets, many moons and
asteroids -- have ephemerides and models in CASA ({\tt setjy
  standard='Butler-JPL-Horizons 2012'} \citealt{Butler2012}).  These
are generally very accurate with a few caveats for insolation-dependent
flux densities and atmospheric lines.  When setting time-dependent model such as these in {\tt
  setjy} always use {\tt usescratch=True}.  For early-cycle
observations the task {\tt fixplanets} was needed early in processing
to apply the ephemeris to a Solar System target.  Having attached or applied the ephemeris,  a Solar System object, whether observed as a 
calibrator or target, can be self-calibrated mostly like any other
target, apart from any issues with very large or bright objects, discussed below.   If combining multiple executions, use
{\tt concat forcesingleephemfield='Callisto'} (or whatever the object
is). It may not be possible to combine observations separated in time such that the apparent angular size or flux density of the object has changed significantly.

See the help for {\tt tclean} for imaging ephemeris objects;
normally, use {\tt phasecenter='TRACKFIELD'} and, if relevant, {\tt
  specmode='cubesource'}. 
It is a matter of judgement whether to use your image, or a
built-in model as the model for self-calibration; using the latter,
especially for continuum where the objects are very well-studied, at
least for the initial phase calibration, can allow you to go straight
to a shorter solution interval if S/N permits.

Fig.~\ref{missing.png}, left, shows a source with flux continuing to rise on the shortest baselines. It will be necessary to take great care not to include artefacts due to the missing spacings in the self-calibration model, and the S/N needs to be based on the $\sim$0.5 Jy flux on the longest baselines, but self-calibration for phase and amplitude should be possible with the usual S/N considerations. If additional observations sampling the missing spacings are available, see Sec.~\ref{sec-configs}.

Fig.~\ref{missing.png}, right, shows a planet which not only has much missing flux, as shown by the precipitous rise for baselines $<$20 m, but many nulls.  This indicates that not only is there much smooth flux on scales not properly sampled but it has it an angular size probably approaching or exceeding the PB FWHM. 
Phase self-calibration can be possible without necessarily restricting the $uv$ range, usually using the ephemeris model.
If there were only one or two nulls, it could be possible to select a $uv$ range or
ranges in {\tt gaincal} where all antennas have
at least 3 baselines with `enough' S/N for amplitude self-calibration, to a suitable selection of reference antennas.
However, the reduced number of
antennas contributing to some solutions will increase the effective noise and solutions
derived from selected $uv$ ranges should be inspected carefully.
In the case shown in Fig.~\ref{missing.png}, right there are too many nulls and amplitude self-calibration is probably not possible.

Check for lines if the object has an atmosphere; these may be included
in the built-in model but if the object is resolved their location may
be variable and it is best to exclude the affected channels during
self-calibration.  The motion of the emitting regions relative to the
planet centre (due to its solid-body rotation and/or winds) may be
significant, see e.g. \citet{Sault2004} for an example of dealing with
this in VLA data.

 If you are imaging a Solar System object which was originally
 observed as a calibrator and calibrated per execution using the built
 in model, you can continue to calibrate combined data using that
 model.  If you combine data for a Solar System calibrator and make an
 image model and insert this into the MS for self-calibration
 (rather than using the built-in model), the data are then aligned
 with the single image so you may not need to use {\tt 'TRACKFIELD'},
 but just leave {\tt phasecenter} blank.  However, be very careful to
 write a model scratch column (not virtual, see Sec.~\ref{sec-ft}) and check on a sample image.

Obtaining an image with the correct flux scale is  problematic for very bright and/or beam-filling planets, even with an accurate ephemeris value.  If it
makes a significant contribution to $T_{\mathrm{sys}}$,  not only will
the flux scale need correcting for this (Appendix~\ref{sec-avc}) but the internal model
is for the `true' brightness distribution, although the signal imaged
by ALMA in a single pointing will have PB attenuation.  Two
possible solutions are to use a scaled model, or to ignore the flux
scale during processing and apply a global correction factor ({\tt
  immath}) to the final images or measurements.  Emission may extend beyond the
well-characterised part of the PB.

For Solar observations with ALMA, see  \citetalias{ALMA-TH} and the Solar CASA Guide (Appendix~\ref{ap-casaguides}).
Once an image giving a good starting model has been achieved, self-calibration can be used, with the final rounds of calibration normally using very short (per-integration, e.g. 2s) averaging times, see e.g. \citet{Nindos2018}.

\subsection{Baseline-dependent calibration}
\label{sec-baseline}

Baseline-dependent errors are errors which cannot be factored into per-antenna terms.
Such errors provide an additional additive term for just one or some of the terms in square brackets in
the phase closure relationship Eq.~\ref{eq-phclos}. Thus the expression does not sum to zero.
For this reason they are sometimes called `non-closing' errors.  They are usually instrumental in origin (arising after correlation), and not direction dependent (${\mathbf{M}}_{ij}$ in Eq.~\ref{eq-me}), although using a point model for a resolved phase calibrator can produce similar effects (\citetalias{SI99} Ch. 5 Fomalont \& Perley).
For this reason, check if this may be the problem and if so, see Sec.~\ref{sec-nophref} before considering baseline calibration.

The number of baselines contributing to an image is $N(N-1)/2$ where $N$ is the number of antennas, and hence if solutions are estimated per baseline the noise per baseline is much higher than per antenna, giving
$\sigma_{\mathrm{solint, baseline}} = \sigma_{\mathrm{solint, ant}} \times \sqrt{N-3}$ for the same solution interval (see  Sec.~\ref{sec-SN}, Eq.~\ref{noise_ant}).
Equivalently, for a given image S/N, the minimum solution interval for baseline self-calibration is thus increased by a factor of $(N-3)$ relative to the value obtained from Eq.~\ref{min_solint} for antenna-based self-calibration.

For $N=20$  (the example in Sec.~\ref{sec-SN}) the image S/N needed for antenna-based calibration is 35 (Eq.~\ref{eq-minsnr}) but would need to be nearly 150 for baseline-dependent calibration. Each solution is less well-constrained, especially in extended configurations and/or short observations when very few baselines may sample a particular scale in the target, so the
model needs to be close to perfect since there is a higher risk
of obtaining apparently good solutions which are in fact based  on artefacts. Task  {\tt
  blcal} can be used to solve for phase and/or amplitude and even has an
 option of frequency-dependent solutions -- requiring even
greater S/N and model accuracy.

Antenna-based calibration should be completed and applied
first. Baseline-based calibration is most often used when the target
contains a very bright point source plus very weak extended emission
or spectral lines (typically, an AGN at cm wavelengths), so that a
dynamic range $\gg1000$ is needed.

\section{Imaging problems and diagnostics}
\label{sec-errorrecognition}

This section describes some of the problems you might find in an
image, how to track down the causes and possible remedies.  It
concentrates on ALMA problems, and assumes that the observations are
well-designed. For example, time- and bandwidth-smearing are not covered, but the conditions are shown in Appendix~\ref{ap-smear}. More details are given in
\citetalias{SI99} Ch.~18 (Bridle \& Schwab) and Ch.~15 (Ekers) and the
talk on error
recognition by Taylor at the 2017 NRAO workshop (see Appendix~\ref{ap-schools})
for more general coverage.  Not all errors are correctable by
self-calibration, and there may be more than one problem present.
Check whether pipeline or QA2 images show that previous
calibration/flagging needs improving.  Sometimes re-imaging with
parameters more suitable for a particular target brings good improvements.

In many cases it is not possible to distinguish unambiguously between causes of error. In order of decreasing probability (in ALMA data), stripes may be due to an amplitude calibration error (solvable by self-calibration, Sec.~\ref{sec-ampscal}) or due to bad data (needing flagging, see Sec.~\ref{sec-flag}) or due to confusion (requiring better imaging, Sec.~\ref{sec-confusion}). However, you may have to test different solutions by trial and error.

\begin{figure}
\includegraphics[width=8.4cm]{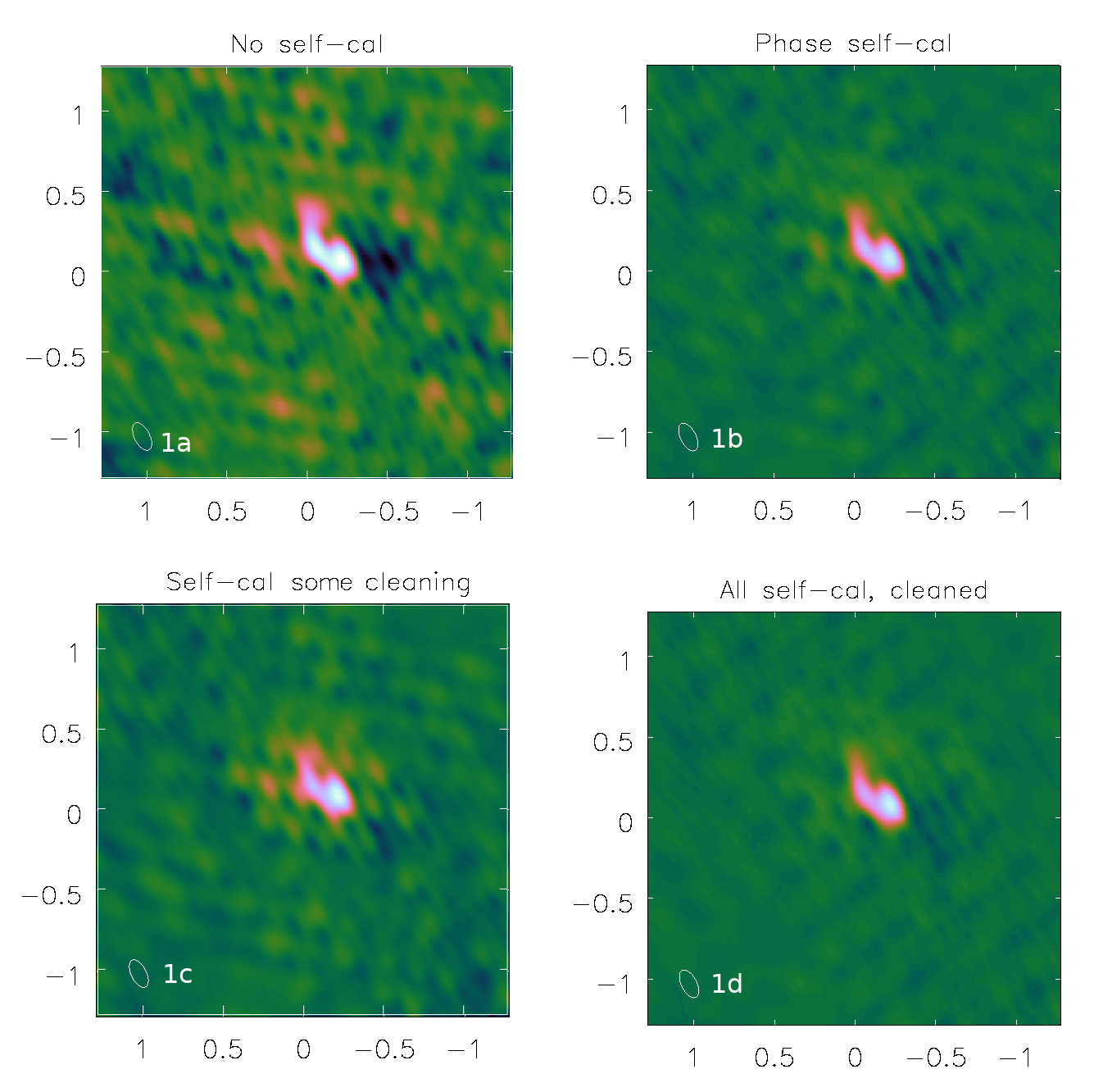}
\includegraphics[width=7.8cm]{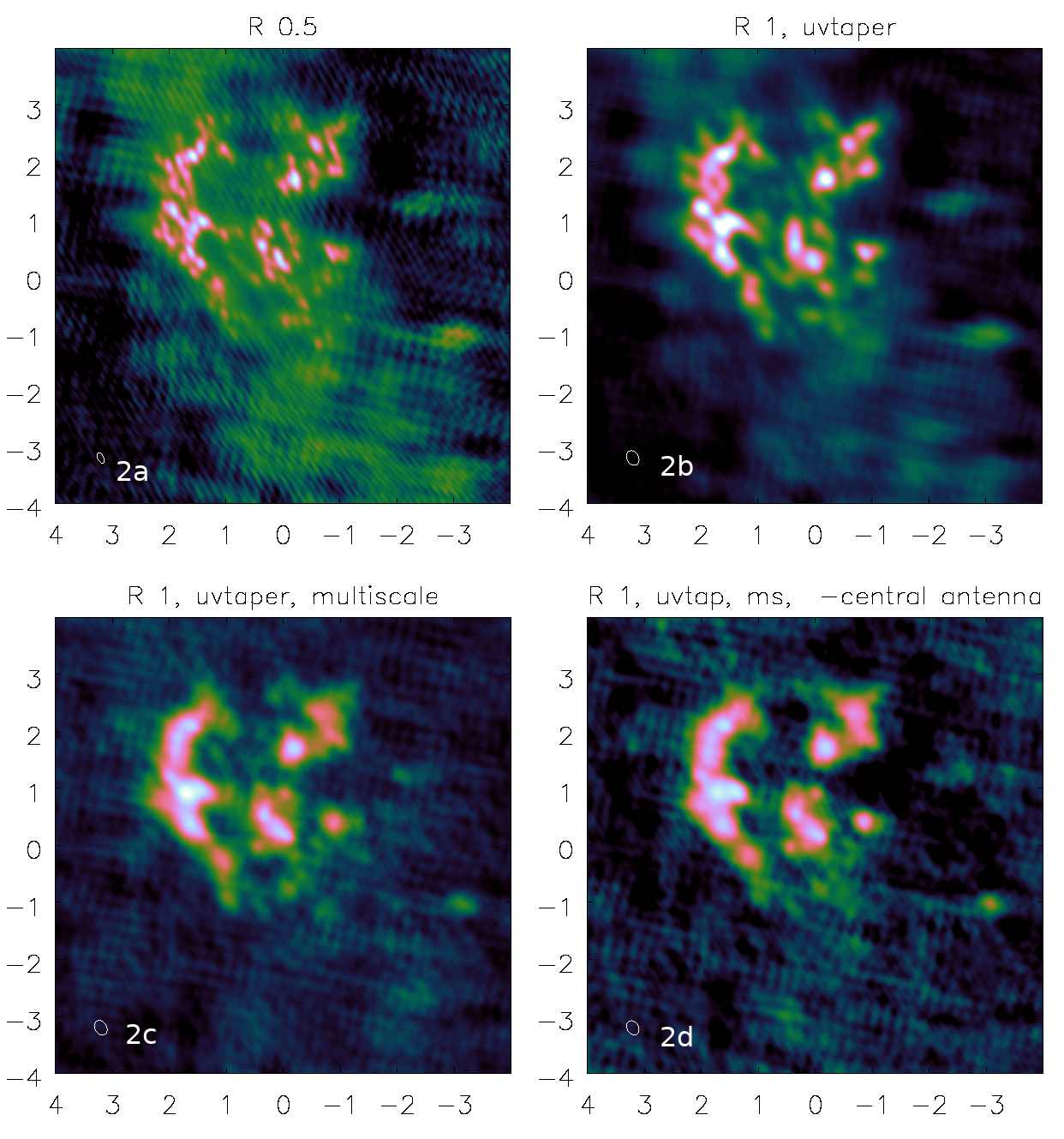}
\caption{\small 1a: Single, peak channel of the VY CMa data to be used for self-calibration,
  phase calibrator solutions only. 1b: Phase self-calibration. 1c:
  Phase then amplitude self-calibration but too few clean
  iterations. 1d as 1c but cleaned fully. 2a: Different VY CMa channel,
  with extended emission, fully calibrated, full resolution, {\tt robust=0.5}. 2b: {\tt
    uvtaper='0.1arcsec', robust=1}. 2c: as 2b but also multiscale
  clean. 2d: as 2c but omitting a central antenna.}
\label{Multiscale_etc.png}
  \includegraphics[width=7.5cm]{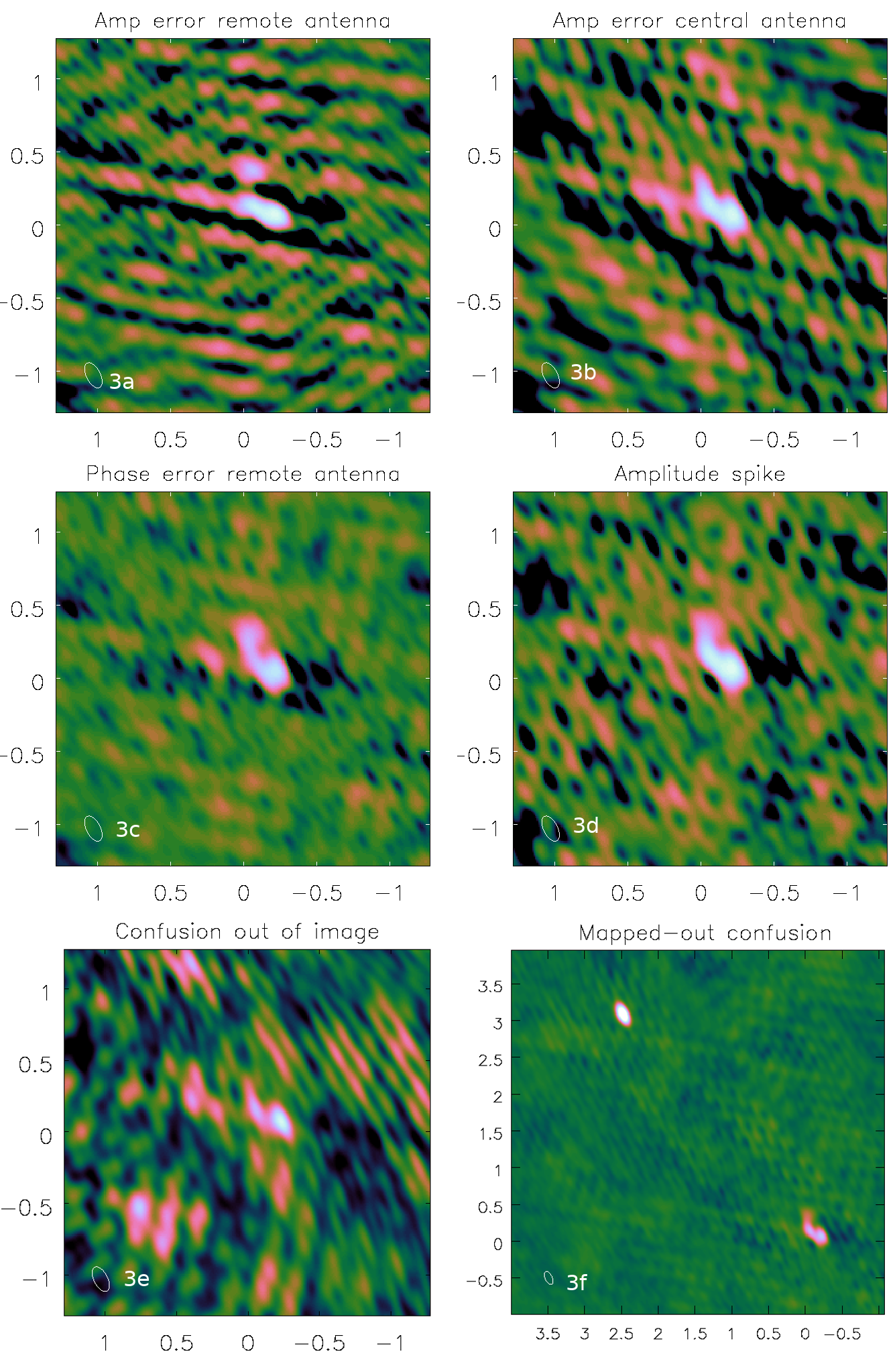}
\includegraphics[width=9cm]{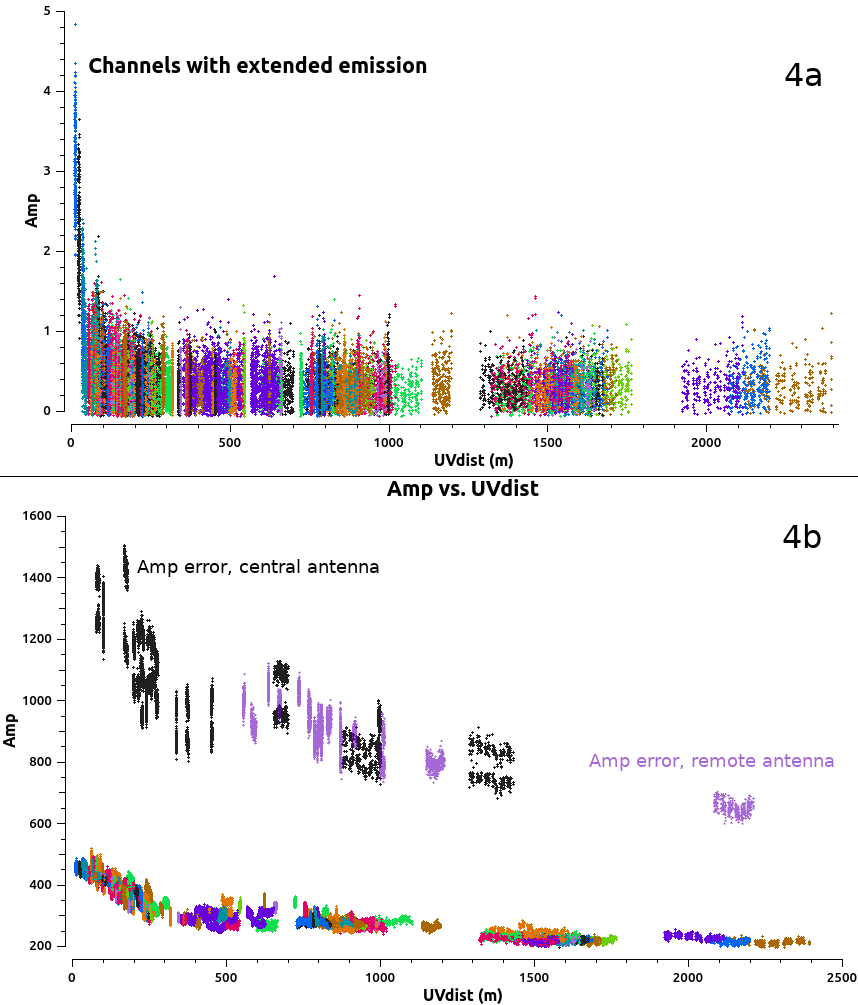}  
\caption{\small 3a--d: VY CMa peak channel as in Fig.~\ref{Multiscale_etc.png} 1a with
artificial  errors added. 3a (3b): Amplitude error on an antenna with mostly long (short) baselines. 3c: Phase error  on an antenna with mostly long baselines. 3d: Amplitude error on many baselines for a short time (few integrations). 3e: Confusing source outside image, not included in
  cleaning (target at (0,0)). 3f: 3e but with confusing source included in cleaning
  (note different angular size of image).
4a: Visibility amplitudes for Fig.~\ref{Multiscale_etc.png} 2a--c,
showing extended flux on antenna removed for 2d. 4b: Visibility
amplitudes showing mis-scaled data causing errors in 3a (mauve) and 3b (black).
}
\label{AmpPhaseConfusion.png}
\end{figure}

\subsection{Stripes and other structured artefacts}
\label{sec-stripes}

It may be easier to see the errors illustrated in Figs.~\ref{Multiscale_etc.png} and ~\ref{AmpPhaseConfusion.png} on-screen. In real life you can change the colour scaling in the CASA viewer.
 Fig.~\ref{n_allerrors.png} showed  examples taken from very early ALMA data where having only 6 antennas meant errors had a very obvious effect.  Similar examples are given here for some spectral lines taken from the VY CMa data.

In theory you can use
the artefact spacings in the image to identify the baseline lengths
responsible for errors but this is very difficult for many antennas, although it
can give some guidance. Artefacts with a spacing similar to the
synthesised beam suggest that one or more antennas with long baselines
are affected; the larger the image spacing of the artefacts, the
shorter the baselines which are implicated.  It is usually easier to spot errors in the visibilities, e.g. use the {\tt plotms} selections to scroll through all baselines to the reference antenna and use the {\tt locate} facility.

 Fig~\ref{Multiscale_etc.png} 1a shows anti-symmetric errors typical of
widespread phase errors (Sec.~\ref{sec-phase-fidelity}) which
initially dominate a target image with only phase calibrator (and earlier)
solutions applied.  If only one antenna is affected, the errors are elongated as in Fig.~\ref{AmpPhaseConfusion.png} 3c,
especially for a distant antenna in a short observation, where most of the baselines have a
similar direction. After phase errors are corrected by self-calibration, more symmetric amplitude errors are revealed,
Fig~\ref{Multiscale_etc.png} 1b.  These are usually not so serious but sometimes one antenna is
dramatically mis-scaled, as in Fig.~\ref{AmpPhaseConfusion.png} 3a and
3b. The longer the baselines, the
narrower are the stripe spacings in the image plane. Fig.~\ref{AmpPhaseConfusion.png} 4b shows the
baseline ranges of the bad data, which can help pin down the antenna(s) responsible. Such errors can
occur in real life due to $T_{\mathrm{sys}}$ errors, for example -- see Sec.~\ref{sec-point} for amplitude errors without associated phase errors.

Fig.~\ref{AmpPhaseConfusion.png} 3d shows the similar effect of a
short burst of bad amplitudes on all or many antennas but just for a few integrations. Plotting visibility
amplitudes against time can show the bad data.  This is less common in ALMA data but typical of
interference at cm wavelengths, where the affected data also have bad
phase.  Usually such data must be flagged.  For ALMA, if a lot of data seem to be affected, check the raw visibilities (i.e. before any calibration at all) in case the target data are fundamentally OK but there is a problem with $T_{\mathrm{sys}}$ or the phase calibrator (as in Sec.~\ref{sec-selective}, Fig.~\ref{CWLeo.png}).

Dynamic range errors (Sec.~\ref{sec-tderror-DR}) are usually dominated
by amplitude errors as in Fig~\ref{Multiscale_etc.png} 1a or resemble
Fig~\ref{Multiscale_etc.png} 1c and can similarly be tackled by self-calibration (starting with
phase as usual, being careful to restrict the model to emission you are sure is real).  If the signs of phase or amplitude errors persist
for a very bright target, reduce the solution interval as much as is
possible for the S/N (Sec.~\ref{sec-SN}).  Check whether the weights
need optimising (e.g. {\tt statwt}, Sec.~\ref{sec-weights})
or even consider baseline calibration (Sec.~\ref{sec-baseline}).

Errors due to a specific time or antenna where the amplitude and/or phase are anomalous, but the phase
is coherent and amplitudes are not much noisier, can usually be
corrected by self-calibration. However, for very short-timescale errors, it might be necessary to flag the data if the S/N is too low for a short enough solint.
You might have to exclude the
affected data to get a starting model (Sec.~\ref{sec-selective}). 
If you have trouble identifying the antenna(s), scans, channels
responsible for bad data, check the weights for anomalies and try
no averaging and different averaging intervals -- remember that bad
phases depress amplitudes; the affected data will have lower
amplitudes the more you average bad phase.

A persistent sidelobe pattern as in Fig~\ref{Multiscale_etc.png} 1c or worse (often looking like
spokes radiating from the peak) even after attempting
full calibration and cleaning can indicate a time-variable source
(Sec.~\ref{sec-variable}).

Fig~\ref{Multiscale_etc.png} 2a shows the `basket weaving' effect due to
gaps in the visibility plane coverage on longer baselines whilst Fig~\ref{Multiscale_etc.png} 2d
shows the effects of missing short spacings.  These may be mitigated
by changing imaging parameters (Secs.~\ref{sec-optimise},
\ref{sec-missing}) but not by calibration.

Fig.~\ref{AmpPhaseConfusion.png} 3e shows curving artefacts typical of
sidelobes of a confusing source to the left of the target at
(0,0). The solution is simply to make a larger image and mask all the
sources in cleaning, as in Fig.~\ref{AmpPhaseConfusion.png} 3f (confusing source in the NE), see
Sec.~\ref{sec-confusion}.  All sources must also be included in models
for self calibration.

\subsection{Optimise imaging}
\label{sec-optimise}

Before self-calibration, check for problems which may be mitigated by
more suitable imaging parameters. This will also help to make sure you
have good calibration models (although see Sec.~\ref{sec-imstart_cont} about not
downweighting baselines during calibration if possible).

Imaging
weights are applied in addition to sensitivity-based weights
applied during calibration (Sec.~\ref{sec-weights}). The 
combination affects the final image resolution (for example, if
the most distant antennas are less perfectly calibrated then
sensitivity-base weighting in calibration will coarsen resolution, but using a smaller {\tt tclean robust} value can mitigate this albeit probably increasing the noise). Both need
careful tuning if you have to optimise for a specific resolution or surface brightness sensitivity
rather than best S/N.

For ALMA, {\tt tclean weighting='briggs', robust=0.5} usually gives
the best S/N. Negative robust values give finer resolution but tend to
increase the noise and so do not necessarily improve measurement
precision (see Sec.~\ref{sec-gaussfit}).  Higher robust values may be useful if the observations are
affected by missing spacings as in Fig.~\ref{Multiscale_etc.png} 2a
and 2d, see Sec.~\ref{sec-missing}.  For sparse arrays/VLBI, {\tt
  weighting = 'natural'} may make the best use of a small number of
antennas. 
Where appropriate, use  spectral
index and/or multiscale imaging (as noted in Secs.~\ref{sec-imstart_cont}
and~\ref{sec-mfs}); Fig.~\ref{Multiscale_etc.png} 2c shows an example of multiscale imaging, which also speeds up cleaning.  The models produced by both methods can be used in self-calibration.

\begin{figure}
 \includegraphics[width=9.2cm]{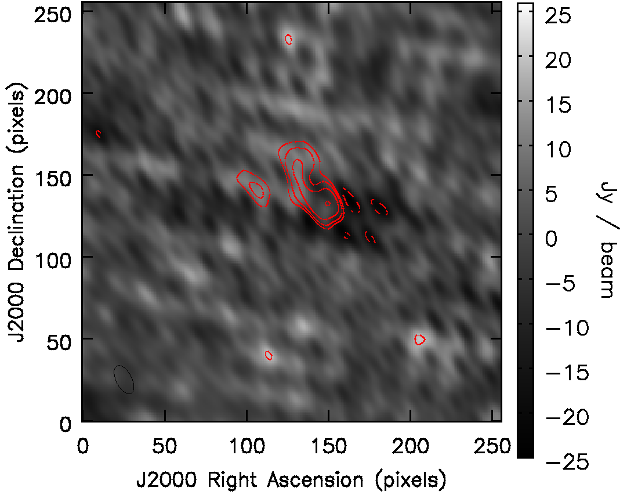}
 \includegraphics[width=7.2cm]{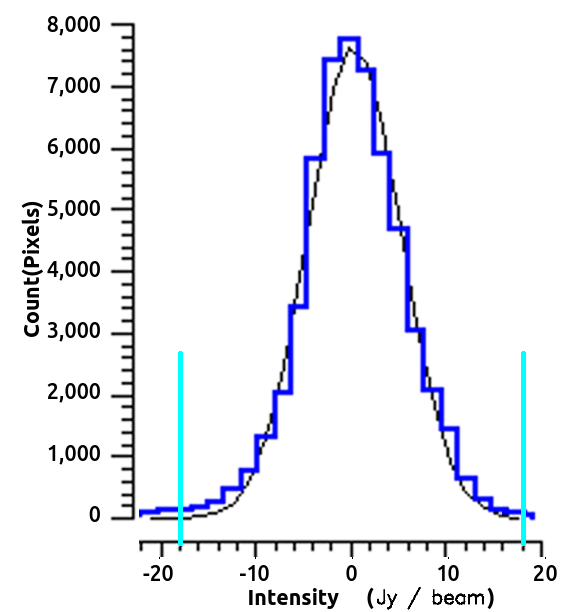}
 \caption{\small Left: Residual  (in greyscale) with  image contours (at [-1,1,2,4,8,16]$\times$  $3\sigma_{\mathrm{rms}}$ ($\sim$18 Jy beam$^{-1}$) overlaid, for the image in Fig.~\ref{Multiscale_etc.png}.  Note that although residual images are normally not convolved with a restoring beam, i.e. flux units Jy pixel$^{-1}$, they have been converted to the equivalent in Jy beam$^{-1}$ for the image beam, for easy comparison. 1b.  Right: Histogram of the distribution of pixel intensities in the residual (from the CASA viewer).  The black line shows a Gaussian distribution fitted to the noise, with cyan lines marking the $\pm3\sigma_{\mathrm{rms}}$ values.
 }
\label{fig-hist} 
\end{figure}

Clean masks should be neither too large nor too small -- ideally
roughly just enclosing the emission which is brighter than anything
not known to be real, which may increase in area in successive clean
cycles.  Don't clean too deeply such that CC are made from noise (Fig.~\ref{PBim.png}) nor too
shallowly. Fig.~\ref{Multiscale_etc.png} 1c shows an undercleaned image with
sidelobes still visible. 
Check the residual map, which should not show any traces of
the target above a few times the rms noise.
  For a perfectly Gaussian noise distribution
in an image (without PB correction), about 3
out of every 1000 pixels are likely to have values outside
$3\sigma_{\mathrm{rms}}$, which is 200 pixels in a $256\times256$
image (although the chances of being correlated over a beam is lower);
less than 1 pixel (on average) exceeds $4.5\sigma_{\mathrm{rms}}$. The image noise may
be non-Gaussian before calibration is perfect but there should be
approximately equal distributions of positive and negative artefacts.
Fig.~\ref{fig-hist} left shows the residual for the image in Fig.~\ref{Multiscale_etc.png} 1b, with the image contours overlaid. A histogram of the noise distribution is shown on the right. It can be seen that a few hundred pixels lie outside the cyan lines marking $\pm3\sigma_{\mathrm{rms}}$ values ($\sigma_{\mathrm{rms}}\sim$6 Jy beam$^{-1}$) but in this $256\times256$ image none exceed $|4\sigma_{\mathrm{rms}}|$.

\subsection{Missing spacings}
\label{sec-missing}
 Fig~\ref{Multiscale_etc.png} 2d shows a small effect of missing short
 spacings.
If you have negative regions surrounding brighter
parts of the image, (the `cereal bowl' effect) a probable cause is
resolved-out, large scale emission.  This occurs when a source has
structure which is smooth on scales similar to and larger than the
resolution of the shortest baselines present, bright enough to be detected
but not imaged properly, so the sparse sampling creates large-scale artefacts.
If so, a plot of amplitude v. $uv$
distance will reveal a steep increase in flux on short spacings e.g. Fig.~\ref{missing.png}.
In such cases, really good results are only likely to be
possible by adding in shorter spacings or total power data.
This is particularly true if the field contains emission on all
scales (possibly extending outside the total field of view), e.g. a
large planet or star forming region with bright, extended emission.

If you can't add in more data, you can either emphasise the extended
emission or minimise its impact, depending on what aspect of the data
you are interested in.  If only the shortest baselines (partially)
detect the extended emission whilst you are mainly interested in
compact emission, you can try just excluding the shortest
baselines. The extended emission may then be completely resolved out
or below the surface brightness threshold.  Conversely, if the
extended emission is mainly on scales which are poorly sampled but not
much larger, try weighting down the longest baselines using {\tt
  uvtaper} as in Sec.~\ref{sec-optimise} to increase the synthesised
beam.  In either case, removing the contribution of a large fraction
of baselines will increase the noise.

Images are also affected by missing intermediate spacings, which tends to affect ALMA configurations with 
 baselines $\gtrsim5$ M$\lambda$.
Fig.~\ref{Multiscale_etc.png} 2a shows that the default {\tt tclean}
deconvolver and weighting ({\tt robust=0.5}) can leave a criss-cross pattern sometimes
called `basket weaving', in this case due
to the missing spacings around 1250 and 1850 m. This is seen in
Fig.~\ref{AmpPhaseConfusion.png} 4a, restoring beam size
(204$\times$101) mas, and the fine details cannot be trusted.
Applying a {\tt uvtaper='0.1arcsec'} and  {\tt robust=1.} gives a beam (274$\times$205) mas and
Fig.~\ref{Multiscale_etc.png} 2b shows that although the artefacts can
still be seen faintly in the noise, the cleaned emission is no longer
distorted. 

Avoid using $uv$ tapers or other downweighting of specific baselines during calibration if possible.  The longest baselines are likely
to have greater phase errors, and if you can make a reasonable model
and self-calibrate without a $uv$ taper this will allow better
calibration for the most distant antennas; just apply the taper for
the final images if necessary, as explained in  Sec.~\ref{sec-weights}.
If,
however, the most distant antennas are so badly calibrated initially
that a good image cannot be made, exclude these to make a starting
model and then gradually increase the $uv$ range as self-calibration
progresses, as in Sec.~\ref{sec-selective}.

\subsection{Confusion}
\label{sec-confusion}
Confusion is the presence of bright sources offset from the pointing
centre  but within the PB. 
This can be due to continuum sources affecting the whole bandwidth or in some cases due to spectral lines, affecting just some channels. 
If these are not
included in clean masking, artefacts as in Fig.~\ref{AmpPhaseConfusion.png} 3a are seen.
This is commonest at longer wavelengths, so check ALMA data for amplitude calibration errors or bad data first,  but high dynamic-range ALMA observations of
crowded fields e.g. star clusters may be affected.  Confusion by point sources creates
a sidelobe pattern, and can be cured by imaging the culprits.
Confusion due to extended emission appears as a steeply rising excess
on short baselines (similar to Fig.~\ref{AmpPhaseConfusion.png} 4a) but
a compact source more distant than the largest angular scale sampled
will not show up like this.

All emission in the field used for a self-calibration model -- even if it is not your science target --  should be included in any self-calibration model. This
may even (occasionally, for ALMA) allow  self-calibration fields containing targets which
are themselves too faint.  In wide-field imaging at lower frequencies, they
may require direction-dependent calibration, Sec.~\ref{sec-wide}.

The primary planet can
cause confusion in ALMA moon observations and, often extending through
a range of PB sensitivities, adds diffuse noise  and is
harder to deal with. When planning observations,
remember to allow for the diameter of the planet, not just its nominal
central coordinate.

Spectral cubes may be affected by CO and other contamination from the
Milky Way or nearby galaxies, or in planetary atmospheres.  This may
be on scales too large to image but large enough to increase the noise
and create artefacts, so check catalogues etc. for observations in
galactic planes.

\subsection{Noisy images and non-noisy blurring}
\label{sec-imnoise}

If images are noisier than expected but without distinctive
artefacts, inspect the visibilities as a function of time
for obvious errors e.g. scans which are complete noise (but see Sec.~\ref{sec-selective} before flagging).
If all target data are noisier than expected check the conditions during observations, see Secs.~\ref{sec-seeing} and ~\ref{sec-prenoise}.

Another possibility is frequency-dependent errors, which can blur continuum images and thus degrade the accuracy of their models, also affecting calibration applied to lines. Check the line cube
or visibility spectrum in case the continuum channel selection
contains line contamination (even in spw designated continuum by the
PI). Exclude more channels if necessary.  If there are likely to be
many lines, check cube spectra using different size apertures and/or
inspect the visibility spectrum with averaging for sensitivity in
different $uv$ ranges. More lines may become significant as
calibration improves the S/N.

Are there atmospheric lines causing very noisy regions of the
visibility spectrum? These can be overplotted in {\tt plotms} or
identified using the ALMA atmospheric
model (see Appendix~\ref{ap-other} for link).
The $T_{\mathrm {sys}}$ correction usually flattens the
bandpass (since Cycle 1) and can lower the weight for affected channels, mitigating
the effect in imaging.
If there are important spectral lines, predicted to be detectable in a noisy spectral region, see Sec.~\ref{sec-transfer}.

If there is  no hope of retrieving any emission of interest, it may be better to
exclude affected channels from calibration and imaging or flag them.
Since sensitivity $\sigma_{\mathrm{rms}}
\propto \sqrt{\Delta \nu}$, where ${\Delta \nu}$ is the bandwidth,
excluding e.g. half of one spw, out of four of equal width, will only increase
the noise by a factor of $\sqrt{4./3.5}$, i.e. an increase $\lesssim7$\%.

Spectral problems may be due to previous calibration issues, Sec.~\ref{sec-previous}.

\subsection{Errors during observations or previous calibration}
\label{sec-previous}

\subsubsection{Delay and phase errors and the effects of antenna position uncertainty}
\label{sec-antpos}

In a spectral cube, bands of higher and lower noise across the whole image, which shift in
position from channel to channel, suggest errors in the delay (phase
as a function of frequency). 
These may also cause small position errors.
In making continuum images of different bandwidth (e.g. for the phase calibrator), you should see $\sigma_{\mathrm{rms}} \propto \sqrt{\delta \nu}$ but bandpass errors (phase or amplitude) will decrease  the noise by a smaller factor than expected.
Such errors may be due to poor bandpass calibration, which usually affects the amplitude corrections more than phase, see Sec.~\ref{sec-point} and/or to earlier delay errors.

The cause of delay errors is often antenna position uncertainty which causes direction-dependent errors in calculating the bulk delay and so phase corrections, (as a function of time and of frequency), derived for one calibrator, will be inaccurate when applied to another source (Sec.~\ref{sec-antposerr}).  If the phase-reference or target is bright enough to see a pattern when plotting phase as a function of frequency this will produce a slope in the corrected continuum visibilities for the affected antenna, i.e. a delay error.  The task
{\tt gencal} uses antenna position updates to derive delay corrections for the affected antennas. This is part of normal pipeline and manual calibration. If more accurate positions become available later (see Analysis Utility  {\tt checkAntennaPositionFiles}, Appendix~\ref{ap-casaguides},  requiring internet access), re-generating  and re-running the reduction pipeline  will use the best current corrections; however you are likely to need the help of an ARC to modify the calibration pipeline.  Alternatively, the antenna correction entries can be edited in manual scripts. In either case it is then necessary also to repeat all bandpass, phase referencing and related calibration. The worst effects of residual antenna position errors can be removed by self-calibration but the frequency-dependence of the phase error
will remain.

From Eq.~\ref{eq-antpos} a 2 mm antenna position error
will cause a change in the phase error $\phi_{\epsilon,
  {\mathrm{antpos}}}$ of e.g. 5\% across a sideband span of 12 GHz
around 230 GHz. For a large
$\phi_{\epsilon,\mathrm{antpos}}=50^{\circ}$ this is  $2.5^{\circ}
\sim 0.04$ rad. In principle this leads to an image position error
$\sim 0.04 \theta_{\mathrm B}$ (equivalent to the error in Gaussian fitting for S/N $\sim25$, Sec.~\ref{sec-gaussfit}).
In reality, on the one hand many
antennas may be affected, but on the other hand the baseline angles
and thus the phase slopes will usually not all be in the same sense. 
This will blur (rather than elongate) continuum sources and cause jitter in the positions of
spectral lines.

The line peak in the VY CMa data used here is 271 Jy, S/N 1330 and
$\theta_{\mathrm B}\sim 200$ mas so the noise-based position uncertainty $\sigma_{\mathrm{pos}} =
\theta_{\mathrm B}/$(S/N) = $0.00075\theta_{\mathrm B}
\sim$0.15 mas. The relative positions of maser components in successive
channels show a drift of $\sim$0.5 mas, or 0.0025 $\theta_{\mathrm  B}$, which is notionally
significant.  If the drifts are in different directions for different sets of spectral components that is probably genuine but if the position-velocity gradient is always the same it is probable that delay errors are the cause.

Delay errors in other data e.g. VLBI may be due to clock errors etc. (see VLBI links in Appendix~\ref{ap-schools}) or to antenna position errors. 

\subsubsection{Noise/artefacts due to phase-reference and other problems}
\label{sec-prenoise}

Although self-calibration can solve some phase-referencing errors, 
the
target noise may not reach the expected level or there may be
persistent artefacts.
If you examine the target corrected visibilities in {\tt plotms} after applying phase
reference solutions, changes due to astrophysical structure should vary smoothly with time, but jumps in amplitude or phase  on individual antennas or larger deviations on short baselines than on long suggest prior errors. 
 This may arise if the initial errors are so
great that a good starting image is hard to make.   If the target is bright
enough you can optimise making the first image with no or poor
phase-referencing, Sec.~\ref{sec-nophref}.

If the phase-calibrator is weak, failed  solutions can lead to excess flagging of good target data.
Check the pipeline weblog or qa2 report plots of calibration tables (see
Sec.~\ref{sec-goodbad}) and the calibrated data for calibration sources. It may be possible to improve the phase calibrator calibration,
see Sec.~\ref{sec-transfer} for some tactics, in particular to reduce the number of failed solutions or improve the amplitude solutions (which requires adequate previous phase solutions). This is especially important if the target is too weak for amplitude self-calibration.
Amplitude errors usually change slowly but one antenna may be mis-scaled, so once the calibrator bandpass and phase are corrected more than a scan can be averaged if necessary.

Excess image noise may be due to short-timescale noise, too rapid to correct by self-calibration.
Occasionally, time-dependent phase errors may originate from WVR measurements. 
The pipeline or QA2 plots and reports show whether applying PWV
corrections improved phase coherence or not.   If the PWV is low (e.g. $<$1 mm
at Band 3) corrections may do more harm than good. Conversely, for PWV $\gtrsim$2 mm, the conventionally-derived corrections may be inadequate due to
clouds, see link  in Appendix~\ref{ap-other} for the correction recipe {\tt Remcloud}. This may also be noted in the QA0
report. 

Tweaking manual QA2 scripts is relatively straightforward. If the data
were originally pipelined, target flagging and continuum channel selection can be modified following the CASA Guide to the ALMA Imaging Pipeline (see Appendix~\ref{ap-casaguides}).
 You will probably need expert help from an ARC to rectify problems with phase-referencing or earlier calibration, either by re-pipelining or by generating manual scripts.

\subsection{Amplitude errors due to $T_{\mathrm{sys}}$, antenna pointing or tracking errors}
\label{sec-point}

If images and calibrated visibilities  appear to show amplitude errors even though the phases and delays look good, this could be due to inconsistent $T_{\mathrm{sys}}$ or bandpass calibration for the target and phase calibrator or  to antenna(s) pointing incorrectly, either systematically or eratically.

If   $T_{\mathrm{sys}}$ tables look mis-scaled but with a reasonable relative S/N and no spikes or large jumps (more than few tens percent, except at edges) these can be rescaled  (see Appendix~\ref{ap-casaguides} link for Analysis Utilities) and then the remaining calibration repeated (as in the pipeline or QA2, but see Sec.~\ref{sec-prenoise} above). Otherwise, to tackle this just by self-calibration, you can try excluding the affected antenna(s) to make a starting model  as in
Sec.~\ref{sec-selective}, and if possible do amplitude self-calibration including all antennas. You can average over the whole time affected if necessary for S/N.
Check for spikes or other  $T_{\mathrm{sys}}$ anomalies. If maverick $T_{\mathrm{sys}}$ channel(s) are present for
only one of the target or the bandpass calibrator you may be able to remove the discrepancy by editing
the $T_{\mathrm{sys}}$ table (see Appendix~\ref{ap-casaguides} link for Analysis Utilities).

Check if the bandpass calibrator corrected bandpass looks noisier than typical amplitude excursions ($<$5\% in lower ALMA bands,  \citetalias{ALMA-TH}).
Bandpass calibration might be improved by flagging bad data or
changing the solution interval in prior phase calibration of the bandpass
calibrator or the spectral averaging interval in {\tt bandpass}, or
perhaps using a different source with better S/N. As long as
$T_{\mathrm{sys}}$ has been applied to take care of short-term
spectral amplitude calibration, the instrumental causes of bandpass errors are
usually stable for hours or even days  so it may be possible to copy
a better bandpass table from another EB taken using the same frequency,
spectral and array configuration.
Occasionally, very noisy spectral regions are present (Sec.~\ref{sec-imnoise}) and can be excluded from the spw selection during time-dependent calibration of the bandpass calibrator, phase calibrator etc. even if you want to include them in bandpass calibration.

An uncorrected antenna pointing offset will skew the primary beam response. \citetalias{ALMA-TH} describes ALMA's pointing strategy and other arrays e.g. VLA at high frequencies use `pointing up' but in some arrays, especially VLBI, antennas take different times to slew and some arrive late. If just the first or last few visibilities of every scan are bad see {\tt flagdata mode='quack'}. 

Normal self-calibration for compact targets may reduce  small pointing/tracking errors but if all sources are affected, you may need to redo earlier calibration from the derivation of the flux scale onwards: exclude the affected antenna(s) in deriving the phase calibrator flux and then set this ({\tt setjy}) and make a good phase-reference amplitude correction table for all antennas.  Applying this plus the phase and other corrections to the target may be good enough (although the image will inevitably be a bit noisier due to the lower sensitivity of badly-pointed antennas to off-centre targets).
However, for Alt-Az mounts such as ALMA uses, the beam will rotate relative to the sky over the course of long observations. If a target field has components  offset from the nominal pointing centre the pointing errors cause them to fall in different regions of primary beam senstivity during the observations. To self-calibrate, make an initial image model excluding the affected antennas, as above for  $T_{\mathrm{sys}}$ errors, and see Sec.~\ref{sec-selective}. The amplitude solution interval needs to be less than the interval over which the apparent amplitudes change significantly.

\subsection{Target Variability}
\label{sec-variable}
A very severe sidelobe pattern radiating outwards from bright sources can be caused by variability
during the observation.
In such a case, phase solutions should be applied normally but do not perform amplitude self-calibration on the whole data set -- if you do, it usually makes the image worse, often how one discovers unexpected variability.
Usually variable sources are point-like, and if you just want a light
curve it might be best to obtain this from the visibilities. If
the variable source is not at the centre of the field,  shift the phase centre to its position using the transform tab in {\tt plotms} or {\tt mstransform};
for large shifts use the task {\tt phaseshift}, in CASA 6.3 and
later. Alternatively, you can make images of each time segment long
enough to produce a reasonable image but short enough to minimise
variability and measure the variable source in each one.

In order to make a combined image of any
non-variable emission in the field,  you need to subtract the variable component. This is possible if you can make good enough image models for time segments over which the variablity has little effect,
masking just the
variable source (unless there are other sources in the field bright enough for their sidelobes to be a problem in which case subtract first or at the same time). Divide the data into the segments and, for each segment, clean to the noise or non-variable background but not too
deeply and check that the model is saved (Sec.~\ref{sec-ft}).  {\tt uvsub} can then be used  to
subtract the model of the variable source. Then recombine the subtracted data sets and image.
You can alternatively subtract a simulated model from each segment.
If there are also antenna-dependent amplitude problems you could try
self-calibration for each time segment before subtracting the variable
source but this runs the risk that the amplitude scale may diverge
between segments.

\bibliographystyle{aa} 
\bibliography{cse2021}

\begin{appendix}
  \section{Abbreviations and symbols}
  \label{acronyms}
  \begin{longtable}{ll}
  \small 
  $A12$            &`true' amplitude on baseline between antennas 1, 2 (etc.)\\
 $A_{\mathrm{eff}}$ & Effective antenna diameter\\
$A_{\epsilon, \mathrm {baseline}}$  & amplitude error on a single baseline\\
$A_{\epsilon1}$   & amplitude error on antenna 1   (etc.)\\
ACA              & Atacama Compact Array\\
ALMA             & Atacama Large Millimetre and sub-millimetre Array\\
AVC              & Automatic Voltage Correction (measure of source contribution to $T_{\mathrm{sys}}$)\\
$B$                & baseline length\\
$B_{1/2}$           & baseline length where visibility amplitude is at half maximum power due to errors\\
$B_{\mathrm{max}}$           & maximum baseline length\\
$B_{\mathrm{min}}$           & minimum baseline length\\
$C$                & multiplier for noise-based position errors for sparse $uv$ coverage\\
CC  & Clean Components\\
$D_{\mathrm {antenna} A}$  & dynamic range limitation due to single antenna amp error\\
$D_{\mathrm {antenna} \phi}$  & dynamic range limitation due to single antenna phase error\\
$D_{\mathrm {baseline} A}$  & dynamic range limitation due to single baseline amp error\\
$D_{\mathrm {baseline} \phi}$  & dynamic range limitation due to single baseline phase error\\
$D_{\operatorname{all-antenna} A}$  & dynamic range limitation due to same all-antenna amp error\\
$D_{\operatorname{all-antenna} \phi}$  & dynamic range limitation due to same all-antenna phase error\\
$D_{\operatorname{all-baseline} A}$  & dynamic range limitation due to random all-baseline amp errors\\
$D_{\operatorname{all-baseline} \phi}$  & dynamic range limitation due to random all-baseline phase errors\\
DGC        & Diff Gain Cal\\
$\mathbf{E}$ &  `Electric' Antenna voltage  pattern including PB response (as used in Measurement Equation)\\
  FDM & Frequency Domain Multiplexing\\
  FWHM               & Full Width Half Maximum\\
$\boldsymbol{I}$& Sky brightness distribution (as used in Measurement Equation)\\
$\mathbf{J}$ & Collective errors affecting recorded visibility (as used in Measurement Equation)\\
$\mathbf{G}$ & Gain errors (non-direction-dependent, as used in Measurement Equation)\\
  $g$              & factor to convert from Jy to Kelvin (inverse of ALMA SEFD)\\
  $K$                & Kolmogorov coefficient\\
$k$                & coefficient for Jy/Kelvin, efficiency etc. in conversion from $T_{\mathrm{sys}}$ to image noise \\
$lm$              & Image plane coordinates in formalism such as Measurement Equation\\
$M$               & number of intervals of independent tropospheric effects\\
$\mathbf{M}$ & Baseline-dependent errors (as used in Measurement Equation)\\
MFS               & Multi Frequency Synthesis (often, the special case of fitting a spectral index in imaging)\\
  MS                & Measurement Set\\
$N$                & number of antennas\\
$N_{\mathrm{indep}}$ & number of independent antennas\\
$N_{\mathrm{pol}}$    & number of polarisations\\
$P$                & Peak flux density\\
  $\mathbf{P}$ & Parallactic angle rotation (as used in Measurement Equation)\\
  PB & Primary Beam \\
  PWV                & Precipitable Water Vapour\\
  QA0               & ALMA Quality Assurance immediately after observations\\
  QA2               & ALMA Quality Assurance for calibration and imaging\\
  QSO             & Quasi Stellar Object\\
  SEFD           & System Equivalent Flux Density, system temperature conversion factor\footnote{ALMA usage, see Appendix~\ref{tsystoJy}} K Jy$^{-1}$ \\
$S_0$             & flux density at frequency 0 (etc.)\\
  $S_{\mathrm{app}}$   & apparent target flux density\\
  $S_{\mathrm{tar}}$   & target flux density\\
  S/N                & Signal to Noise ratio (usually, $P/\sigma_{\mathrm{rms}}$ of image)\\
   S/N$_{\mathrm{sc}}$ & Signal to Noise ratio required for self-calibration\\
spw & spectral window\\
TDM & Time Domain Multiplexing\\
TEC                & Total Electron Content of ionosphere\\
$\mathbf{T}$ & Gain errors (direction-dependent as used in Measurement Equation)\\
$T_{\mathrm{Rx}}$     & receiver temperature\\
$T_{\mathrm{sky}}$   & sky temperature \\
$T_{\mathrm{sys}}$    & system temperature\\
$uv$               & visibility plane coordinates\\
$V$                & complex visibility\\
$V_0$              & amplitude\\
$\boldsymbol{V}$ &Complex visibility  (matrix form as used in Measurement Equation)\\
$V_{\mathrm{LSR}}$    & velocity with respect to the Local Standard of Rest (aka LSRK)\\
VLBI               & Very Long Baseline Interferometry\\
WVR       & Water Vapour Radiometry\\
$\alpha$           & spectral index\\
$\alpha_{\mathrm K}$           & Kolmogorov exponent\\
$\Delta \mathrm{TEC}$       & difference in TEC above each antenna in a baseline\\
$\Delta \nu$                & useful bandwidth (excluding gaps/deselected ranges)\\
$\Delta t$                  & time on source (excluding gaps)\\
$\Delta_{\mathrm{pcal-tar}}$ & phase calibrator -- target angular separation \\
$\delta \nu$               & useful frequency interval per solution\\
$\delta\nu_{\mathrm{spw}}$  & line-free continuum bandwidth per spw\\
$\delta t$                  & solution interval\\
$\delta t_0$                & minimum solution interval for required S/N\\
$\delta\phi$       & magnitude of rapid phase fluctuation\\
  $\theta_{\mathrm B}$ & synthesised beam (FWHM) \\
$\lambda$          & wavelength\\
$\nu$             & frequency\\
$\nu_0$           & frequency of measurement 0 (etc.)\\  
$\phi$             & phase\\
$\phi12$            &`true' phase on baseline between antennas 1, 2 (etc.)\\
$\phi_{\epsilon 1/2}$ & phase error for $B_{1/2}$\\
$\phi_{\epsilon, \mathrm{short}}$   & phase error due to rapid fluctuations\\
$\phi_{\epsilon,  \mathrm{angsep}}$ & target phase error due to target-phase calibrator angular separation\\
$\phi_{\epsilon,  \mathrm{transfer}}$ & total error in applying phase calibrator solutions to target\\
$\phi_{\epsilon, \mathrm{baseline}}$ & phase error on a single baseline\\
$\phi_{\epsilon, \mathrm{target,ant}}$ &total error in phases for each antenna\\
$\phi_{\epsilon, \mathrm{target}}$ & phase error for whole target data set\\
$\phi_{\epsilon, \mathrm{time}}$ & target phase error due to time separation between target and phase calibrator scans\\
$\phi_{\epsilon,\mathrm{antpos}}$& target  phase error due to antenna position error\\
$\phi_{\epsilon,\mathrm{ion}}$  & phase error due to ionosphere\\
$\phi_{\epsilon,\mathrm{trop}}$  & phase error due to tropospheric refraction\\
$\phi_{\epsilon1}$   & phase error on antenna 1   (etc.)\\
$\phi_{\mathrm{rms \mu}}$       & tropospheric effective path length fluctuation \\
$\phi_{\mathrm{rms}}$ & phase noise\\
$d\phi_{\mathrm{atm/min}}$    & phase change per minute\\
$\sigma_{\mathrm {pos Target}}$  & target astrometric uncertainty due to phase calibrator solution transfer errors\\
$\sigma_{\mathrm {pos fit}}$   & relative position uncertainty of fitted 2D Gaussian due to noise\\
 $\sigma_{\phi \mathrm{rate, antpos}}$ & phase error rate due to antenna position error\\
$\sigma_{\mathrm{antpos}}$     & antenna position uncertainty\\
$\sigma_{\mathrm{rms}}$          & image noise\\
$\sigma_{\mathrm{solint, ant}}$      & rms noise per antenna in solution interval\\
$\sigma_{\mathrm{solint, baseline}}$   & rms noise per baseline in solution interval\\
$\tau_{\mathrm{scan}}$ & Length of a scan (interval between source changes)\\
\end{longtable}
{Many of the concepts behind these terms are explained more in the Technical
  Handbook  \citetalias{ALMA-TH}.}

\section{Online resources}
\label{resources}
\setlength\itemsep{1em}
\subsection{Schools and workshops}
\label{ap-schools}
{\bf Schools focusing on Self-Calibration}
These provide an
introduction to the CASA tasks and links to the VY CMa data
and scripts
used for examples in this manual:
\vspace*{-0.1cm} \begin{itemize}
\item[]
Self-calibration and advanced imaging, Bologna 2017\\
\url{http://www.alma.inaf.it/index.php/Self-calibration\_and\_advanced\_imaging\_workshop}
\item[]
I-TRAIN 6: Improving image fidelity through self-calibration\\
\url{https://almascience.eso.org/euarcdata/itrain06/}. 
\end{itemize}

{\bf{NRAO  Synthesis Imaging summer schools}}
\vspace*{-0.1cm} \begin{itemize}  
\item[]Synthesis~Imaging~in~Radio~Astronomy~II, the 1998 proceedings (\citetalias{SI99}) are online at\\
\url{https://www.aspbooks.org/a/volumes/table\_of\_contents/?book\_id=292}\\
\url{http://www.phys.unm.edu/~gbtaylor/astr423/s98book.pdf}
\item[]
Material from the 2017 school is available at\\
\url {https://science.nrao.edu/science/meetings/2018/16th-synthesis-imaging-workshop}\\
including Error Recognition\\
\url{https://science.nrao.edu/science/meetings/2018/16th-synthesis-imaging-workshop/talks/Taylor\_Error\_Recognition.pdf}
and see also \citet{Brogan18}.
\end{itemize}
{\bf{ERIS and DARA}}
ERIS (European
Radio Interferometry Schools) and DARA (Development in Africa with Radio Astronomy) have much online
material including on self-calibration, including data sets and step by step guides aimed at beginners. 
 DARA      covers e-MERLIN and (simple) VLBI data reduction and self-calibration as well as preparation for the AVN; ERIS also covers some ALMA examples.
 \vspace*{-0.2cm}
 \begin{itemize}
\item[]
  ERIS 2019\\
  \url{https://www.chalmers.se/en/researchinfrastructure/oso/events/ERIS2019}\\
  including T7 Self-calibration (e-MERLIN continuum, suitable for CASA 5) at:\\
  \url{https://www.chalmers.se/en/researchinfrastructure/oso/events/ERIS2019/Pages/Software-packages-and-datasets.aspx}\\
 From 2022 onwards, ERIS will be run by ORP (\url{https://www.orp-h2020.eu}).
  
\end{itemize}

\begin{itemize}
\item[]
  DARA 2021\\
  Workshops for tutorials,  as of 2021 in CASA 6, at:\\
      \url{http://www.jb.man.ac.uk/DARA/unit4/unit4.html}\\
      EVN continuum covers all stages including self-calibration for a basic VLBI data set. Spectral Line covers processing target data after applying phase calibrator etc. corrections, including self-calibration. 3C277.1 is the CASA 6 equivalent of the ERIS 2019 data. 
\end{itemize}  

\noindent
{\bf{Other schools and resources for other arrays}}\\
\noindent
    {\bf VLBI}
    \vspace*{-0.1cm} \begin{itemize}
  \item[]
VLBI calibration is described in more detail at the 2020 CASA-VLBI  Workshop\\
 \url{https://www.jive.eu/casa-vlbi2020/}
\item[]
 Other links are provided from\\
 \url{https://casaguides.nrao.edu/index.php?title=VLBI_Tutorials}
 \end{itemize}

   \vspace*{0.1cm}
  \noindent
{\bf Wide-field/low frequency} guidance can be found in the VLA P-band CASA guides (although not covering self-calibration); for LOFAR see
\vspace*{-0.1cm} \begin{itemize}
  \item[] LOFAR data reduction\\
\url{https://support.astron.nl/LOFARImagingCookbook/} Ch. 20.
\end{itemize}
\vspace*{0.1cm}
\noindent
    {\bf NOEMA} data reduction including self-calibration (in GILDAS)
     \begin{itemize}
  \item[] 2018 IRAM Summer School\\
\url{https://www.iram-institute.org/EN/content-page-399-7-67-367-399-0.html} (and see IRAM pages for future events).
\end{itemize}

     \subsection{CASA Guides}
     \label{ap-casaguides}
It is worth browsing \url{https://casaguides.nrao.edu/} to check for recent versions and additions; these are just some current examples:
\vspace*{-0.1cm} \begin{itemize}
\item[]
A basic guide to parameter settings for ALMA is provided by the CASA
introduction to self-calibration\\
\url{https://casaguides.nrao.edu/index.php?title=First\_Look\_at\_Self\_Calibration}.
\item[]
VLA self-calibration guide\\
\url{https://casaguides.nrao.edu/index.php?title=VLA_Self-calibration_Tutorial-CASA6.2.0}
\item[]
For Solar data reduction, see the CASA Guide for sunspots \\
\url{https://casaguides.nrao.edu/index.php?title=Sunspot\_Band6} and for self-calibration see  \citet{Nindos2018}.
\item[] Data weighting\\
  \url{https://casaguides.nrao.edu/index.php?title=DataWeightsAndCombination}
\item[] Reimaging pipelined ALMA data by modifying pipeline commands\\ 
  \url{https://casaguides.nrao.edu/index.php?title=ALMA\_Imaging\_Pipeline\_Reprocessing}).
\item[] Flagging (at cm wavelengths)\\
  \url{https://casaguides.nrao.edu/index.php?title=VLA\_CASA\_Flagging-CASA6.2.0}.
\item[] Analysis Utilities\\
  \url{https://casaguides.nrao.edu/index.php?title=Analysis\_Utilities}
\item[] Combining arrays\\
  \url{https://casaguides.nrao.edu/index.php?title=M100_Band3}
\end{itemize}

\subsection{CASA Documentation}
\label{ap-casadoc}
Help (task) provides inline help but more details are available from CASA documentation \url{https://casadocs.readthedocs.io/en/stable/index.html}. The most relevant here are 
\vspace*{-0.1cm} \begin{itemize}
\item[] {\tt tclean}\\
\url{https://casadocs.readthedocs.io/en/stable/api/tt/casatasks.calibration.tclean.html}
\item[] {\tt gaincal}\\
\url{https://casadocs.readthedocs.io/en/stable/api/tt/casatasks.calibration.gaincal.html}
\item[] {\tt applycal}\\
\url{https://casadocs.readthedocs.io/en/stable/api/tt/casatasks.calibration.applycal.html}
\end{itemize}
(also see below Appendices~\ref{ap-gaincal} and~\ref{ap-applycal}

\subsection{Sensitivity calculators}
\label{scs}
\begin{itemize}
\item[]
  ALMA
  \url{https://almascience.eso.org/proposing/sensitivity-calculator} (also built in to the Observing Tool)
\item[]  
  VLBI (including EVN, VLBA and global VLBI)
  \url{https://www.evlbi.org/capabilities}
\item[] GMVA (3mm VLBI)
  \url{https://www3.mpifr-bonn.mpg.de/div/vlbi/globalmm/sensi.html}
\item[]
  e-MERLIN
  \url{http://www.e-merlin.ac.uk/calc.html}
\item[]
  VLA
\url{https://obs.vla.nrao.edu/ect/}
\end{itemize}

\subsection{Other resources}
\label{ap-other}
\vspace*{-0.1cm} \begin{itemize}
\item[]{\bf NRAO Template script for ALMA imaging and self-calibration} {\url{ https://github.com/aakepley/ALMAImagingScript/blob/master/scriptForImaging\_template.py}}

\item[]{\bf NRAO  Science ready data products} provided on request for selected ALMA and VLA observations, including restored ALMA pipeline calibrated MS  {\url{https://science.nrao.edu/srdp}}

\item[]
{\bf Atmospheric model}
\url{http://www.apex-telescope.org/sites/chajnantor/atmosphere/transpwv/index_ns.php}
This plots transmission as a function of frequency for selected PWV using the ATM libraries, \citet{Pardo2019} for the APEX site (which is very similar to ALMA).
At present this is only available specifically for  ALMA   inside CASA, via Analysis Utilities as \url{https://safe.nrao.edu/wiki/bin/view/ALMA/PlotAtmosphere} (see Appendix~\ref{ap-casaguides} and \url{https://help.almascience.org/kb/articles/how-can-i-plot-the-atmospheric-transmission-opacity-or-sky-temperature-for-alma}). It may be restored to the ALMA Science Portal in time.  The transmission for a specific MS can be overplotted in {\tt plotms}.

\item[] {\bf Clouds} Scaling the WVR-derived correction in the presence of clouds: {\tt Remcloud}
\url{https://help.almascience.org/kb/articles/what-is-remcloud-and-how-could-it-reduce-phase-rms}

\item[] {\bf Spectral contribution to $T_{\mathrm{sys}}$}
Checking for flux scale errors due to large scale spectral flux contribution to $T_{\mathrm{sys}}$\\
\url{https://help.almascience.org/kb/articles/what-are-the-amplitude-calibration-issues-caused-by-alma-s-normalization-strategy}.
As of 2021,  ALMA procedures were being updated and this correction may already have been included so check whether it is needed for observations processed more recently.

\end{itemize}

\section{Guidance for self-calibration parameters in {\tt gaincal} and {\tt applycal}}
\subsection{\tt gaincal}
\label{ap-gaincal}

More details to guide the use of  {\tt gaincal}, see Sec.~\ref{sec-gaincal} and the full CASA documentation, Appendix~\ref{ap-casadoc}.
 The comments here are applicable for phase and for amplitude self-calibration unless otherwise stated.

\begin{description}
\item[{\tt vis, field}] Input MS and target
\item[{\tt caltable}] Output calibration table. Normally, delete any previous table of the same name and close any plotting window displaying it. 
\item[{\tt spw}] This selection should contain only channels where all the emission should be described by the model.  For example, you can make a continuum image model from a channel selection for one spw and select the continuum channels of another spw near in frequency (e.g. in the same EB) but do not use a continuum model to derive solutions for line channels with different structure or flux density (although you can apply the continuum solutions to these).
\item[{\tt selectdata}] has various sub-parameters occasionally needed if part of the data are used to build up a model before calibrating all data,  e.g. {\tt uvrange} (see Sec.~\ref{sec-uvrange}) or {\tt scan} (see Secs.~\ref{sec-selective}, \ref{sec-mosaic})
\item[{\tt solint,}]{\tt combine} See Sec.~\ref{sec-SN} for setting the solution interval. Note that {\tt solint='inf'} means average within each scan by default. Use {\tt combine}  to average over more than one scan or spw (etc.) in low S/N cases, see Sec.~\ref{sec-lowsn}
\item[{\tt refant,}]{\tt refantmode} See Sec.~\ref{sec-refant}
\item[{\tt minblperant}] The default is to require solutions for at least four baselines to each antenna in order to calculate the per-antenna solutions. As explained in Secs.~\ref{sec-closure} and~\ref{sec-qpr},  4 degrees of freedom are needed for amplitude calibration and 3 for phase. In practice, more baselines will give better-constrained solutions for complex models, but for a point source, with good S/N, even a single baseline can be used -- rarely necessary for ALMA except possibly for badly-shadowed ACA data or some VLBI observations.

\item[{\tt minsnr}] The minimum S/N per solution is 3 by default. However,  there is not a literal equivalence between
visibility plane calculations, which are affected by the goodness of
the model as well as the data, and the calculations used here (Sec.~\ref{sec-SN}) where the solint needed for S/N of 3 per antenna is derived from image-plane measurements.  
Whilst lower minimum S/N  will lead to less well-constrained solutions, values as low as {\tt minsnr=1.5} can be acceptable for data which have already had phase reference solutions applied, see Sec.~\ref{sec-goodbad}. 
Greater care is needed for amplitude self-calibration, since bad solutions can lead to severe image artefacts.

\item[{\tt solnorm}] This is False by default so the corrected data will be rescaled directly by the amplitude solutions, required  in earlier calibration when calibrator solutions are applied to scale from correlator counts to Jy. Normalisation of the solutions, {\tt solnorm=True}, means that
the solutions are squared and then are divided through by the mean for that antenna, spw,
polarisation, so that the product of all solutions is unity.  Thus,
fluctuations from solint to solint will be corrected but the overall
flux scale does not change. This is usually what is wanted in self-calibration but see Sec.~\ref{sec-ampscal}.
\item[{\tt gaintype}$^*$] The default, {\tt 'G'}, solves separately for each
  polarisation, which is usually best for phase calibration. In low S/N cases, {\tt 'T'} can be used to average the polarisations or even {\tt GSPLINE}  (Sec.~\ref{sec-lowsn}). For ALMA, the errors affecting X and Y are normally similar and so if separate X and Y solutions derived from the phase calibrator have been applied, it is usual to  average X and Y in target self-calibration, {\tt gaintype='T'}.
  Moreover,  the effects of source polarisation and
parallactic angle rotation can distort the target solutions if X and Y
are separated (even for dual-polarisation data).  It is especially important to use {\tt 'T'} for full polarisation ALMA data,
see Sec.~\ref{sec-polarisation}.  If you think X and Y may have different amplitude errors, in ALMA data this is often due to problems with the $T_{\mathrm{sys}}$ correction, which may be tackled using Analysis Utilities (see Appendix~\ref{ap-casaguides}).
For circular
(L and R) feeds, parallactic angle rotation is not normally an issue (but see Sec.~\ref{sec-astrometry})
and if, for example, one hand of polarisation suffers instrumental
mis-scaling, they can be separated in amplitude self-calibration ({\tt 'G'}), if the source is not circularly polarised.
\item[{\tt smodel}] By default this is ignored and the model column of the MS is used. If you have not inserted a specific model (in {\tt tclean} or {\tt ft}) the default of a 1 Jy source at the phase centre is used.
\item[{\tt calmode}] The default is {\tt 'ap'} which for modern arrays you probably do not want (Sec.~\ref{sec-ampscal}) so always specify {\tt 'p'} or {\tt 'a'} for phase or amplitude.

\item[{\tt solmode}] The default is least squares fitting to derive per-antenna corrections from per-baseline comparisons (Sec.~\ref{sec-qpr}). Alternative methods including least difference usually lead to worse, wild solutions for ALMA data but can be useful e.g. for VLBI where the S/N is high but the data errors are large.

\item[{\tt gaintable,}] {\tt interp, spwmap} List any previous calibration tables to apply.  {\tt gaincal} uses the DATA column, so even if you have run {\tt applycal} previously for the same MS you still need to list the tables you want to apply here, e.g. the phase-only calibration when you are now solving for amplitude. (This is not the case if you have split out the corrected data as in Sec.~\ref{sec-alt}.) 
When applying phase-reference solutions to the target, or applying
target solutions to itself, the default linear interpolation ({\tt
  interp='linear'}), as shown in Fig.~\ref{VYCont_phref.png} bottom
left, is usually the most accurate.  However, if you are applying
solutions to a source not bracketed by calibration table entries (see
Sec.~\ref{sec-transfer}) use ({\tt interp='nearest'}), which can
extrapolate. This is required if applying corrections from one EB to another, for example. 
Use the {\tt 'PD'} suffix for transfer across large frequency differences. By default, {\tt spwmap} will apply the calibration derived for each spw to itself, but you need to specify this if transferring between (possibly averaged) spw, see Secs.~\ref{sec-transfer} and~\ref{sec-cross-cal}.
\end{description}
$^*$
\begin{minipage}[t]{16.7cm}{\small The parameters {\tt 'T'} and {\tt 'G'} as used in CASA denote solving for separate and averaged polarisations, but contain no assumptions about the direction-dependence of gains within a target pointing; whilst direction dependent solutions are not directly implemented in CASA, either parameter can be used as required in techniques for anisoplanatic fields such as peeling, Sec.~\ref{sec-wide}. On the other hand the customary definitions  in the Measurement Equation (Eq.~\ref{eq-me})  associate only $\mathbf{T}$ with direction-dependent effects (and use additional terms to cover polarisation-dependent errors), whilst   $\mathbf{G}$ is used for instrumental, non-direction-dependent effects. }
\end{minipage}
\vspace{0.3cm}

If you are working on data where the model was originally inserted into the MS using an early version of CASA, gaincal may fail with errors such as {\tt RecordInterface: .... is unknown}. Try removing the model with {\tt delmod}, {scr=True} and replacing the mode with {\tt ft} or {\tt setjy} as needed and re-run gaincal.

\subsection{\tt applycal}
\label{ap-applycal}
More details to guide the use of  {\tt applycal}, also see the full CASA documentation, Appendix~\ref{ap-casadoc}.

The parameters most relevant for ALMA self-calibration are:
\begin{description}
\item[{\tt vis,}] {\tt field} The correction tables are divided into the DATA column of the specified MS using the task which writes out the CORRECTED column (replacing any previous corrected visibilities) for the specified field.
\item[{\tt spw}] The spw to apply calibration to, which does not have to be the same as that used in {\tt gaincal}. For example, narrow spw may not be included in a continuum-only selection.
 Corrections are independent of target structure, determined only by atmospheric and other error causes.  See Sec.~\ref{sec-transfer} and {\tt spwmap} below.
\item[{\tt selectdata}] has various sub-parameters occasionally needed if part of the data are used to build up a model before calibrating all data. By default the solutions will be applied to all scans etc. but you can restrict this (see Secs.~\ref{sec-uvrange}, \ref{sec-selective}, \ref{sec-mosaic}).
\item[{\tt gaintable,}] {\tt interp} Following use of  {\tt gaincal}, list the calibration table generated as {\tt caltable}, plus any given as {\tt gaintable} in   {\tt gaincal}. See Appendix~\ref{ap-gaincal} for notes on  {\tt interp}.
\item[{\tt spwmap}] See Appendix~\ref{ap-gaincal} and the example in Sec.~\ref{sec-transfer}. You must have one spwmap list for each table to be applied; if in some cases this is simply applying corrections for each spw to itself this can be blank; when any spw is specified there must be a value for each spw in the MS (appearing in {\tt listobs} output) even if some are not used for self-calibration.
\item[{\tt calwt}]  The default of {\tt True} will downweight data with noisier solutions (Sec.~\ref{sec-weights}) but in early rounds of self-calibration, especially for long baselines or low S/N (see Sec.~\ref{sec-lowsn}) it is safer to use {\tt
  calwt=False}.
\item[{\tt applymode}] The default, {\tt 'calflag'}, will flag data for which solutions have failed.  However, solutions may fail due to a poor model or unsuitable parameters in {\tt gaincal}. It is safer to use {\tt 'calonly'}
which does not flag any data, unless you know all the
data in a failed solint for a given antenna are really bad, see Sec~\ref{sec-goodbad}.
\end{description}

\section{Example script fragments}
\label{script-examples}
These are some snippets which can be edited as required and included in CASA scripts.

\subsection{Aligning multiple data sets and other position changes}
\label{fixvis}
As explained in Sec.~\ref{sec-configs}, if the position of the target, relative to the pointing centre, is consistent between multiple observations  {\tt concat} can be used to combine the data if the pointing centre changes represent the target proper motion accurately.  
However, if the proper motions are not accurate or for any reason 
you need to change the position of a source in visibility data, you can use the task {\tt fixvis} or, since CASA 6.3, {\tt phaseshift},  to shift the phase centre to the actual target peak position, followed by {\tt fixplanets}  (for any source, not just planets) to transform the phase centre to the required, common position. Most ALMA data are in the ICRS coordinate frame, but {\tt fixplanets} only recognises `J2000' (actually FK5, epoch 2000). The simplest procedure is to label the coordinates J2000 fictitiously if necessary and then change back. This example provides a template. See Sec.~\ref{sec-align} for how and why to set {\tt freqtol}.

{\begin{verbatim}
target  = 'target'
msin1   = 'target1.ms'                      # original ms 1 
msout1  = 'target1-shift.ms'
pos1    = 'J2000 01h23m45.67s -10d09m08.7s' # target position measured from ms 1 image
msin2   = 'target2.ms'                      # original ms 2 (add more as needed)
msout2  = 'target2-shift.ms'
pos2    = 'J2000 01h23m45.67s -10d09m08.0s' # target position measured from ms 2 image
alignedpos  = 'J2000 01h23m45.67s -10d09m08.7s' # maybe one target position or an average...
cvis    = 'target-concat.ms'                # combined ms
\end{verbatim}}
\noindent \# shift phase centre to actual target position
  {\begin{verbatim}
os.system('rm -rf '+msout1)
fixvis(vis=msin1
       outputvis=msout1
       datacolumn='corrected',              # or as appropriate 
       reuse=False,
       field=target,
       phasecenter=pos1)
\end{verbatim}}
\noindent \#{\textbf{\textit {Or}}}, as of CASA 6.3, use 
{\begin{verbatim}
os.system('rm -rf '+msout1)
phaseshift(vis=msin1
       outputvis=msout1
       datacolumn='corrected',              # or as appropriate 
       field=target,
       phasecenter=pos1)
\end{verbatim}}
\noindent \# transform coordinates to same centre for all data sets
  {\begin{verbatim}
fixplanets(vis=msout1
       	   field=target,
           direction=alignedpos)
\end{verbatim}}
\noindent \# repeat fixvis/phaseshift, fixplanets for further input ms 2 etc.\\
  \# Concatenate all shifted, centred data sets
  {\begin{verbatim}
os.system('rm -rf '+cvis)
concat(vis=[msout1, msout2],               # or more as needed  
         concatvis=cvis
         freqtol='3MHz',                   # or as needed, check other parameters
         copypointing=False,
         visweightscale=[1,1.,1.])     
\end{verbatim}}
  \noindent
Finally, change labelling back to ICRS,
see\\\url{https://help.almascience.org/file.php/165YGANGGQKYW16499778A9E0/relabelmstoicrs.py}\\
This uses python command {\tt xrange}; in python 3 / CASA 6, replace this with {\tt range}.
To change the actual coordinate frame (not just the labelling), use the {\tt refcode} parameter in {\tt fixvis} and enter the converted coordinates; in CASA 6.3 and later use task {\tt phaseshift}, see the Help. Analysis Utilities 
(Appendix~\ref{ap-casaguides}), or a package such as astropy, can be used to convert FK5 (J2000) to ICRS or etc.

\subsection{Deriving flux scale from $T_{\mathrm{sys}}$ and adjustments to visibilities in {\tt gencal}}
\label{tsystoJy}

The system temperature of ALMA (or other) antennas is often used to scale the correlator counts, which  provides an alternative method to set the flux scale.  The  System Equivalent Flux Density (SEFD) is needed.  
For ALMA bands 3 -- 7, the SEFD is $\sim$35 K Jy$^{-1}$  for 12-m antennas, see the latest Technical Handbook  \citetalias{ALMA-TH}, (sec. 9.2.1 in 2021)s\footnote{for other arrays, SEFD$_{\mathrm{other}}$ may be used to describe the total in Jy (\citetalias{SI99} ch 9 Wrobel \& Walker), e.g. if the actual system temperature is 70 K, by the alternative definition  SEFD$_{\mathrm{other}}$ = 2 Jy for this example.}.  {\tt gencal} is used to calculate conversion factors per baseline, so the SEFD is squared, and the inverse is taken because CASA divides corrections into the data.  This is a simple example, but see the help for {\tt gencal} for how to provide different corrections for different antennas. 
\begin{verbatim}
SEFD=35
SEFDfactor = SEFD**-0.5
msin='mstoscale.ms' 
os.system('rm -rf '+msin+'.rescale')
        gencal(vis = msin,
               caltable = msin+'.rescale',
               parameter = SEFDfactor,
               caltype = 'amp')
\end{verbatim}
Apply {\tt mstoscale.ms.rescale} in applycal like any other gaintable.
 
 This technique can be used to scale visibility data for any reason, and {\tt gencal} offers many other options for manipulating visibility data.  Check first in case there is already an Analysis Utility to perform what you need (Appendix~\ref{ap-casaguides}).
 If that is insufficient, the CASA table ({\tt tb}) tool can be used to edit MS or calibration tables directly, see the CASA help pages or ask an expert.  

\subsection{Flux scale correction for very bright continuum sources}
\label{sec-avc}

Please note that as of 2021, a more accurate ALMA observing procedure was under development for known very bright continuum targets so check in case the correction below is not necessary for recent observations. 
ALMA measures $T_{\mathrm{sys}}$ in 128 channels per 2 GHz spw, and uses this to normalise the raw spectra, so there can be a contribution from lines, see Appendix~\ref{ap-other}.

The net  correlated signal 
is made up from contributions from the receiver system (and the rest of the instrumental signal path) $T_{\mathrm{Rx}}$, the sky (and other non-astronomical emission like spillover) $T_{\mathrm{sky}}$ and the sources in the field. For ALMA,  $T_{\mathrm{sys}}$ is usually measured in a direction slightly offset from the target field, close enough to correct for the same sky background, so the measured $T_{\mathrm{sys}}=T_{\mathrm{Rx}}+T_{\mathrm{sky}}$.
The ratio of the `true' on-source system temperature to the measured  $T_{\mathrm{sys}}$ is historically known as the AVC or Automatic Voltage Correction (assuming no frequency dependence within the measurements):

\begin{equation}
\label{eq_avc}
\mathrm{AVC} = \frac{T_{\mathrm{Rx}} + g S_{\mathrm{tar}} +
  T_{\mathrm{sky}}}{T_{\mathrm{sys}}}
\end{equation}
where $S_{\mathrm{tar}}$ is the source flux (total, as detected by each antenna, not per synthesised beam) and
$g$ is the conversion factor  in K Jy$^{-1}$ (the inverse of the SEFD, Sec.~\ref{tsystoJy}). 

Using ALMA  at Band 6 as an example, the (off-source)
$T_{\mathrm{sys}}$ is typically 100 K  and $g\sim$0.03   \citetalias{ALMA-TH}, so for continuum sources of less than a couple of hundred Jy, their contribution to $T_{\mathrm{sys}}$ is $\lesssim5$\%,  less than the uncertainty in the flux scale.  
In order to derive accurate flux scale corrections for bright sources, measurements of the correlator units before scaling, for both the target and a standard source are needed, but reasonable estimates can be made from a-priori information or images. 

The total flux density of many Solar System objects is  well-modelled. The Analysis Utilities function {\tt planetFlux}  (Appendix~\ref{ap-casaguides}) provides accurate flux densities from \citet{Butler2012}.  The continuum image total apparent flux density (including PB correction) can be measured and then scaled using {\tt immath} which will bring any minor contributions (e.g. spectral lines, moons in the field) onto a more accurate flux scale.  Alternatively you can scale the visibilities for compact objects (Sec.~\ref{tsystoJy}) but if emission extends outside e,g, 90\% of PB sensitivity, the visibility amplitudes are not PB corrected, so comparison with a model of the `true' total flux will not be accurate.

If the exact flux density is unknown but the apparent flux density $gS_{\mathrm{app}}$ (in K) is, roughly, less than half $T_{\mathrm{sys}}$ then, since  $S_{\mathrm{tar}}=\mathrm{AVC}S_{\mathrm{app}}$, Eq.~\ref{eq_avc} can be rearranged to give AVC $\sim T_{\mathrm{sys}}/(T_{\mathrm{sys}}-g S_{\mathrm{app}})$. Thus for a $S_{\mathrm{app}}$=1000 Jy source at Band 6, AVC =1.43 and $S_{\mathrm{tar}}\sim1428$. For relatively brighter sources this crude method progressively overestimates the `true' flux density.  

 Note that for Solar observations, system temperature measurements are made towards the Sun and a different procedure is used, see link to Sunspot CASA Guide, Appendix~\ref{ap-casaguides}.

At cm wavelengths,  $T_{\mathrm{sys}}$ is usually lower and $g$ is higher so many compact objects such as QSO contribute significantly and it is usually easier to make a calculation including the contribution of calibrators if you have access to the raw data. If the average  $T_{\mathrm{sys}}$ is measured across each observing band or spw, even very bright lines are unlikely to contribute significantly.
For example, the primary calibration source is 3C286,
flux density $S_{\mathrm{3C286}}$. The correlation coefficients for
3C286 and for a source of unknown flux density are
$\rho_{\mathrm{3C286}}$ and $\rho_{\mathrm{tar}}$.  If the target flux density is within a factor of a few of that of 3C286 (or the flux standard used), use an
approximate, linear estimate of the target source flux density,
 to estimate its AVC factor, and the factor for 3C286, using Eq.~\ref{eq_avc} and use this to scale the correlation coefficients to obtain a more accurate $S_{\mathrm{tar}}$. For very bright sources this can be done iteratively to improve  $S_{\mathrm{tar}}$ until the changes are negligible.
\begin{equation}
S_{\mathrm{tar}} = S_{\mathrm{3C286}}\frac{\rho_{\mathrm{tar}}}{\rho_{\mathrm{3C286}}}\frac{\mathrm{AVC_{tar}}}{\mathrm{AVC_{3C286}}}
\end{equation}

\subsection{Measuring image noise ({\tt imstat})}
\label{imstat}

It is easy to choose a peak or emission free regions by eye in a single image, but it is useful to script the procedure for multiple images or exact comparison during self-calibration. This example assumes a square, continuum image which has not been PB corrected, so the noise everywhere outside the sources should be similar. The rms noise is measured in four corners in the hope that one is emission free.
Change the precision in the print statement etc. as needed. This could be further modified to loop through planes of a cube, etc.
\begin{verbatim}
imin = 'continuum.image'
# get image size
i=imhead(imagename=imin,mode='get',hdkey='crpix1')
# all except edges to find peak
maxreg='box[['+str(int(i*0.05))+'pix,'+str(int(i*0.05))+'pix],\
            ['+str(int(i*0.95))+'pix,'+str(int(i*0.95))+'pix]]'
# four corners to find noise
noiseregs=['box[['+str(int(i*0.05))+'pix,'+str(int(i*0.05))+'pix],\
                ['+str(int(i*0.25))+'pix,'+str(int(i*0.25))+'pix]]',
           'box[['+str(int(i*0.05))+'pix,'+str(int(i*0.75))+'pix],\
                ['+str(int(i*0.25))+'pix,'+str(int(i*0.95))+'pix]]',
           'box[['+str(int(i*0.75))+'pix,'+str(int(i*0.05))+'pix],\
                ['+str(int(i*0.95))+'pix,'+str(int(i*0.25))+'pix]]',
           'box[['+str(int(i*0.75))+'pix,'+str(int(i*0.75))+'pix],\
                ['+str(int(i*0.95))+'pix,'+str(int(i*0.95))+'pix]]']
immax=imstat(imagename=imin, region=maxreg)['max'][0]
noises=[]
for n in range(len(noiseregs)):
    noises.append(imstat(imagename=imin, region=noiseregs[n])['rms'][0])
imrms=min(noises)
print(imin+' peak %6.3f Jy/beam, rms %5.3f Jy, S/N %6.1f' % (immax, imrms, (immax/imrms)))
# which returns
> continuum.image peak 23.321 Jy/beam, rms 0.233 Jy, S/N    100.0
\end{verbatim} 

\subsection{Using AIPS images as models}
\label{ap-CC}
Cleaned images made using {\sc AIPS} and exported as FITS may be used to provide models for self-calibration, etc.  These contain a list of clean components (CC) as an extension table, which must be converted to a CASA {\tt .model}. In future CASA releases a utility {\tt aipsCC2model} will be available, or use {\tt wget http://almanas.jb.man.ac.uk/amsr/aipsCC2model.zip} to obtain scripts for CASA 5 or 6 (courtesy of K. Golap, NRAO).

\section{ALMA smearing limits}
\vspace*{-0.2cm}
\label{ap-smear}
\begin{figure}[!h]
  \includegraphics[width=8cm]{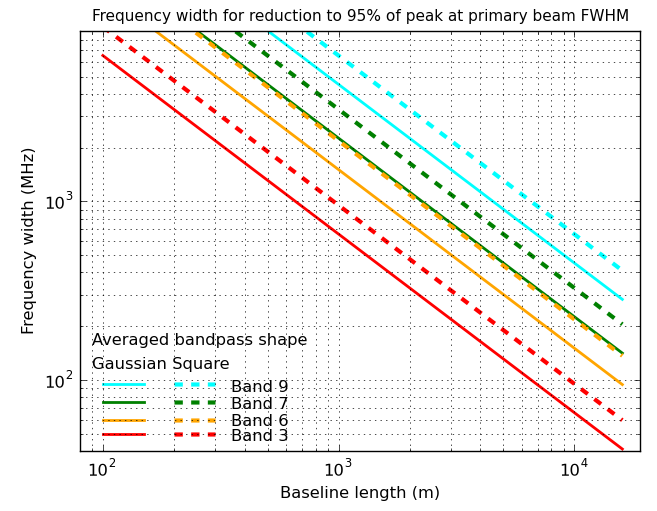}
  \includegraphics[width=8cm]{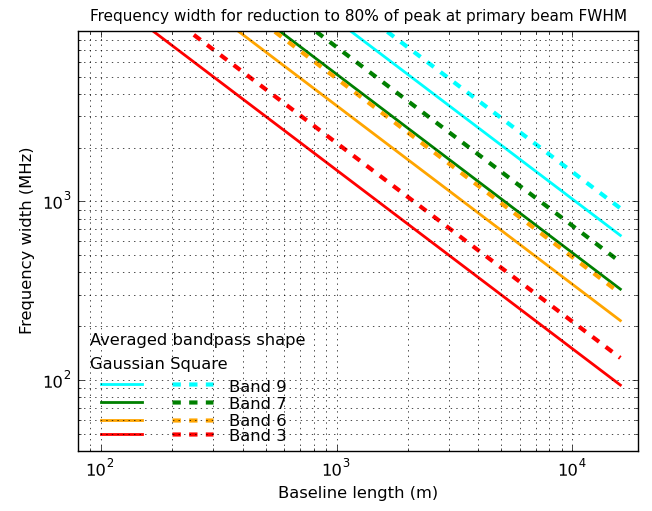}
   \includegraphics[width=8cm]{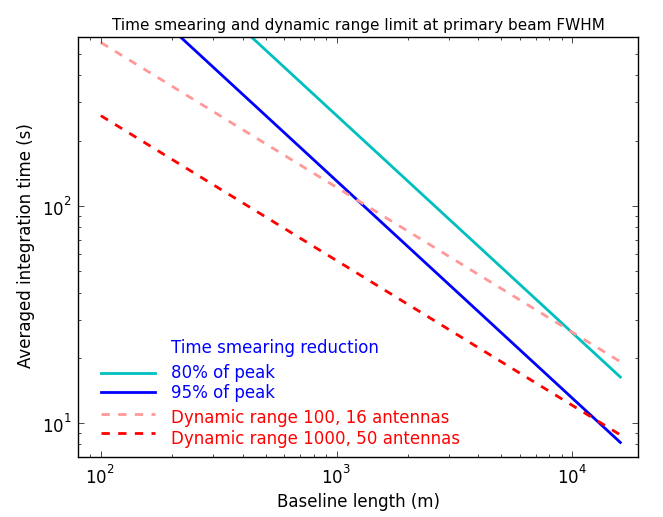}
  \hspace*{1cm}
    \includegraphics[width=8cm]{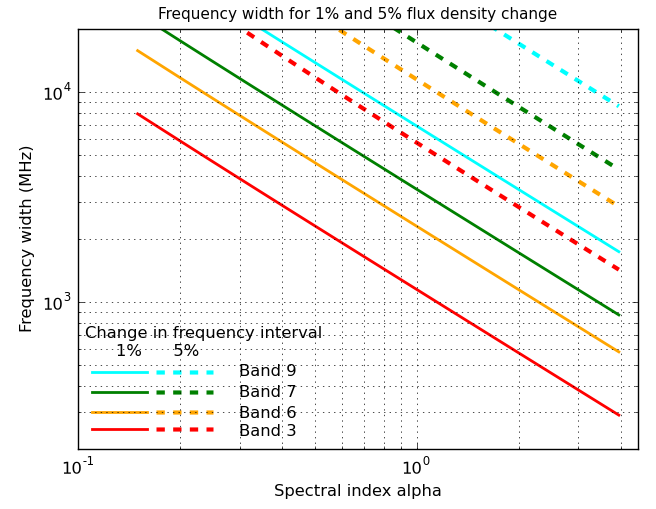}
    \caption{\small Approximate limits to averaging intervals to prevent smearing of sources at the 50\% sensitivity point in wide field imaging (single pointing), and potential for spectral index errors (x-axis shows absolute value of $\alpha$).  The dynamic range and spectral index limits will only be apparent for bright, well-calibrated sources.}
    \end{figure}
\end{appendix}
\end{document}